\documentclass{aa}  
\usepackage{graphicx}
\usepackage[dvipsnames]{xcolor}
\usepackage{txfonts}
\usepackage{xfrac}
\usepackage[colorlinks=true,linkcolor=blue,citecolor=blue,urlcolor=blue]{hyperref}
 \usepackage{natbib}
\usepackage{amsmath}
\usepackage{amssymb}	% Extra maths symbols
\usepackage[flushleft]{threeparttable}
\usepackage{rotating}
\usepackage{enumitem}
\usepackage{enumerate}
\usepackage{xspace}
\usepackage{svg}
\setlist{leftmargin=0.4cm}

\newcommand{\orcid}[1]{\href{https://orcid.org/#1}{\includesvg[width=10pt]{orcid}}}

\begin{document} 

   \title{The Next Generation Fornax Survey (NGFS).VIII. A Support Vector Machine Approach for Disentangling Globular Clusters\\ from other Sources}\titlerunning{SVM Approach for Disentangling Globular Clusters from other Sources}

   \author{Yasna~Ordenes-Brice\~no
          \inst{1}\orcid{0000-0001-7966-7606}
          \and
          Thomas~H.~Puzia\inst{2}\orcid{0000-0003-0350-7061}
          \and
          Paul~Eigenthaler\inst{3}\orcid{0000-0001-8654-0101}
          \and
          Matias~Bla\~na\inst{4}\orcid{0000-0003-2139-0944}
          \and
          Juan~P.~Carvajal\inst{2}\orcid{0000-0001-6584-7104}
          \and
          Matthew~A.~Taylor\inst{5}\orcid{0000-0003-3009-4928}
          \and
          Bryan~W.~Miller\inst{6}\orcid{0000-0002-5665-376X}
          \and
          Rohan~Rahatgaonkar\inst{2}\orcid{0000-0002-5350-0282}
          \and
          Evelyn~J.~Johnston\inst{1}\orcid{0000-0002-2368-6469}
          \and
          Prasanta~K.~Nayak\inst{2}\orcid{0000-0002-4638-1035}
          \and
          Gaspar~Galaz\inst{2}\orcid{0000-0002-8835-0739}
          }\authorrunning{Y. Ordenes-Briceño et al.}
          
          \institute{Instituto de Estudios Astrof\'isicos, Facultad de Ingeniería y Ciencias, Universidad Diego Portales, Av. Ejército Libertador 441, Santiago, Chile.\\
              \email{yasna.ordenes@mail.udp.cl}
          \and 
          Instituto de Astrof\'isica, Pontificia Universidad Cat\'olica de Chile, Av. Vicu\~na Mackenna 4860, Santiago, 7820436, Chile.
          \and
          Instituto de Astrof\'isica, Universidad Andres Bello, Fernandez Concha 700, Las Condes, Santiago, Chile.
          \and
          Vicerrector\'ia de Investigaci\'on y Postgrado, Universidad de La Serena, La Serena 1700000, Chile.
          \and 
          University of Calgary, 2500 University Drive NW, Calgary, Alberta, T2N 1N4, Canada.
          \and
          International Gemini Observatory/NSF NOIRLab, Casilla 603, La Serena, Chile. 
          }

   \date{Received July 17, 2025; accepted \today}

% \abstract{}{}{}{}{} 
% 5 {} token are mandatory
 
\abstract
{Wide‑field, multi‑band surveys now detect millions of unresolved sources in nearby galaxy clusters, yet separating globular clusters (GCs) from foreground stars and background galaxies remains challenging. Scalable, automated classification is therefore essential to convert the forthcoming data from facilities such as the Vera C.~Rubin Observatory’s Legacy Survey of Space and Time (LSST) and the Nancy Grace Roman and Euclid space telescopes into robust constraints on galaxy assembly and dark‑matter halos.}
{We introduce a supervised classification method to separate GCs, stars, and galaxies based on their locations in color–color diagrams. The main objective is to recover a clean GC sample for future scientific analysis. The method exploits broad spectral energy distribution (SED) coverage, deep photometry, and is optimized for next-generation survey volumes.}
{We use the central 3 deg$^2$ of the Next Generation Fornax Survey (NGFS), which images the Fornax cluster in $u'g'i'$ (BLANCO/DECam) and $JK_s$ (VISTA/VIRCAM). We build a Support Vector Machine (SVM; \texttt{svm.SVC}, \texttt{scikit-learn}) using 15 features: all color combinations of $u'g'i'JK_s$ and basic morphological parameters (e.g. FWHM, ellipticity). Spectroscopically confirmed sources define the training classes.}
{Color pairs connecting near-UV/optical to near-IR - particularly ($u'\!-\!g'$) vs. ($g'\!-\!K_s$) - yield the strongest discrimination between classes. The full 15 feature model achieves 97.3\% accuracy and a pruned 7 feature model built from the most informative, least correlated features achieves 96.6\% accuracy. Misclassifications amount 8.4\% and 10.4\%, respectively. Omitting the $u'$ or/and near-IR bands degrades performance. Emulating LSST filters with NGFS $u'g'i'$ and Dark Energy Survey (DES) $r'z'Y$ shows that $u'$ and $Y$ bands are crucial, but models lacking NIR remain suboptimal. The final 7 feature $u'g'i'JK_s$ classifier will underpin forthcoming NGFS GC studies.}
{Combining broad SED coverage with simple morphological parameters enables precise, scalable separation of unresolved sources. Including NIR bands significantly improves GC classification, and joining LSST with forthcoming Euclid and Roman data will further enhance machine-learning frameworks for extragalactic surveys.}


\maketitle

%%%%%%%%%%%%%%%%%%%%%%%%%%%%%%%%%%%%%%%%%%%%%%%%%%%%%%%%%%%%%%%%%%%
\section{Introduction}
\label{sec:intro}

Globular clusters (GCs) occupy a singular niche in astrophysics: they are among the oldest {\it and} simplest stellar systems in the Universe \citep[e.g.][]{Vandenberg1996, Willman2012}. They encode the earliest star formation and assembly events of their host galaxies \citep[e.g.][]{ashman92, brodie2006, chen2025}. Systematic studies over the past decades have shown that essentially every galaxy more massive than $\sim\!10^{8}\,\mathrm{M_\odot}$ (stellar mass) hosts a GC system whose chemical compositions, kinematics, sizes, and spatial distribution mirror the host's evolution and merger history \citep[e.g.][]{Forbes97, Puzia2005, Puzia2006, Peng2006, chiessantos2022}. Because GC colors, age - metallicity scaling relations, and specific frequencies correlate tightly with host halo mass \citep[e.g.][]{Georgiev2010, harris2013, forbes2018}, they have become widely used tracers of dark matter assembly on galactic and cluster scales \citep{Cooper2025, chen2025}. In dense environments such as galaxy clusters, GCs also contribute to understanding environmental effects on galaxy evolution \citep[e.g.][]{Smith2015, Lim2024} and the intracluster light (ICL) formation \citep[e.g.][]{peng2011, alamo2013, madrid2018, kluge2025}. 

Accurate identification of GCs serves as an additional means to comprehend the formation and evolution of galaxies. Deep imaging from state-of-the-art ground-based and space-based observatories has significantly improved the photometric characterization of GCs. Despite the depth and quality of modern imaging, separating GCs from stars and background galaxies remains challenging due to overlapping photometric and morphological properties \citep[e.g.][]{puzia2014}. Traditional selection criteria based on color cuts or structural parameters fail to disentangle these various stellar systems properly, producing contamination fractions of $30$--$70\,$\% in purely optical photometric samples \citep[e.g.][]{powalka2016}, unless sample statistics is overwhelmingly in favor of GCs, e.g. in brightest cluster galaxies \citep{Harris2024}. Decontamination becomes particularly challenging in the central regions of galaxy clusters or in the vicinity of their massive and regular galaxies, where their light profile affects the detection and photometry of stars (foreground), GCs (cluster) and galaxies (background), which can significantly reduced a GC catalog purity, complicating the interpretation of GC samples \citep[e.g.][]{durrell2014,Lim2025}.

\cite{Munoz2014} have shown that by incorporating a broader wavelength baseline, particularly including the $u'$ and $K_s$-band, the separation between GCs and other sources becomes clearer, as these bands enhance the contrast in the spectral energy distribution (SED) properties between the stellar systems.
The scientific return from extragalactic GC studies has grown in lock‐step with advances in wide‐field, optical/near‐IR imaging \citep[e.g.][]{Munoz2014, Taylor2017}. Surveys such as the Next Generation Virgo Survey \citep[NGVS,][]{Ferrarese2012}, the Next Generation Fornax Survey \citep[NGFS,][]{Munoz2015,eigethaler2018, Ordenes2018}, PHANGS--\textit{HST} \citep[e.g.][]{Maschmann2024}, and, more recently, deep \textit{Euclid} pilot programs now detect thousands of GC candidates per pointing \citep{Saifollahi2025}, probing to larger galactocentric radii and lower surface‐brightness regimes than ever before. Spectroscopic studies are infeasible for the millions of candidates predicted by modern cosmological simulations \citep[e.g.~E--MOSAICS, see][]{Pfeffer2018} and for the 350,000 estimated GCs in the \textit{Euclid} footprint \citep{Euclid2025} and $\gtrsim 4\times 10^{6}$ GCs forecast to be visible in the Vera C. Rubin--LSST imaging \citep{Ivezic2019,usher2023}.

These data volumes demand scalable, fully automated classification pipelines. Machine-learning (ML) methods have already demonstrated superior performance over traditional approaches in many domains of astronomy \citep[e.g.][]{Angora2019, Saifollahi2021, Barbisan2022, chiessantos2022, Ting2025}. Among "classical" algorithms, the support vector machine (SVM) remains attractive because it yields interpretable decision boundaries \citep{Cortes1995, platt99a,crammer2001}, is robust against the "curse of dimensionality" \citep[e.g.][]{Joachims1998, Guyon2002}, and can be tuned efficiently with modest training sets \citep{Chapelle2002}. SVMs have been applied successfully for classification to galaxy morphology \citep[e.g.][]{huertascompany2008, vavilova2021}, stellar objects and transients \citep{Li2025}, spectral line classification \citep{shi2015} and for identifying active galactic nuclei (AGNs), galaxies, and stars, with $\lesssim\!5\,$\% cross-contamination \citep{Malek2013,Cenarro2019,Wang2022a}.

In this work we advance these efforts by developing a supervised SVM pipeline using the NGFS's ultra-deep, broad spectral baseline imaging - spanning $u'g'i'JK_s$ and morphological parameters, to classify point‐like sources in GC, star, and galaxy categories. We use Python with its \texttt{sklearn} library \footnote{https://scikit-learn.org/stable/} \citep{pedregosa2011}. Our SVM classifier is immediately applicable to forthcoming data releases from Vera C. Rubin--LSST, \textit{Euclid}, and the Nancy Grace Roman Space Telescope, where rapid and reliable GC identification will be critical for studies of galaxy assembly, stellar‐population gradients, and GC formation efficiencies across diverse environments. 

The structure of this paper is organized as follows. Section~\ref{sect:data} summarizes the sNGFS data and catalog construction. Deep color–color diagrams (cc-diagrams) are presented in Section~\ref{sect:deepCC}. Section~\ref{sec:met} describes the SVM methodology, feature selection and hyper-parameter optimization. Results for the full and reduced filter sets are given in Section \ref{sect:performance_results}. In Section~\ref{sect:lsst} we assess expected performance for LSST-like photometry, and Section~\ref{sect:con} summarizes our conclusions and outlines future applications.

%%%%%%%%%%%%%%%%%%%%%%%%%%%%%%%%%%%%%%%%%%%%%%%%%%%%%%%%%%%%%%%%%%%%%%%%%%%%%%%%%%%%
\section{Data}
\label{sect:data}
This work makes use of data from the NGFS, a deep, multi-wavelength survey of the Fornax galaxy cluster that extends out to a projected radius of 1.4 Mpc, which enclosed a total cluster mass of $7 \times 10^{13}\,M_\odot$ \citep{drinkwater2001}. Optical photometry was obtained using the Blanco 4-meter telescope equipped with the Dark Energy Camera (DECam; \citealt{Flaugher15}), covering a total of 19 tiles ($\simeq 57\ deg^2$) complete in three bands: $u'$, $g'$, and $i'$, where one DECam tile is $2.2\ deg$ Field of View (FoV). Additionally, NGFS includes a near-infrared component observed with the VISTA telescope using the VISTA InfraRed CAMera (VIRCam; \citealt{Sutherland15}, now decommissioned), providing $J$ and $K_s$ band data over 12 cental tiles ($\simeq 20\ deg^2$), with a VIRCam tile FoV of $1.6\ deg$. The specific preparation of the NGFS optical dataset in $u'g'i'$ for the 19 tiles will be presented in a dedicated paper (NGFS et al. {\it in prep.}), which includes a detailed description of data reduction, photometry calibration, completeness tests, final photometry, among others. 

\begin{figure*}[ht]
    %trim=left bottom right top
     \centering
     \includegraphics[trim=0.1cm 0.1cm 0.1cm 0.1cm,clip,width=0.9\linewidth]{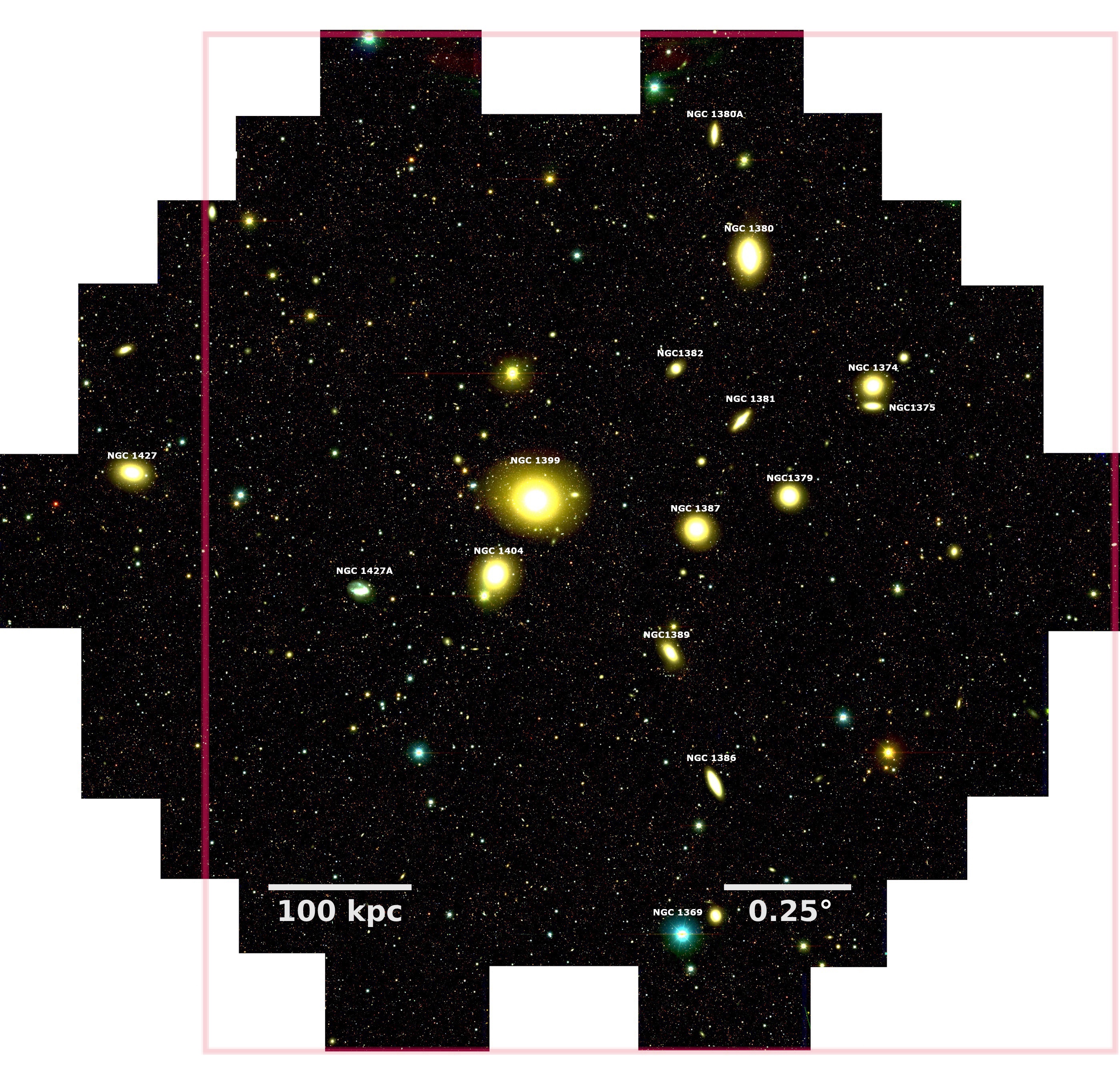}
     \caption{RGB composite image of NGFS Tile 1, constructed using DECam filters ($i'$ in red, $g'$ in green, and $u'$ in blue). The field of view corresponds to a single DECam tile, with a radius of $1.1^\circ \approx 370$ kpc at the distance of the Fornax cluster, D = 19.3 Mpc \citep{Anand2024}. The NIR imaging FoV is shown with a red unfilled rectangle, see Sect \ref{sect:data} for details. The names of the main galaxies are shown in the image, with the cD galaxy, NGC 1399, located near the image center. Angular and physical scales are indicated: a white line at the bottom right represents $0.25^\circ$, and the line at the bottom left corresponds to 100 kpc.}
     \label{fig:t1_rgb}
\end{figure*}

This work focuses on the central region of the Fornax Cluster, from now on called NGFS-T1, composed of T1 DECam ($u'g'i'$) and T1,T2 VIRCam ($JK_s$), their pixel scale in arcseconds is 0.263 and 0.339, respectively. The data reduction pipeline for NGFS-T1 is described in \cite{ordenes2018_thesis} PhD thesis, but we include a brief description about the data reduction and photometric calibration in the Appendix Section \ref{sect:A1_phot_cal}. The VIRCam central tile is composed of two tiles, however the output science image is one single stack image for each band ($1.6\deg\ \times \ 2.2\deg$), see Figure \ref{fig:t1_rgb} where the FoV is shown with a superposed red rectangle. We have revised the photometric calibration of the complete survey, see Figure \ref{fig:stars_calib}. This includes the five filter scientific images of NGFS-T1, spanning from near-UV to near-IR: $u'g'i'JK_s$. For instance GC candidates at the Fornax distance D = 19.3 Mpc \citep{Anand2024} are seen as unresolved sources due to the spatial resolution limitation of DECam ($1\,{\rm pix}=24.46$\,pc) and VIRCam ($1\,{\rm pix}=31.62$\,pc) instruments.

The central region of the Fornax cluster has a high galaxy density, and the extended surface-brightness profiles of these galaxies hinder source detection, with many faint GCs being obscured by their diffuse light. We implement a Point like source detection image, the procedure is explained in Section \ref{sec:point_source_detection} and see Figure \ref{fig:GC_detection} illustrating the result. Photometry was performed with \texttt{Source Extractor} \citep[\texttt{SExtractor}][]{Bertin96}, using the detection catalog as a prior to obtain the cleanest possible photometry for sources projected behind the galaxies. We constructed a Point Spread Function (PSF) model with \texttt{PSF Extractor} \citep[\texttt{PSFex}][]{Bertin2011}, which accounts for PSF variations across the detectors. The final photometry catalogs have magnitudes ($PSF$, $APER$, $AUTO$) in the optical pass-bands in the AB system and the NIR magnitudes were transformed from the Vega to the AB system using $K_s(m_{\rm AB}\!-\!m_{\rm Vega})\!=\!1.85$\,mag and $J(m_{\rm AB}\!-\!m_{\rm Vega})\!=\!0.91$\,mag \citep{Blan07}. Magnitudes are corrected by galactic extinction towards the Fornax galaxy cluster: $A_{u'}$=0.054, $A_{g'}$=0.041, $A_{i'}$=0.020, $A_J$=0.009, $A_{K_s}$=0.004 \citep{Fitz99,Schl98,Schf11}, and shown in color-magnitud diagrams (CMDs) and color-color diagrams (cc-diagrams), using the $\_ 0$ subscript for both magnitudes and colors.

%Figure 1aq ssssweedr

The primary master catalog for NGFS-T1 consists of PSF photometry with complete SED cross-matching in the $u'$, $g'$, $i'$, $J$, and $K_s$ bands, containing a total of 65,581 sources. To ensure photometric quality, we use the \texttt{SExtractor} output parameter called $FLAGS$ to select sources with no issues for extracting the photometry ($FLAGS = 0$), which guards against the presence of bad-pixels, close neighbors, deblended or saturated sources (i.e. $FLAGS>0$). After applying our selection to all filters, the final sample is composed of 62,416 sources. 

For this article, a completeness analysis is not required, but our depth is defined by the limits of $u'$ and $K_s$ bands, which are the shallowest in the NGFS observations.~The faintest objects in the 5 filter matched catalog with magnitudes of $u'=28.05$ with a mean of $u'=23.96$, ${g'}=26.46$ with a mean of $g'=23.17$, ${i'}=25.37$ with a mean of $i'=21.92$, $J=24.9$ with a mean of $J=21.23$, $K_s=23.63$ with a mean of $K_s=21.23$, all in the AB magnitude system. The completeness analysis will be included in the follow-up article of this work.

For the subsequent analysis, we construct colors using PSF magnitudes in all filters. We adopt colors instead of magnitudes because the latter are distance-dependent, while colors remain unaffected by distance and are thus more robust and appropriate for relative comparisons and source classification. For this specific work, we used the following parameters of the master catalog: Coordinates, colors, Full Width at Half Maximum (FWHM), Flux Radius (FR), SPREAD\_MODEL (SM), ellipticity ($e$) and concentration index parameter ($C_{\lambda}$). 

The spread model parameter is an output of the photometry with \texttt{SExtractor}, when using a PSF model created with \texttt{PSFex}.~It compares the profile of the source with the PSF model, thus describing whether the source is more like a point source or an extended source. Other structural parameters in the catalog are: flux radius (FR) at 50\% as the radius (in pixels) containing 50\% of the total flux (brightness) of the source, FWHM as the width of the source brightness profile at half its maximum intensity, ellipticity $e$ = $1 - a/b$, with $a$ as the semi-major axis and $b$ as the semi-minor axis. An extra parameter is the concentration index $C_{\lambda}$, estimated as $C_{\lambda}$= MAG\_APER (2pix) -  MAG\_APER (8pix) \citep{powalka2016}. The $C_{\lambda}$ parameter indicates that, for a small difference, light is mostly in the PSF core, i.e. compact source (e.g. star), and for a large difference, light extends outward, i.e. the source is more diffuse (e.g. galaxy). We use these parameters in the $i'$-band photometry, as it has the best image quality compared to $u'$, $g'$, $J$ and $K_s$ bands, which includes seeing and depth. The average FWHM for point sources in the $u'-$, $g'-$ and $i'-$ bands is 6.4 pixels (1.68"), 4.2 pixels (1.08") and 3.6 pixels (0.95"), respectively.

NGFS-T1 cover a radius of $r = 1.1^\circ \simeq$ 370 kpc in the DECam FoV and $r = 0.8^\circ \simeq$ 268 kpc in the VIRCam FoV, which are centered on the giant elliptical or central dominant (cD) galaxy NGC 1399 (see Fig. \ref{fig:t1_rgb}). This is the ideal tile to contrast and test this ML method, because of the high density of galaxies and GCs, thus making it more challenging to separate the different stellar systems.

%%%%%%%%%%%%%%%%%%%%%%
\section{Deep color-color diagrams}
\label{sect:deepCC}
\begin{figure*}[htbp]
    %trim=left bottom right top
     \centering
     \includegraphics[trim=4.6cm 2.3cm 4.6cm 2cm,clip,width=\textwidth]{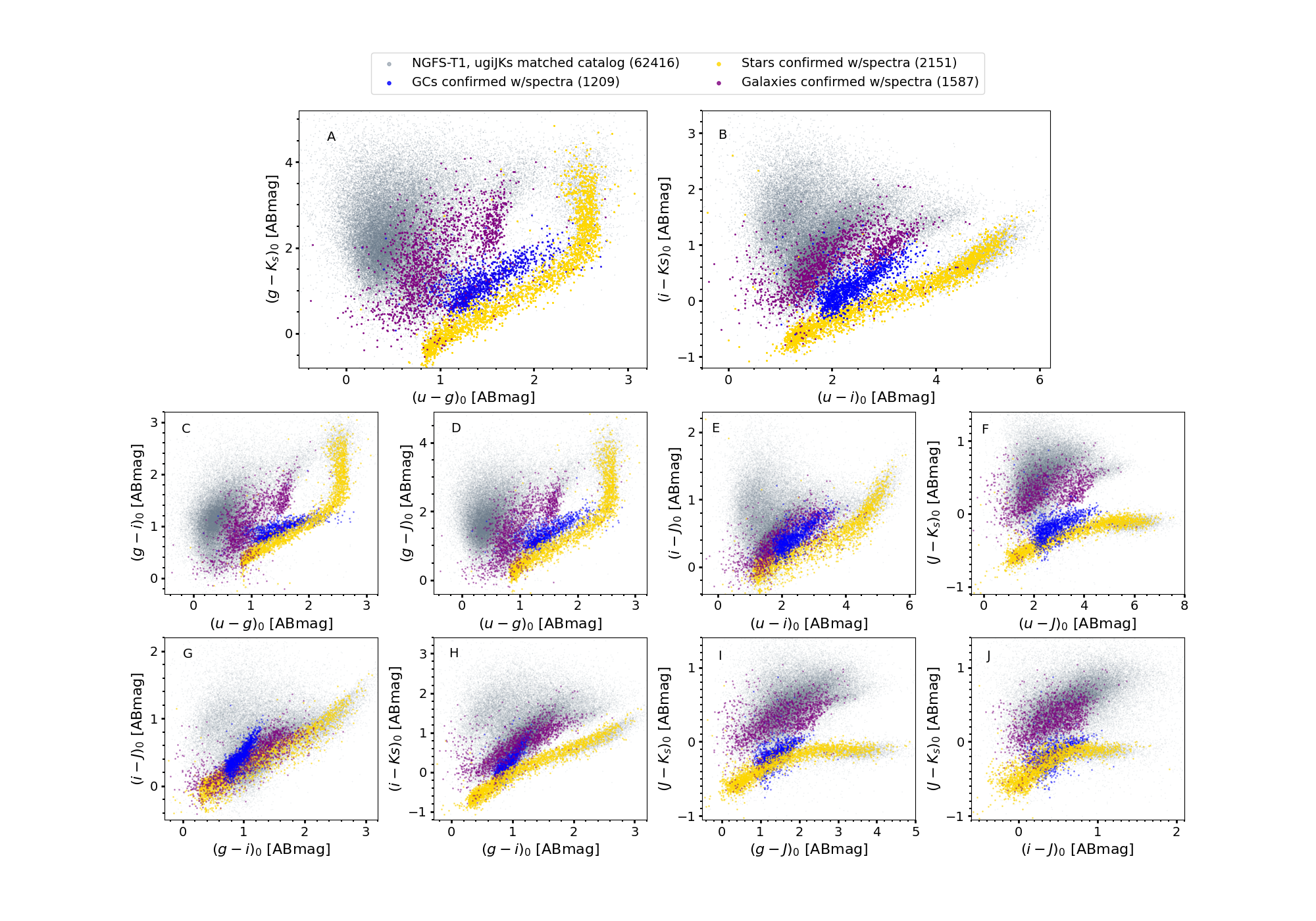}
     \caption{Color-color diagrams for all sources with multi-wavelength photometry in the core region of the Fornax galaxy cluster, shown as gray dots. Spectroscopically confirmed samples are shown for GCs (blue), stars (golden) and galaxies (purple), which are used as labeled samples for the \texttt{svm.SVC} model (see Section \ref{sect:traindataset}).~Note that the different diagrams present the same source sample for which photometric information was obtained in the master cross-matched catalog between $u'g'i'JK_s$ filters.}
     \label{fig:f_cc_ugiJKs}
\end{figure*}

From the NGFS-T1 master catalog with PSF photometry in the $u'$, $g'$, $i'$, $J$, and $K_s$ bands, we construct 10 cc-diagrams using different combinations of filters, always ordered from blue to red wavelengths (see Fig.~\ref{fig:f_cc_ugiJKs}). Each panel shows the NGFS-T1 sources as grey dots, while spectroscopically confirmed GCs, stars and galaxies are shown in blue, gold and purple colors, respectively. Further details about these confirmed sample are provided in Section~\ref{sect:traindataset}.

%Figure 2

Figure~\ref{fig:f_cc_ugiJKs}, top-left panel (panel A) shows the $(u'- g')$ vs.$(g' - K_s)$ diagram, hereafter referred to as $u'g'K_s$, and this naming convention is also applied to other filter combinations. These cc-diagrams demonstrate the diagnostic utility of combining near-UV and near-infrared filters to distinguish between different stellar populations and object types in deep, wide-field imaging.

In diagrams such as $u'g'K_s$ (panel A), $u'i'K_s$ (panel B), and $u'JK_s$ (panel F), four main populations can be identified: background galaxies at various redshifts, passive early-type galaxies, Fornax cluster GCs, and foreground Milky Way stars \citep[see also][]{Munoz2014}. In contrast, these groups are less clearly separated in diagrams lacking the $u'$ band or where the filter wavelengths are more closely spaced, e.g., in the third row, from $g'i'J$ to the rightmost panels (panels G to J). In the $u'g'i'$ diagram (panel C), only three distinct sequences are clearly visible. Here, the GC sequence overlaps with the bluer region of the foreground stellar sequence. However, this diagram still retains diagnostic power to help distinguish between unresolved point sources and more extended objects (see also Fig. \ref{fig:f_cc_wModels}).

Choosing the right color combinations - for example, the $u'i'K_s$ diagram \citep{Munoz2014} - maximizes the separation between GCs and stars-galaxies in color-color space. An interesting point is that, when one single color-color plane is used to select a sample, this could still generate a considerable amount of contamination. For example, the work of \cite{Gonzalez2017} used the $u'i'K_s$ diagram to select GCs around NGC 4258, a spiral galaxy, with 39 GCs identified as candidates. After completeness correction, GC luminosity function extrapolation, correction for spatial coverage and assuming 5\% contamination, they calculate a total of $N_{GC}=144\pm{31}_{-36}^{+38}$ (random and systematic uncertainties, respectively). In an spectroscopic follow-up of 23 GC candidates \citep{Gonzalez2019}, 70\% were confirmed as GCs and the other 30\% were contamination, thus correcting the total number of GCs $N_{GC}=105\pm{26}$ (only random uncertainties), which is still a high \% of contamination.

To further minimize uncertainties in the selection of candidates, we expand the feature space (e.g., by including additional colors) and implement a supervised machine learning algorithm to provide a data-driven approach for modeling complex, non-linear decision boundaries in a multidimensional feature space. We include a summary of other ML techniques applied for similar objectives in Section \ref{sect:ML_literature}. The strategy chosen for this work enhances the statistical robustness of the population separation by reducing subjective biases inherent in manual or threshold-based classification methods. The following section presents the methodological details.

%%%%%%%%%%%%%%%%%%%%%%
\section{Method: Support Vector Machine}
\label{sec:met}
 
Figure \ref{fig:f_cc_ugiJKs} shows the complexity of disentangling different stellar systems from deep cc-diagrams. Although NUV-optical-NIR filter combinations are very helpful, there is still much superposition of sources in some areas, making it difficult to select one particular population and being certain of having small contamination. Therefore, our classification strategy should be able to manage the available inputs to find the most suitable solution for our dataset to select three astrophysical sources: GCs, foreground stars and background galaxies.

\subsection{Support Vector Machine algorithm}
\label{sect:svm}
SVM is a machine learning method mainly used for classification, but it can also be applied to detect outliers and perform regression (predicting values). In this work we use SVM with Support Vector classification, \texttt{sklearn.svm.SVC} from \texttt{sklearn} \citep{pedregosa2011}, hereafter \texttt{svm.SVC}. In simple terms, \texttt{svm.SVC} works by finding the best decision boundary (called a hyperplane) that separates different groups (classes) of data points based on their characteristics (features). The goal is to find the boundary that maximizes the margin — the widest possible gap between the groups. Only the data points closest to this boundary, known as support vectors (x and x' in the kernel equations), are used to define it. These support vectors are critical to the model, as they determine the position and orientation of the hyperplane and ultimately influence how new data points are classified \citep{platt99a,Platt1999b,crammer2001,pedregosa2011}.

In cases where the classes are not perfectly separable due to overlapping distributions or the presence of noise in the data, the SVM employs the soft margin technique \citep{crammer2001}. This approach introduces a degree of tolerance for misclassified samples while still aiming to maximize the margin between classes. In doing so, the model achieves a balance between model complexity and classification accuracy, enhancing the robustness to outliers.

Thus, SVM benefits from projecting the input data into a higher dimensional space where the classes could be linearly separable, through the use of kernel functions. This kernel-based mapping allows SVM to effectively capture complex class boundaries. The mathematical formulations of the kernel functions commonly used in SVM are:
\begin{align}
&\text{Linear:}      & f & =\langle x , x' \rangle \label{eq:kernlin} \vspace{-0.1cm}\\
&\text{Polynomial:}  & f & =\left(\gamma\langle x , x' \rangle + r \right)^d \label{eq:kernpol}\\
&\text{Radial basis function (RBF):} & f & = \exp\left(-\gamma || x , x' ||^2\right)\label{eq:rbf}\\
&\text{Sigmoid:\hspace{1.8cm}} & f & =\tanh\left(\gamma\langle x , x' \rangle + r \right)\label{eq:kernsig}
\end{align}
Where x and x' are input vectors, $r$ is a constant, $d$ is the degree parameter in the polynomial kernel. The parameter $\gamma$ controls the influence of individual training examples: a small $\gamma$ implies a larger influence region (smoother decision boundary), while a large $\gamma$ implies a narrower influence region, which can lead to overfitting by capturing noise in the data.

The \texttt{svm.SVC} is strongly influenced by the regularization parameter $C$, although it is not explicitly shown in the kernel equations. This parameter controls the trade-off between achieving a low training error and a large margin, which is fundamental to the soft margin SVM approach. For example, a small $C$ value seeks to find a larger margin hyperplane, even if that leads to more misclassifications. On the other hand, a larger $C$ value forces the model to fit the training data more strictly to minimize training errors, but potentially leading to overfitting, which is more likely if there is noise or outliers. Therefore, $C$ acts as a penalty parameter for misclassified points. It modifies the optimization objective by adding a penalty term for margin violations. 
For weighting a specific group or class, we use the \texttt{class\_weight} parameter. Specifically, we use \texttt{class\_weight}='balanced', for adjusting the penalty applied to each class based on their frequency in the training set.

%%%%%%%%%%%%%%%%%%%%%%%%%%%%%%%%%%%%%%%%%
\subsection{Classes, Training \& Test dataset}
\label{sect:traindataset}

For the training and testing of SVM models, we require a confirmed sample for each class, which we call the labeled sample, as they are spectroscopically confirmed sources. In the following, we give a description of where the labeled sample comes from.

To compile the sample, we used two reference studies to identify all radial velocity confirmed objects within the area covered by NGFS-T1: \citet{chaturvedi22}, which focuses on the GC population within 0.7 Mpc of NGC 1399, and \citet{maddox2019}, which provides a catalog of all sources in the Fornax region (out to 1.4 Mpc, encompassing the main Fornax cluster and the Fornax A subgroup), including stars, GCs, and both cluster and background galaxies. The above studies enclosed the following individual spectroscopic catalogs: \cite{ferguson1989,Kissler-Patig1999, hilker1999, hilker2007, drinkwater2000,drinkwater2001,mieske2002,mieske2004,mieske2008,bergond2007,gregg2009,schuberth2010,chilingarian2011,pota2018,fahrion2020}.

\begin{figure}[htbp]
    %trim=left bottom right top
     \centering
     \includegraphics[trim=1.6cm 0cm 0cm 0.6cm,clip,width=\linewidth]{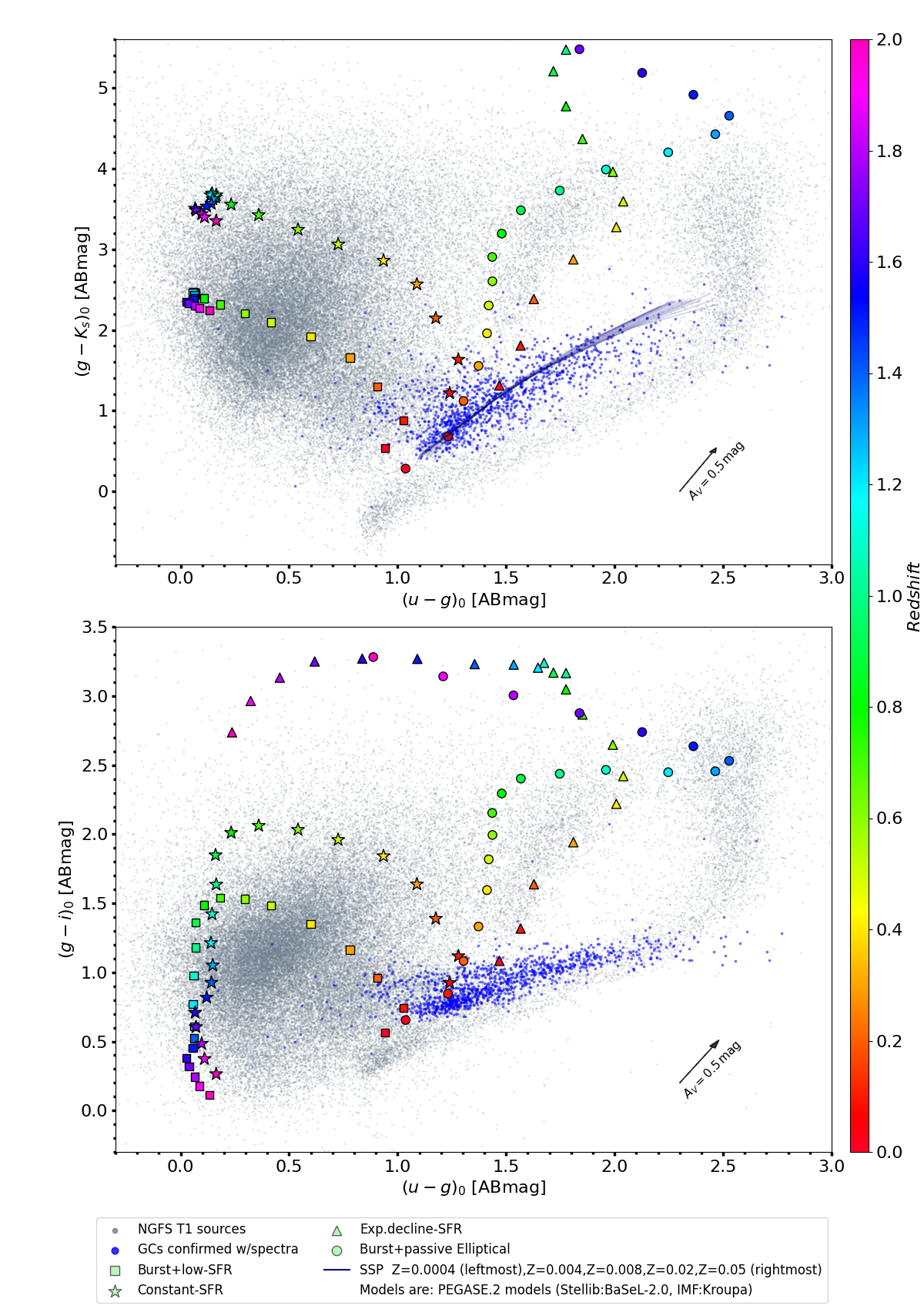}
     \caption{Color-color diagram $u'g'K_s$ (top-panel) and $u'g'i'$ (bottom-panel) with PEGASE.2 population synthesis models \citep{fioc97}. In addition to the same layout as in Figure \ref{fig:f_cc_ugiJKs}, the finely blue lines in the top-panel cc-diagram represent sequences of old single-age stellar populations (at redshift zero), the metallicity range is indicated in the legend. The large colored symbols represent the evolutionary paths of observed colors for four galaxies formed at redshift 3, with different star formation histories: burst+low SFR (squares), constant SFR (stars), exponential decline SFR (triangle) and burst+passive Elliptical (circle).}
     \label{fig:f_cc_wModels}
\end{figure}

We do not apply any parameter cut to the final catalog of RV-confirmed objects. However, when cross-matching the photometric and spectroscopic catalogs, some confusion may arise due to projection effects in the images. For instance, when sources have nearby companions or exhibit extended morphologies, leading to less reliable PSF photometry. This confusion can manifest in the color–color diagram as specific objects labeled as GCs appearing outside the typical GC locus, instead falling within the galaxy region. When applying this methodology to large survey datasets, even in the presence of contaminant inputs in the SVM model, the script must remain as automated as possible, minimizing manual intervention. Taking that into account, we divide the objects in three main classes, as follows
\begin{itemize}
\item \textbf{Class 1: Globular Clusters} - This class includes 1,209 objects with radial velocity confirmation as GCs from the catalogs explained above. Within an area of 0.7 Mpc, \cite{chaturvedi22} estimated to be around 2,300 sources, but the cross-match between the complete GC RV sample and the NGFS-T1 ($<$ 0.34 Mpc) catalog yields half of this population.\\

\item \textbf{Class 2: Foreground Stars} - The \cite{maddox2019} compilation reports 9,483 stars with radial velocity confirmation within the Fornax Cluster area of $<1.4$ Mpc. In the same region, the Gaia mission in the EDR3 distance catalog \citep{gaia2021} detected and characterized, a total of 9,595 stars. Cross-matching with the NGFS-T1 master catalog results in a final sample of 2,151 stars.\\ 
\item \textbf{Class 3: Galaxies} — The galaxy RV confirmed sample from \cite{maddox2019} compilation has a total of 6722 galaxies coming from the Fornax Cluster and the majority as background galaxies. Cross-matching with the NGFS catalog, the final sample consists of 1,587 galaxies within NGFS-T1 area. We decided to keep the objects classified as ultra-compact dwarf galaxies (UCDs) within class 1 for GCs. Although we acknowledge the uncertain nature of these objects, distinguishing between truly massive GCs and stripped galaxy nuclei is beyond the scope of this paper. An approximate number of UCDs in the RV catalog is 100 - 120 sources, using only the criteria of $mag_i < 20$ \citep{mieske2002}. In our \texttt{svm.SVC} model, we do not separate galaxy subtypes (e.g., QSOs), as it is a complex task (See Figure \ref{fig:f_cc_wModels}) for an area of 0.34 Mpc. \cite{maddox2019} report 264 objects marked in the catalog as QSOs for a total of 6334 background galaxies, i.e. $4.2\%$ for 1.4 Mpc. Therefore, for the NGFS-T1 field of view (0.34 Mpc), the presence of QSOs represent less than 1\% of the sample. A similar estimation was also found by \cite{cristiani2001}.
\end{itemize}

Figure \ref{fig:f_cc_wModels} shows the color-color diagrams ($u'g'K_s$ and $u'g'i'$; same layout as Panels A and C in Fig. \ref{fig:f_cc_ugiJKs}, respectively). In the $u'g'K_s$ diagram (top panel), the blue lines represents old single-age stellar population (SSP) models at redshift zero, which closely follow the locus of the Fornax Cluster GC population (blue circles, corresponding to the 1,209 RV-confirmed GCs in Class 1). In addition, we plot the redshift evolution for four prototypical SEDs of galaxies born at $z = 3$ with different types of star formation history. These tracks were computed using DECam $u'g'i'$ and VIRCam $JK_s$ filter throughput curves to obtained the observed colors with the population synthesis code PEGASE.2 \citep{fioc97}. The four galaxies with different path histories are: (a) galaxy that formed most of their stars at high redshift but maintained a low, constant star formation rate (squares); (b) galaxy with a constant star formation rate (stars); (c) galaxy with an exponentially declining SFR (triangles); and (d) “red and dead” galaxy that formed all their stars at redshift 3 and evolved passively thereafter (circles). This figure highlights the diversity and complexity of possible star formation histories for galaxies present in a deep color-color diagram.

%Figure 

With the three confirmed training samples established, we split the complete NGFS-T1 catalog into two subsets: a labeled sample (used for training and testing) and an unlabeled sample (used for classification). The labeled sample includes objects with known classifications: GCs, foreground stars and background galaxies, while the unlabeled sample consists of sources with no prior classification.

The training and testing workflow consists of the following essential steps:
\begin{enumerate}[label=\roman*.]
\item Split the labeled sample randomly into training subsets and testing subsets using \texttt{train\_test\_split}. In order to define the percentage of the total sample for training and for testing, we perform several combinations: 80\% to 20\% or 80/20, 70/30, 60/40 and 50/50. We show the results in Section \ref{sect:impl_svm_model}.
\item We undersample the training set using \texttt{RandomUnderSampler} to address class imbalance, specifically the over representation of galaxies compared to stars and GCs. This step reduces the sample size in the majority class, helping the model learn more effective decision boundaries across all classes.
\item Scale the resampled training set using \texttt{StandardScaler}. Since \texttt{svm.SVC} relies on distance metrics, feature scaling is particularly important when using the RBF kernel. It rescales each feature so that it has zero mean and unit variance, taking in consideration a ($\mu$) as the mean and ($\sigma$) as the standard deviation of that feature in the training set. This ensures that all features contribute on a comparable scale to the SVM optimization, preventing features with larger numerical ranges from dominating the model.
\item Scale the test set using the same scaler fitted to the training data to maintain consistency between them.
\item Train the classifier using \texttt{svm.SVC}, setting \texttt{class\_weight}='balanced' to account for any remaining imbalance and evaluate its performance on the test set.
\item Finally, the trained model is applied to classify the remaining unlabeled sources in the catalog.
\end{enumerate}

%%%%%%%%%%%%%%%%%%%%%%%%%%%%%%%%%%%%%%%%%
\subsection{Searching for the best kernel function and parameters}
\label{sect:gridsearch}

In Section \ref{sect:svm}, we describe the different kernels available and the parameters in Equations (1) - (4), that can be adjusted to influence how the \texttt{svm.SVC} model learns. The efficiency of the model depends on the kernel choice and the structure of the data, whether it lies in a linear or non-linear space.

To identify the most suitable combination of kernel and hyperparameters for our dataset, we use \texttt{GridSearchCV} in \texttt{scikit\-learn.model\_selection}, which performs an intensive search over a specified grid of parameters by training and validating the model on all possible combinations of the given hyperparameter values. For each combination, it evaluates the performance using cross-validation and then selects the one that yields the best score. We also set the random state to a fixed value ($RS=42$), to ensure reproducibility of the results. We set the following \texttt{param\_grid}:
\begin{itemize}
\item RBF kernel: $\gamma$=$['scale', 1., 0.1, 0.01, 0.001, 0.0001]$.
\item Poly Kernel: $degree$=$[2, 3, 4, 5]$.
\item Sigmoid kernel:  $\gamma$=$['scale', 1., 0.1, 0.01, 0.001, 0.0001]$. 
\end{itemize}

For each of these kernels we use a regularization parameter grid of $C$=$[0.1, 1, 10, 100, 1000]$. We perform this search after step i) to iv) from the previous section, on the training scaled sample. The optimal parameters are C=10 and $\gamma$= 0.1 or "scale" found in Section \ref{sect:performance_results} and its figures, with titles containing this information.

Therefore, \texttt{GridSearchCV} procedure optimize the SVM hyperparameters ($C$, $\gamma$, and kernel type), systematically exploring the parameter space through cross-validation. This process ensures that the selected kernel and parameters correspond to the best-performing configuration.

%%%%%%%%%%%%%%%%%%%%%%%%%%%%%%%%%%%%%%%%%
\subsection{Implementation of \texttt{svm.SVC} model}
\label{sect:impl_svm_model}

Feature selection in SVM is a critical step in optimizing model performance. Figure \ref{fig:f_cc_ugiJKs} illustrates how different color combinations contribute to separating the various populations present in the area. For the five filters used ($u'g'i'JK_s$), we construct 10 color indices, considering only those formed by the difference in magnitude between a blue and a red filter.

Besides colors, we leverage the morphological parameters of the sources (morpho-parameters for short) such as size, shape and light profile to improve classification. As explained in Sect. \ref{sect:data}, the parameters we use are: spread model, Flux Radius, FWHM, ellipticity, concentration index.

Considering colors and morpho-parameters, we provide in total 15 features (15F) for the \texttt{svm.SVC} model, see Table \ref{tab:15F_description}. 

\begin{table}[ht]
\centering
\caption{Description for the 15 features (15F) provided for the model.}
\begin{tabular}{ll}
\hline
\textbf{Feature} & \textbf{Definition} \\
\hline
\hline
$(u'-g')_0$    & Photometric color.\\
$(u'-i')_0$    & Photometric color.\\
$(u'-J)_0$     & Photometric color.\\
$(u'-K_s)_0$   & Photometric color.\\
$(g'-i')_0$    & Photometric color.\\
$(g'-J)_0$     & Photometric color.\\
$(g'-K_s)_0$   & Photometric color.\\
$(i'-J)_0$     & Photometric color.\\
$(i'-K_s)_0$   & Photometric color.\\
$(J-K_s)_0$    & Photometric color.\\
$SM$         & Spread model, difference between PSF and object\\
             & profile.\\
$C$ & Concentration Index.\\
    & $C_{\lambda}$= MAG\_APER (2pix) -  MAG\_APER (8pix)\\
$FR$ & Flux Radius, containing 50\% of the total flux.\\
$FWHM$ &  Full Width Half Maximum, of the object's profile.\\
$e$ & Ellipticity, $e$ = $1 - a/b$.\\
\hline 
\hline
\end{tabular}
\label{tab:15F_description}
\end{table}

In order to obtain the best validation performance of the model, we use \texttt{GridSearchCV} that cross-validates parameters (see Sect. \ref{sect:gridsearch}) with the training portion of the data and it further splits that training data into multiple cross-validation folds, where for each candidate kernel and parameter combination, it trains on some folds and validates on the others, iteratively. Then, the combination with the best validation performance is chosen and the model with those parameters is retrained on the full training set and evaluated on the testing set.

Another aspect to consider for the best validation, is the split of the labeled sample randomly into training and testing subsets with \texttt{train\_test\_split}. We perform several combinations: 80\% to 20\% or 80/20, 70/30, 60/40 and 50/50. The best result is 70\% train and 30\% test sample and it is shown in Figure \ref{fig:poi_7030split} and \ref{fig:CM_15F_7F}. For reference, the others cases are shown in the Appendix \ref{fig:CM_poi_other_split}.

Therefore, \texttt{GridSearchCV} cross-validates not only by kernel and parameters but also by changing the \% of the train/test dataset. For the \texttt{SVM.svc} model with 15 features, the final combination that gives the best validation performance is: 70\% train and 30\% test, Kernel: RBF with parameters C=10 and $\gamma$=0.1. The performance of the \texttt{SVM.svc} model is presented in Section \ref{sect:performance_results}.

\begin{figure}
%trim=left bottom right top
     \centering
     \includegraphics[trim=0.5cm 0.1cm 0.5cm 0.3cm,clip,width=\linewidth]{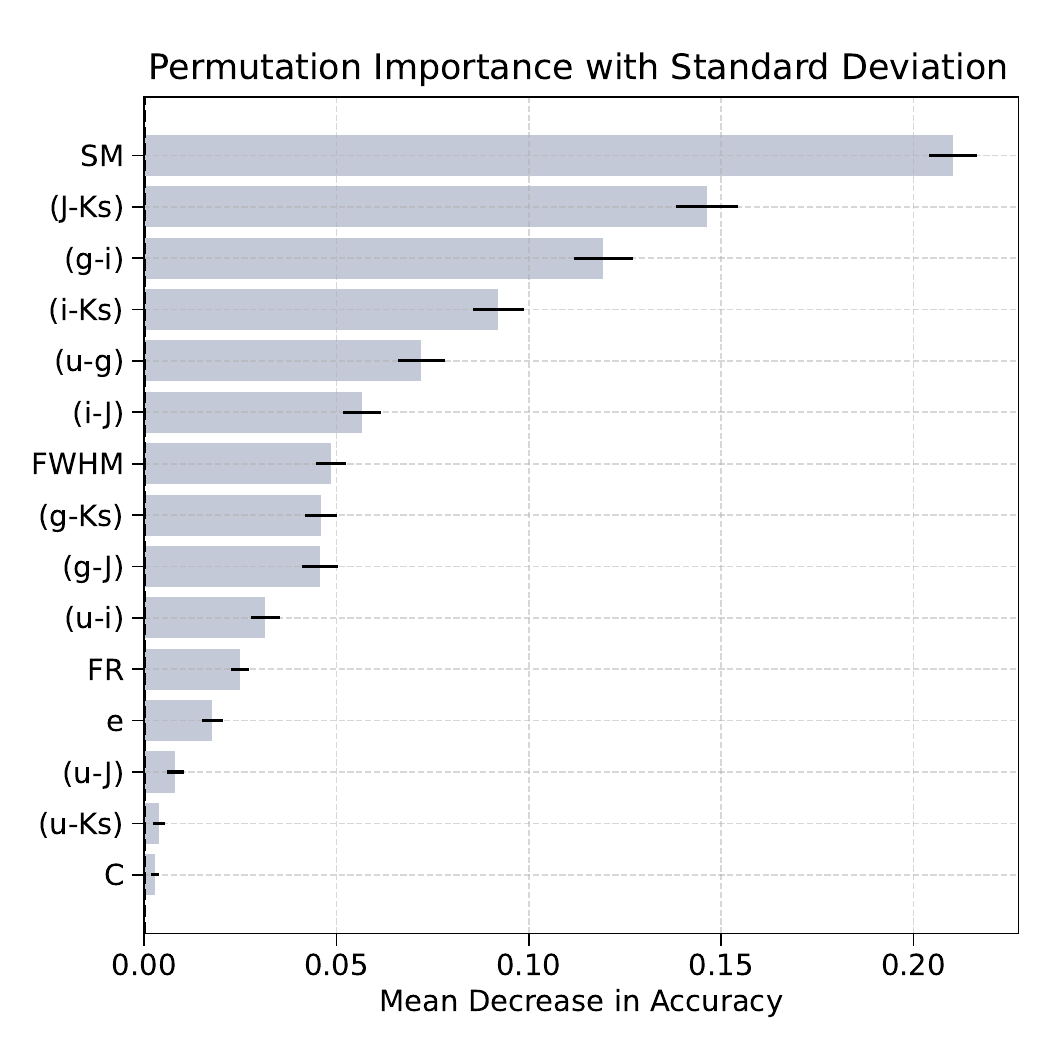}
    \caption{Best outcome model accuracy and scores, with a split in 70\% train and 30\% test and kernel=RBF, C=10 and $\gamma$=0.1. The permutation of importance for the 15 features provided to \texttt{SVM.svc}, ordered from the most to the least important features for the model.}
     \label{fig:poi_7030split}
\end{figure}

\subsection{Testing Relevance and Redundancy of Features}
\label{sect:f_selection}

One of the main challenges in feature selection is to avoid incorporating features that can cause the model to overfit, introduce redundancy (two or more input features provide the same or highly correlated information), or hinder optimization during training \citep{liu2011,pedregosa2011}. 

A key strategy is to distinguish between relevance and redundancy. Although, the 15 feature \texttt{SVM.svc} model already has a high performance, in the following we present two main methodologies used in this work for inspecting the statistical performance contribution of each feature of the fitted model: "permutation feature importance" and the "Correlation Clustermap". 

Permutation importance, as implemented in the \texttt{scikit-learn} library, is used to evaluate the relevance of individual features by quantifying their impact on the performance of the model. It works by randomly shuffling the values of a single feature and measuring the resulting decrease in the accuracy or score of the model, which reveals how much the model depends on that feature \citep{Breiman2001}. This method is particularly useful for non-linear models such as \texttt{SVC(kernel='rbf')}, where traditional feature importance metrics are less interpretable. In addition, we use a correlation clustermap generated with the \texttt{Seaborn} library \citep{Waskom2021} to visualize feature redundancy and inter-feature correlations (method = 'average' and metric='correlation'), aiding in the identification of highly collinear or redundant inputs. It is primarily used to assess feature collinearity and identify potential redundancies by highlighting correlations between features. Although its main purpose is to detect collinearity, it also serves as a powerful tool for examining pairwise feature relationships during feature selection.

\begin{figure*}[ht]
%trim=left bottom right top
     \centering
     \includegraphics[trim=0.1cm 0.1cm 0.1cm 2.8cm,clip,width=0.7\textwidth]{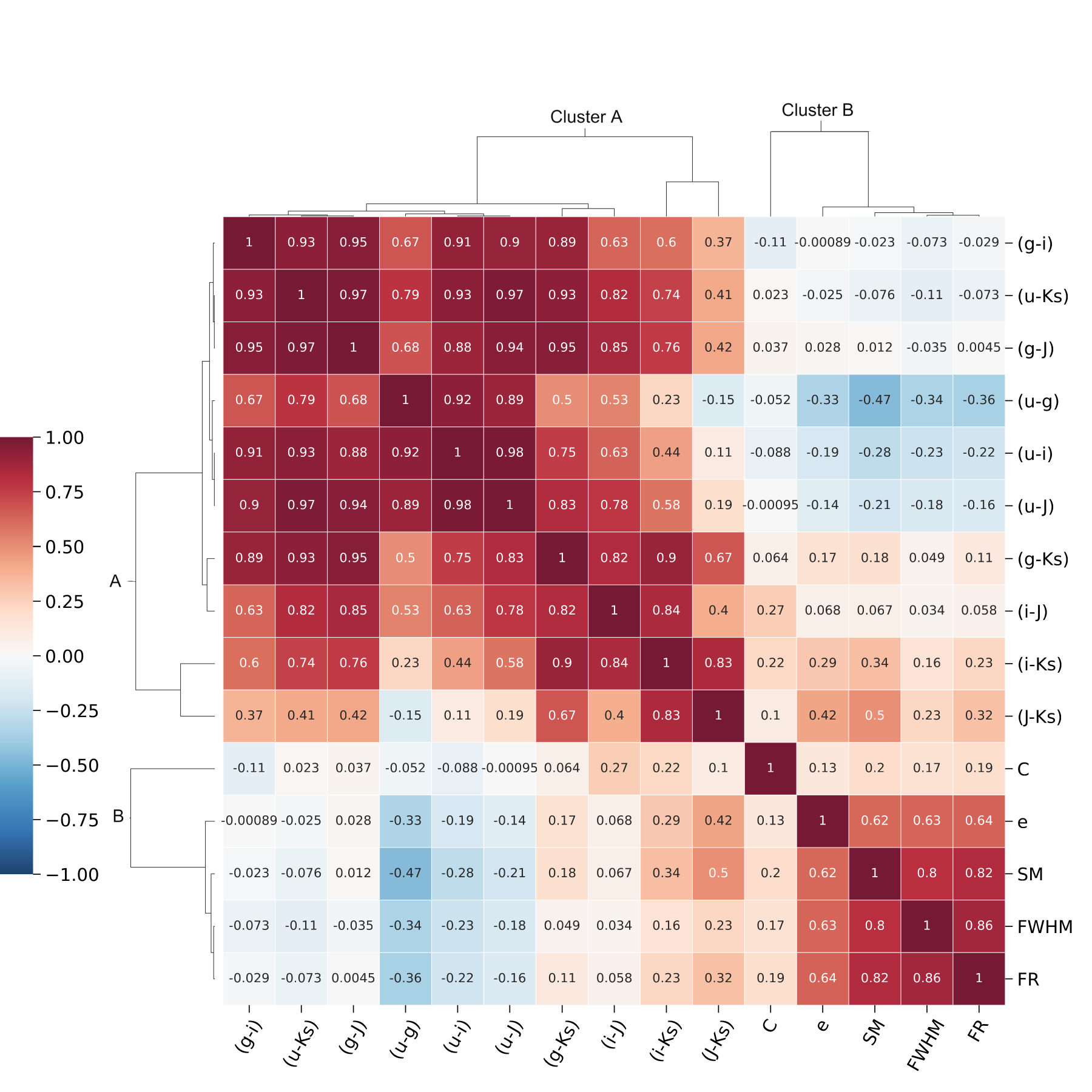}
     \vspace{-0.1cm}
     \caption{Clustermap correlation for 15 features. The correlation matrix displays numerical values in each cell with a heatmap ranging from -1.0 to 1.0, where '0' indicates no correlation (white), '-1' indicates linear anti-correlation (blue), and '1' indicates linear correlation (red). The dendrogram identifies two major feature clusters, a cluster A of color indices and a cluster B of morpho-parameters.}
     \label{fig:clc_f_15F}
\end{figure*}

In Figure \ref{fig:poi_7030split}, the permutation of feature importance is shown for the 15F set with their standard deviation. The SVM model was implemented using the best kernel, which is the RBF kernel, and its parameters for our data (see Section \ref{sect:impl_svm_model}). The two most relevant features are the spread model parameter and the color $(J-K_s)$ and the least important features of the model are the color $(u'-K_s)$ and the concentration index parameter. 

The clustermap\footnote{Figure \ref{fig:clc_f_15F} may be particularly useful for cases with limited filter coverage, as it highlights highly correlated feature pairs (correlation $> 0.9$) and suggests optimal combinations for constructing effective SVM models.} for the 15F set is shown in Fig. \ref{fig:clc_f_15F}, the correlation matrix displays red, blue, and white colors to indicate positive, negative or no collinearity, respectively. The dendrogram (tree structure) identifies two major feature clusters. Cluster A consists of color indices (photometric colors), among which several are strongly correlated. Within cluster A, there are two sub-clusters, one with the colors $(i'-K_s)$, $(J-K_s)$ and the others with the rest. Cluster B comprises the morpho-parameters in two sub-clusters, with the concentration index as one of them, and the rest in the other sub-cluster, with some being highly correlated, but remain largely uncorrelated with the photometric colors. It is worth noting that the apparent anti-correlation between color and morphological size observed when the $u'$ band is included, arises from the larger seeing, PSF variability, and lower S/N (particularly for faint or extended sources) compared to the $g'$ and $i'$ bands. These effects artificially increase the measured structural parameters of the sources, while the mixture of populations (either bright or faint in the $u'$ band) naturally reinforces the observed trend. Notably, the color $(J-K_s)$ shows weaker correlations with other colors, suggesting that it may carry more independent information about the underlying stellar populations. This could be related to the advantages of including NIR filters, where the spectral energy distribution of individual stars or the integrated light of stellar systems (GCs or galaxies) is dominated by intermediate to old age populations, such as cool K and M giant stars, and stars at the tip of the red giant branch or the asymptotic giant branch \citep[e.g.][]{verro2022}.

We use a combined strategy: permutation feature importance for relevance and correlation clustermap for redundancy, to obtain the maximal signal with minimal overlap.  With the result of the permutation feature importance in Fig. \ref{fig:poi_7030split} (bottom-panel), we select the most diagnostic features with a mean decrease in precision greater than $0.025$, thus excluding ellipticity, $(u'-J)$, $(u'-K_s)$ and the concentration index parameter. The clustermap in Fig. \ref{fig:clc_f_15F} allow us to identify possible redundant features.

Among the 11 remaining features and considering the correlations mentioned above, we maintain the first seven parameters: $SM$, $(J-Ks)$, $(g'-i')$, $(i'-K_s)$, $(u'-g')$, $(i'- J)$ and $FWHM$. For the other features, it is advisable to either exclude them as features or assess how model performance is affected when each is removed. It is worth noticing that the four of them are correlated with one (or more) of the seven selected features: $(u'- i')$ with $(g'- i')$ ($corr=0.91$) and $(u'- g')$ ($corr=0.92$); $(g'-K_s)$ with $(i'- K_s)$ ($corr =0.9$), $(g'-J)$ with $(g'- i')$ ($corr=0.95$); and flux radius $FR$ with $FWHM$ ($corr=0.86$) and $SM$ ($corr=0.82$). 

Additionally, we evaluated two reduced sets: 6F - excluding $FWHM$ and 5F - excluding both $(i-J)$ and $FWHM$). The \texttt{SVM.SVC} models using a seven feature configuration showed an improvement in classification, balancing overfitting and scores compared to the other feature models. Therefore, a combination of color indices spanning the near-UV, optical, and NIR regimes, together with two structural parameters ($SM$ and $FWHM$), provides the most effective and efficient class separation. In the next section, we conduct additional tests to further assess the performance of the \texttt{SVM.SVC} model. For the subsequent analysis, we adopt both the 15F and 7F configurations.

%%%%%%%%%%%%%%%%%%%%%%%%%%%%%%%%%%%%%%%%%
\section{Performance and further tests of the \texttt{svm.SVC} Classification Model}
\label{sect:performance_results}

In this section, we present the classification performance for two feature configurations: (i) the full 15 feature (15F) set, and (ii) the reduced 7 feature (7F) set, derived using the combined dimensionality reduction strategy described in Sect. \ref{sect:f_selection}. The goal of this reduction is to enhance model efficiency while balancing overfitting and preserving classification accuracy. Given that GCs, stars, and galaxies exhibit overlapping distributions in both color space and morphological parameters, the classification task is inherently non-linear. In this context, the RBF kernel proves particularly effective, as it can model complex, non-linear decision boundaries. Consistently, the \texttt{GridSearchCV} optimization procedure selects the RBF kernel as the best performing option across all tested configurations. The optimal hyperparameters ($C$ and $\gamma$) for the best-performing models, 15F and 7F correspond to $C=10$ and $\gamma$ equal 0.1 or "scale", respectively.

\subsection{Performance of the \texttt{svm.SVC}}
\label{sect:perfmodel}

It is important to highlight, that our \texttt{svm.SVC} model implementation is designed for deep photometry, using color based features to avoid relying on magnitudes, which depend on distance. To build a robust training and testing sample, the RV confirmed GCs must extend to at least $mag_i \approx 22$ in the $i'-$band, as they are the faintest objects among the three classes to be classified. In the RV confirmed sample, we have 447 GCs with $mag_i \le 21$ mag, 762 with $mag_i \le 21.5$ mag, and 1041 with $mag_i \le 22$ mag. To assess the impact of a shallower labeled sample, we tested the 15F model under the hypothetical scenario in which the labeled data reach only $mag_i = 21$ mag (see Sect. \ref{sect:test_magcut}).

In addition, the code implemented in this study is computationally efficient, completing the entire training, prediction, and classification workflow in less than two minutes on a single CPU for a 3 $deg^2$ sky area.

To evaluate the performance of the \texttt{svm.SVC} models, we use two standard metrics from the \texttt{scikit-learn} library \citep{pedregosa2011}. First, the classification report, which provides per class performance metrics (see Table \ref{tab:metrics}) for GCs, stars, and galaxies. This allows us to compare the performance of the model across classes and assess potential issues with class imbalance. Second, we use the normalized confusion matrix \citep{faw2006}, which shows how the predictions are distributed across the true and predicted classes, highlighting areas where the model tends to confuse one class with another.

\begin{table}[ht]
\centering
\caption{Performance metrics used for model evaluation \citep{sokolova2009}.}
\begin{tabular}{ll}
\hline
\textbf{Metric} & \textbf{Definition} \\
\hline
\hline
Accuracy  & $= (TP + TN) / (TP + TN + FP + FN)$ \\
          & Ratio of correct predictions to the total number\\
          & of predictions. \\
Precision & $ = TP / (TP + FP)$ \\
          & High precision indicates a low number of false\\
          & positives. This term is also called Purity. \\
Recall    & $=TP / (TP + FN)$ \\
          & High recall indicates a low number of false\\
          & negatives. This term is also called Completeness. \\
F1-score & Harmonic mean of precision and recall. \\
         & Provides a balanced evaluation of both metrics. \\
Support & Number of true instances for each class.\vspace{0.1cm} \\
\hline 
\hline
\textbf{Note.} & \( TP \) = true positives, \( TN \) = true negatives,\\
               &\( FP \) = false positives, and \( FN \) = false negatives. 
\end{tabular}
\smallskip
\label{tab:metrics}
\end{table}

Table~\ref{tab:clreport_15f} presents the classification report for the \texttt{svm.SVC} model trained with 15F. In the precision column, the GC, star and galaxy class present 4.6\%, 1.6\% and 2.8\% score for FP, respectively. Looking at the recall column, showing how many objects were misclassified by class, we have 3\% of GCs, 1.9\% of stars and 3.6\% of galaxies. In general, the model correctly classified 97.3\% of the samples (1445 out of 1485).

\begin{table}[ht]
\centering
\caption{Classification Report for the 15 features (15F) model (see Fig. \ref{fig:CM_15F_7F} for the corresponding confusion matrix)}.
\vspace{-0.06\linewidth}
\label{tab:clreport_15f}
\begin{tabular}{|l|c|c|c|c|}
\hline
\textbf{Class} & \textbf{Precision} & \textbf{Recall} & \textbf{F1-score} & \textbf{Support} \\
\hline
GCs       & 0.9539 & 0.9697 & 0.9617 & 363  \\
Stars     & 0.9845 & 0.9814 & 0.9829 & 646  \\
Galaxies  & 0.9725 & 0.9643 & 0.9684 & 476 \\
\hline
\textbf{Accuracy}     & \multicolumn{4}{c|}{0.9731} \\
\textbf{Macro Avg}    & 0.9703 & 0.9718 & 0.9710 & 1485 \\
\textbf{Weighted Avg} & 0.9732 & 0.9731 & 0.9731 & 1485 \\
\hline
\end{tabular}
\vspace{0.05\linewidth}
\caption{Classification Report for the 7 features (7F) model, as the result of applying Permutation importance and clustermap correlation to remove features. See Section \ref{sect:f_selection} and Fig. \ref{fig:CM_15F_7F} and \ref{fig:result_cc_15F_7F}.}
\vspace{-0.01\linewidth}
\label{tab:clreport_7f}
\begin{tabular}{|l|c|c|c|c|}
\hline
\textbf{Class} & \textbf{Precision} & \textbf{Recall} & \textbf{F1-score} & \textbf{Support} \\
\hline
GCs       & 0.9332 & 0.9614 & 0.9471 & 363  \\
Stars     & 0.9828 & 0.9721 & 0.9774 & 646  \\
Galaxies  & 0.9703 & 0.9622 & 0.9662 & 476 \\
\hline
\textbf{Accuracy}     & \multicolumn{4}{c|}{0.9663} \\
\textbf{Macro Avg}    & 0.9621 & 0.9653 & 0.9636 & 1485 \\
\textbf{Weighted Avg} & 0.9667 & 0.9663 & 0.9664 & 1485 \\
\hline
\end{tabular}
\end{table}

For the dimensionality reduction described in Section \ref{sect:f_selection}, we use the 7F set: $SM$, $(J-K_s)$, $(g'-i')$, $(i'-K_s)$, $(u'-g')$, $(i'-J)$, and $FWHM$. Table \ref{tab:clreport_7f} presents the classification report for the \texttt{svm.SVC} model using the 7F set. In the precision column, the FP rates are 6.7\% for GCs, 1.7\% for stars, and 3\% for galaxies. In the recall column, the misclassifications values remain low, but are slightly higher compared to the 15F model: 96.1\% (GCs), 97.2\% (stars), and 96.2\% (galaxies). The overall accuracy is 96.6\%. Although the 15F model yields better results in the classification report compared to the 7F model, we attribute this to model overfitting caused by redundant color indices and structural parameters. A more realistic scenario is represented by the 7F model, as illustrated in Figure \ref{fig:result_cc_15F_7F}.

\begin{figure}
%trim=left bottom right top
     \centering
     \includegraphics[trim=3cm 0.1cm 0.8cm 0.2cm,clip,width=0.85\linewidth]{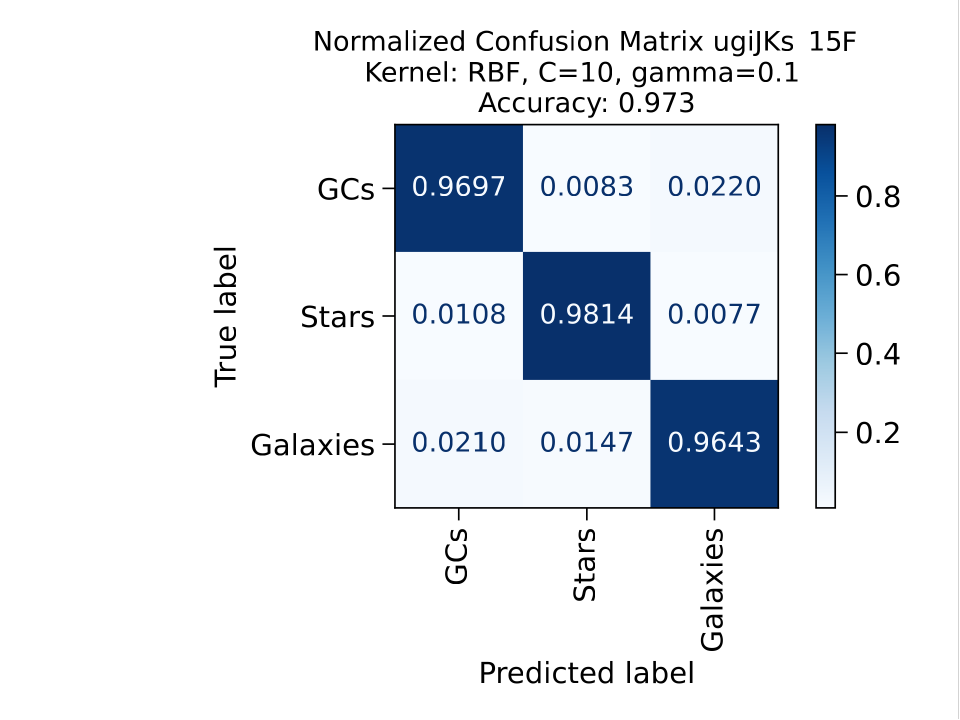} 
     \includegraphics[trim=3cm 0.1cm 0.8cm 0.2cm,clip,width=0.85\linewidth]{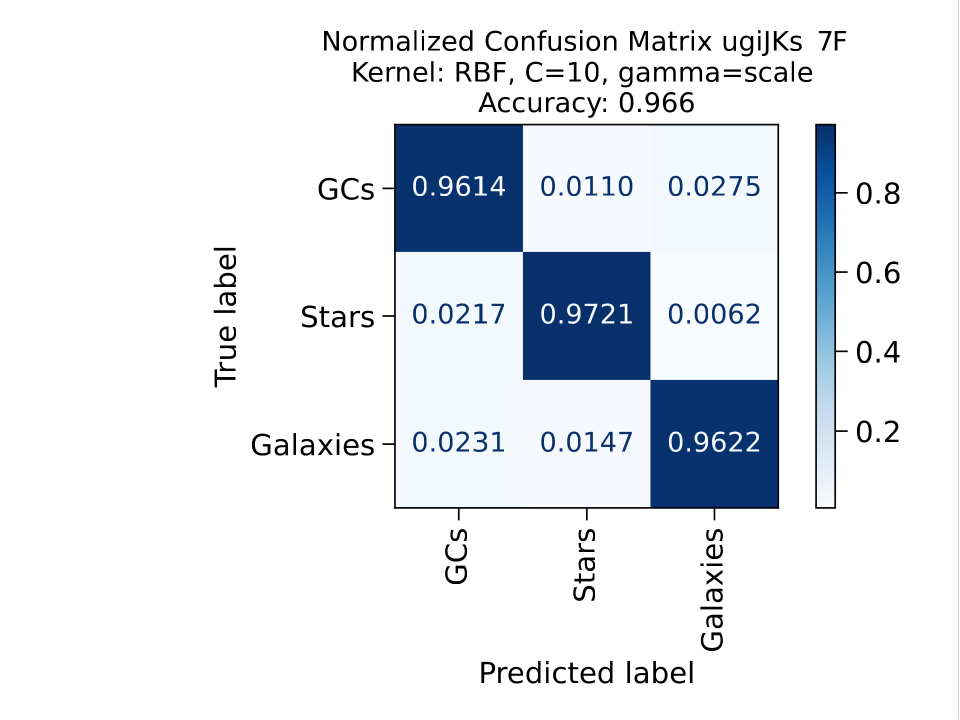} 
    \caption{Normalized confusion matrix, with the True labels (correct label) on the y-axis and Predicted label (by the model) on the x-axis. The \texttt{SVM.svc} results for the 15F and 7F model are shown in the top and bottom panel, respectively.}
     \label{fig:CM_15F_7F}
\end{figure}

The confusion matrices for the 15F and 7F models are shown in Fig. \ref{fig:CM_15F_7F}. Each row corresponds to the true class labels (TP), and each column corresponds to the predicted class labels. The diagonal entries represent correct classifications and off-diagonal entries indicate misclassifications. Specifically, for a given class: false negatives (FN) are instances of that class that were incorrectly predicted as another class (i.e., values in the same row but outside the diagonal) and false positives (FP) are instances predicted as that class but whose true label belongs to another class (i.e., values in the same column but outside the diagonal).

Figure \ref{fig:CM_15F_7F} top-panel, shows the confusion matrix for the 15 feature model. Predictions were correct at 97.0\%, 98.1\% and 96.4\% for GCs, stars and galaxies, respectively. There are FP predictions, misclassifications of GCs as stars (0.8\%) and galaxies (2.2\%), stars misclassified as GCs (1\%) and galaxies (0.8\%) and galaxies misclassified as GCs (2.1\%) and stars (1.5\%), thus, in total a 8.4\% incorrect predictions. Figure \ref{fig:CM_15F_7F} bottom-panel, presents the normalized confusion matrix for the 7F model, which demonstrates lower performance in the classification compared to the 15F. Predictions were correct at 96.1\%, 97.2\% and 96.2\% for GCs, stars and galaxies, respectively. Overall, the model yields a total misclassifications of 10.4\%. These results indicate that the 15F model in terms of scores does better, however according to Section \ref{sect:f_selection}, four features were not relevant for the model and other four were correlated among the most important features for the model. 

\subsection{Testing the \texttt{svm.SVC} Classifier Using a Magnitude Constrained Train/Test Sample}
\label{sect:test_magcut}

This section presents the results of testing the hypothetical scenario in which the labeled sample reaches only $mag_i = 21$ mag. The performance of the 15F model under this magnitude constraint is shown in Fig. \ref{fig:CM_magcut}. After imposing the magnitude cut, the labeled sample contains 447 GCs, 2,150 stars, and 1,539 galaxies, with class 1 (GCs) being the most affected. The key metrics are the per class Precision and Recall (Fig. \ref{fig:CM_magcut}, right panel). The fraction of false positives for GCs increases to 9.3\% (compared to 4.6\% without the magnitude cut), and the misclassification rate also rises, though more moderately, to 5.2\% (from 3\%). In contrast, class 2 (stars) and class 3 (galaxies) show only minor variations, with differences below 1\% relative to the no magnitude-cut case.

In summary, imposing a magnitude limit of $mag_i = 21$ mag leads to a measurable degradation in the ability of the \texttt{svm.SVC} model to identify GCs, reflected in higher FP and misclassification rates, adding an additional $\approx 5\%$ classification error relative to the no magnitude-cut case. In contrast, the performance for stars and galaxies remains nearly unchanged.

This demonstrates that the 15F model relies critically on the inclusion of fainter, RV confirmed GCs to fully capture the structure of the GC class in color space and to maintain robust separability from contaminant classes (stars and galaxies). Consequently, a deeper labeled sample is essential for optimal GC classification performance.

\subsection{Performance of the \texttt{svm.SVC} model in classifying the unlabeled NGFS-T1 catalog}
\label{sect:perf_catcompl}

Following the training and testing the \texttt{svm.SVC} model, it was used to assign predicted class labels to the sources in the NGFS-T1 input catalog. The predicted classes are encoded as follows: GC = 1, Star = 2, and Galaxy = 3. The predictions were obtained using the \texttt{model.predict} method, which returns an array containing the most likely class label for each source in the input dataset. Furthermore, we employed \texttt{model.predict\_proba} to obtain the full class probability distribution for each source. The majority of sources were classified with high confidence: 61\% have a maximum predicted class probability $\ge$90\%, and 92\% exceed probability $\ge$60\%. The final catalog comprising the unlabeled sample provides a classification for every source, determined by the class with the highest predicted probability. Although no probability threshold was imposed to filter classifications, the model performs strongly across most predictions.

\begin{figure*}[ht]
%trim=left bottom right top
     \centering
     \includegraphics[trim=0.3cm 0.7cm 0.3cm 0.5cm,clip,width=0.9\textwidth]{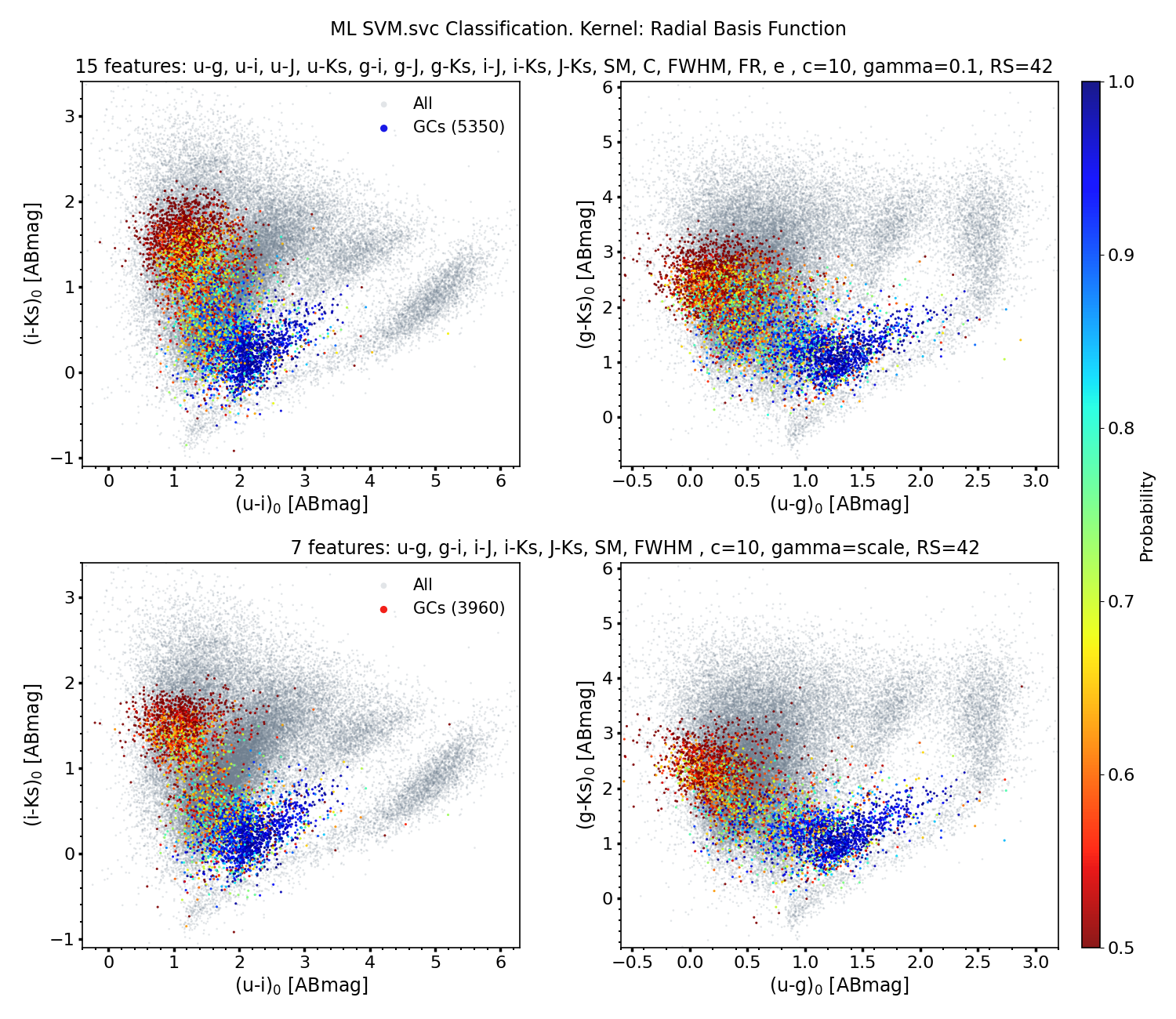} 
     \caption{Result of the \texttt{svm.SVC} model applied to the entire NGFS-T1 catalog for the 15 features (top panels) and 7 features (bottom panels) provided initially (see Sect. \ref{sect:impl_svm_model} and Sect. \ref{sect:performance_results}). Two color-color diagrams composed of $u'i'K_s$ (first column) and $u'g'K_s$ (second column). Color-coded is the output assigned class probability for GCs from 50\% (red) to 100\% (blue). Probabilities are computed in the full feature space, thus these plots serve as a 2D projection for visualization.}
     \label{fig:result_cc_15F_7F}
\end{figure*}

Figures \ref{fig:result_cc_15F_7F}  show the results of the \texttt{svm.SVC} classification for the unlabeled sample from the NGFS-T1 catalog with a total of 57,469 sources. Top panels show the 15F model results and the bottom panels the 7F model, each of them with a title indicating the kernel, its parameters and the features sets used for the specific model. The left column shows the $u'i'K_s$ and the second column the $u'g'K_s$ cc-diagrams. They were chosen because they ideally reveal the three classes for classification: GCs, stars and galaxies. The probability of every source being assigned to a specific class is color-coded from red: 50\% to blue: 100\%. It is important to notice that the underlying probability is calculated in the full feature space. These diagrams only serve to visualize the class separation in the corresponding projections of that space.

The classification results using the 15F model yield 5,350 GCs, 5,646 stars, and 46,473 galaxies. Figure \ref{fig:result_cc_15F_7F} shows in color code the probability of GCs. Sources that exhibit blue colors correspond to regions with high classification confidence. The 7F model shows classification result of: 3,960 GCs, 5,831 stars, and 47,678 galaxies. This reflects a decrease in the number of GC candidates relative to the 15F model. However, the GC selection in the 7F model shows a cleaner probability distribution than the 15F model, particularly with less sources in the range of 70-80\%. This allow that in the cc-diagrams the likely contaminants for the \texttt{svm.SVC} classification are more visible. It is good to note that the GC candidates with lower probability located in the upper part (redder in Y-axis) of the galaxy region, arise by the confusion between galaxies at high redshift (as "compact source"), see the PEGASE.2 Synthesis models in Figure \ref{fig:f_cc_wModels}.

Based on the confusion matrices shown in Figure \ref{fig:CM_15F_7F}, we estimate overall misclassification rates of 8.4\% for the 15F model and 10.4\% for the 7F model. Figure \ref{fig:result_cc_15F_7F} illustrate that despite the 15F model scores presented in Section \ref{sect:perfmodel}, the 7F model that has the most important and the least correlated features seems to be more reliable.

To further refine the GC candidate selection for the 7F model, we apply an additional constraint based on the model-assigned probabilities. As shown in Figures \ref{fig:result_cc_15F_7F}, sources with predicted GC probabilities below 80\% tend to drift away from the GC locus and toward the stellar or galactic regions in color–color space. By applying a conservative probability threshold of 80\%, we retain 1,717, including the 1,209 labeled GCs (with RV), the final GC catalog consists of 2,926 sources.

\subsection{Testing \texttt{svm.SVC} model predictions with fewer filter information}
\label{sect:svm_fewerfilters}

\begin{figure*}[ht]
    %trim=left bottom right top
    \centering
    \includegraphics[trim=0.3cm 0.7cm 0.3cm 0.5cm,clip,width=0.9\textwidth]{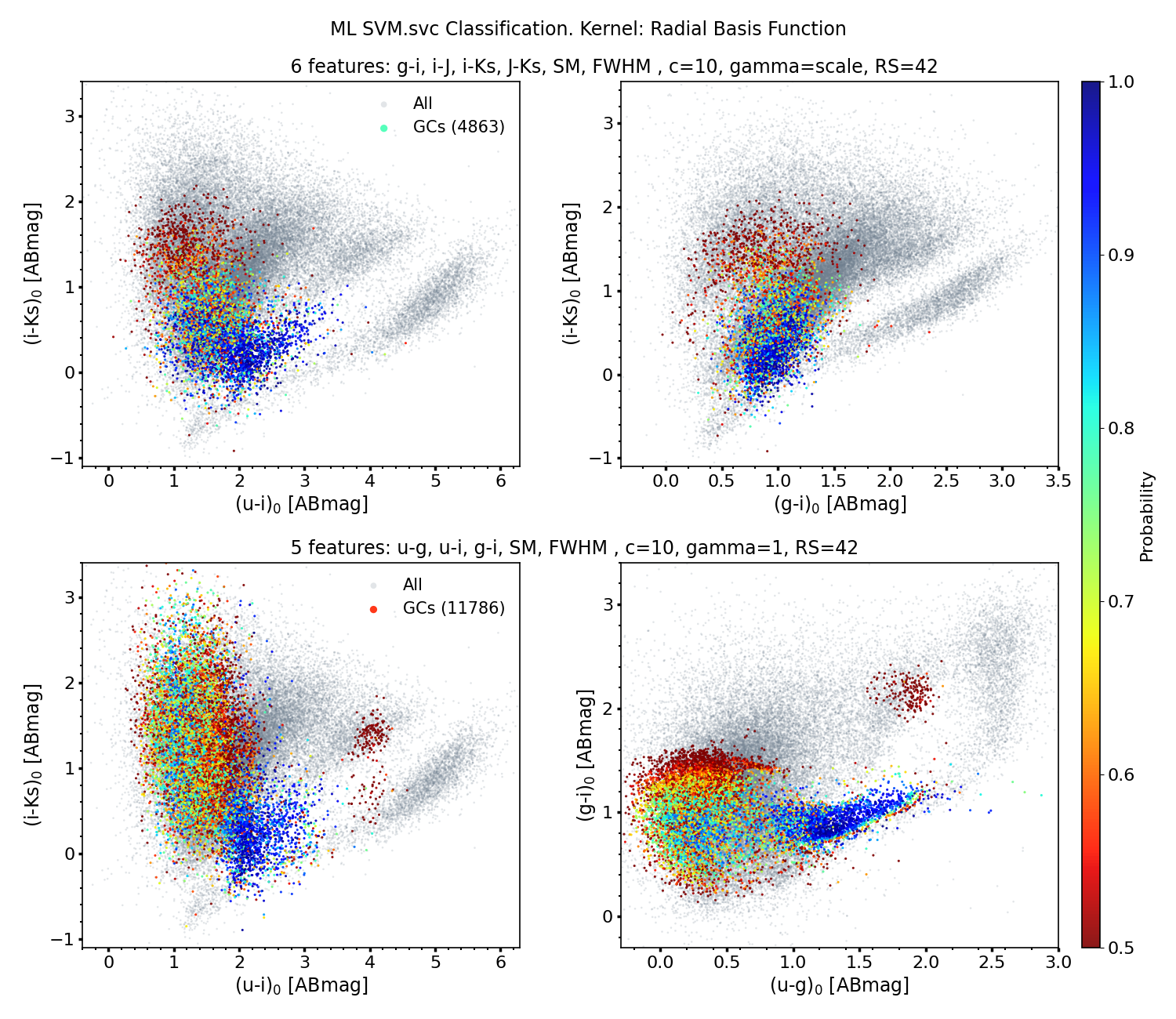}
    \caption{Classification results of the \texttt{svm.SVC} model using different color configurations. Top row: model excluding the $u'$-band. Bottom row: model excluding near-infrared (NIR) bands. The left column panels display the corresponding $u'i'K_s$ color–color diagrams and the right column panels show color–color diagrams tailored to the specific model setup: $g'i'K_s$ for the top and  $u'g'i'$.}
     \label{fig:svm_fewerfilter}
\end{figure*}

The previous section presented results using five filter information ($u'g'i'JK_s$), achieving a great classification performance. In this section, we revisit the model to evaluate its behavior under more limited photometric coverage, specifically, scenarios where the $u'$-band or NIR bands are missing. The outcomes are summarized in Figure \ref{fig:svm_fewerfilter}, which presents the results for: Six features model, with no $u'$-band (top row), and five features, with optical only (no NIR) (bottom row).

Figure \ref{fig:svm_fewerfilter} show the classification output for the unlabeled NGFS-T1 catalog. The left column panels show the $u'i'K_s$ cc-diagram. The right column panels show cc-diagrams based on the specific colors used in their respective model: 6F (no $u'$-band): $g'i'K_s$ and 5F (no NIR): $u'g'i'$. The main results of this test are the following:

\begin{itemize}
    \item 6F model: This configuration omits $u'$-band information, using $(g'-i')$, $(i'-J)$, $(i'-K_s)$, $(J-K_s)$, $SM$, $FWHM$. The classification report (Table \ref{tab:A.1.1_cl_report_6f}) and confusion matrix in Fig. \ref{fig:CM_6F_5F} shows precision and recall of 93.0\% and 95.3\% for GCs, 97.7\% and 96.9\% for stars and 97.3\% and 96.4\% for galaxies, yielding an overall accuracy of 96.4\%. Despite the high performance, the abscence of $u'$-band data increases false positives, false negatives, and the misclassifications across all classes. In Figure \ref{fig:svm_fewerfilter} the top panels show the 6F \texttt{svm.SVC} classification output with GCs as class=1 corresponds to 4,863 GCs, and significant confusion is observed: many high probability GC candidates appear in regions typically occupied by stars or galaxies in color-color space. It is worth noting that if $u'$-band data is unavailable, only the top-right panel (based on available filters) would be used for interpretation.
    \item 5F model: This model excludes NIR filters, using $(u'-g')$, $(u'-i')$, $(g'-i')$, $SM$, $FWHM$. Its classification report (Table \ref{tab:A.1.2_cl_report_5f}) and confusion matrix (Fig. \ref{fig:CM_6F_5F}, bottom-panel) indicate precision and recall values of 83.8\% and 90.9\% for GCs, 95.2\% and 91.6\% for stars and 95.5\% and 94.1\% for galaxies. The overall accuracy is 92.3\%, leading to a substantial increase in confusion between the three classes, as evidenced in the cc-diagrams of Figure \ref{fig:svm_fewerfilter} (bottom panels), and by the 11,786 GCs selected as candidates for class 1 objects.
\end{itemize}

Consequently, the inclusion of both near-UV ($u'$) and near-Infrared photometric data, alongside standard optical bands, is essential for building an effective ML classification method, such as the supervised SVM approach presented in this work. Our most reliable performing model is based on 7 features (5 colors and 2 morpho-parameter), and achieves contamination rates of $\sim10.4\%$.

For comparison, \cite{Gonzalez2019} reported a 30\% contamination rate using only $(u'-i')$ vs $(i'-K_s)$ cc-diagram. In their earlier work  \citep{Gonzalez2017}, they identified GCs (39), and estimated a total GC population of $N_{GC}=144\pm{31}$(random)$\pm{38}$ (systematic) in the galaxy NGC 4258, assuming 5\% contamination rate. However, upon spectroscopic confirmation, the revised estimate decreased to $N_{GC}=105\pm{26}\pm{31}$, reflecting the impact of higher than previously expected contamination on the inferred GC population size.

\subsection{\texttt{svm.SVC} Model Output Catalogs}
\label{sect:cat}
As described in Section \ref{sect:perf_catcompl}, the 7F \texttt{svm.SVC} models yield a total of 3,960 candidates, respectively (see Figure \ref{fig:result_cc_15F_7F}, bottom panels). Incorporating the additional selection criterion of probability $\geq$ 80\% (1,717 GCs) and the spectroscopically confirmed GCs, we have the final GC sample of 2,926 objects for the 7F model.

This final classification catalog produced by our model will serve as the foundation for a comprehensive analysis of the GC population in the NGFS-T1 field (cluster-centric radius $\leq$ 350 kpc), and will be presented in two forthcoming publications. The first will focus on studying GCs associated with the 279 galaxies \citep{eigethaler2018} and with the intra-cluster medium, using magnitudes and colors with full SED and their spatial distribution along the cluster. The second will examine the stellar population properties, luminosity and mass functions of the GC population, and their scaling relations. The full GC catalogs will be made publicly available as supplementary tables in the second work. Catalogs for star and galaxy candidates will also be used in upcoming follow-up studies led by the NGFS collaboration.

\begin{figure*}[ht]
%trim=left bottom right top
 \centering
 \includegraphics[trim=1.5cm 0.6cm 1.5cm 0.6cm,clip,width=\textwidth]{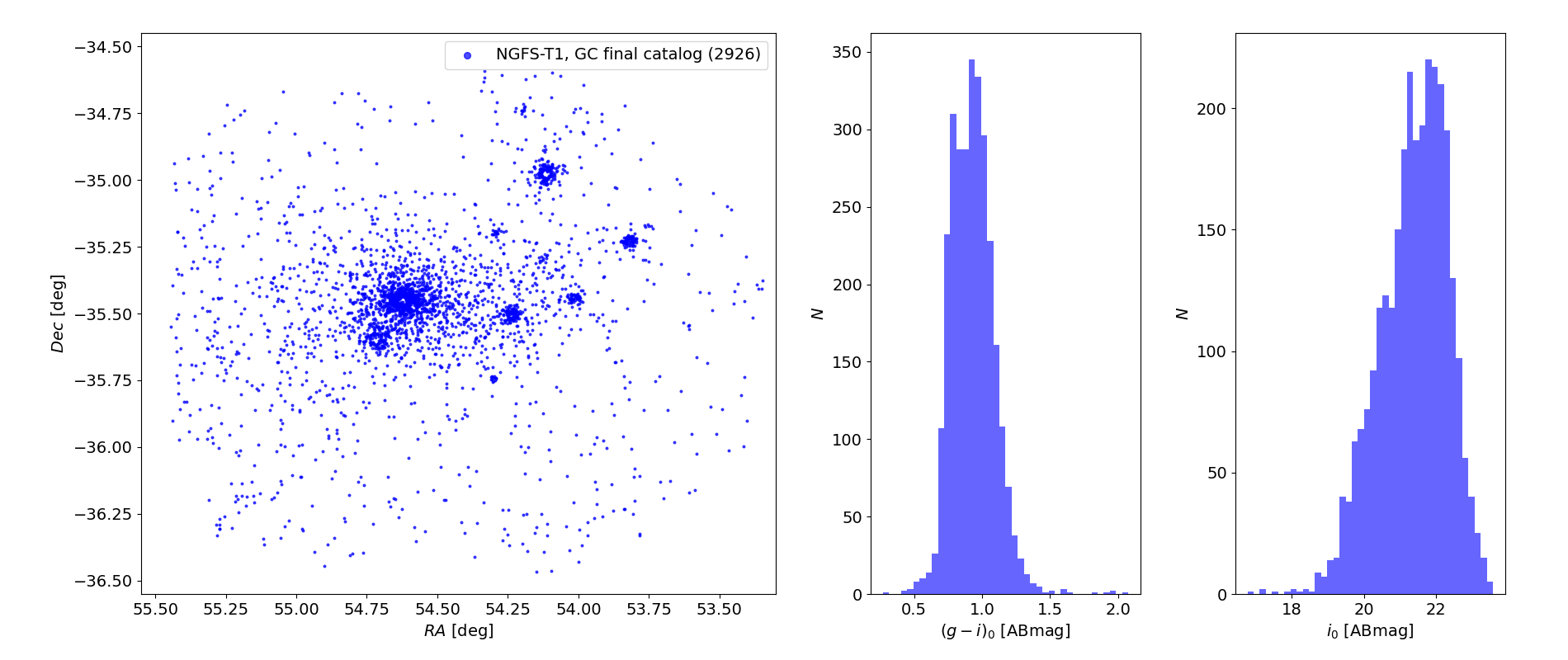} 
 \caption{Final catalog of GCs from the \texttt{svm.SVC} classification model (7F) and RV confirmed in the area. Left panel: GC spatial distribution, with the majority being in the surrounding of NGC 1399 and other massive galaxies. Middle panel: $(g'-i')$ histogram distribution. Right panel: $i'$-band magnitude histogram distribution.}
 \label{fig:GC_fcat}
\end{figure*}

We show in Figure \ref{fig:GC_fcat}, the quality of the output catalog (SVM method and RV confirmed sample). On the left panel we show the GC spatial distribution, and in the middle and right panels the histogram distributions in $(g'-i')$ color and mag $i'$-band, respectively. The GCs density is towards the dominant galaxy NGC 1399, besides some agglomerations around other bright galaxies, and a good amount of intra-cluster GCs. They will be very useful to use them as tracers to understand the stellar assembly of the central region of the Fornax Galaxy Cluster.

In Section \ref{sect:ML_literature}, we summarize other ML methodologies to select GCs and UCDs either in the Fornax area or in other regions of the sky using photometric dataset. Additionally, we show in Section \ref{sect:cat_comparison}, a comparison between our \texttt{svm.SVC} GC sample with existing photometric dataset such as the ACS Fornax Cluster Survey \citep{jordan2015}, the Fornax Deep Survey \citep{cantiello2020} and \citep{Saifollahi2021}.

%%%%%%%%%%%%%%%%%%%%%%%%%%%%%%%%%%%%%%%%%
\section{Testing the \texttt{svm.SVC} Method with the LSST filter system}
\label{sect:lsst}

In this section, we test the performance of the \texttt{svm.SVC} classification model using Vera C. Rubin Observatory Legacy Survey of Space and Time (LSST), which includes the $u'g'r'i'z'Y$. This test is designed to inform potential LSST users about the diagnostic power of cc-diagram in their filter setup and to assess whether reliable classification of GCs in Fornax (or similar environments at comparable distances), stars and galaxies is feasible using LSST photometry alone.

For this experiment, we use the NGFS data, which provides deep imaging in the $u'$-, $g'$-, and $i'$-bands. To complete the LSST-like filter set, we incorporate additional data from the Dark Energy Survey (DES) DR2 catalog \citep{DESdr2}, which was obtained using the BLANCO/DECam instrument, the same facility used for NGFS. DES offers deep coverage in the $g'r'i'z'Y$, though it lacks $u'$-band observations. By combining NGFS $u'g'i'$ and DES $r'z'Y$ data, we construct a synthetic LSST-like dataset with full six-band coverage. 

\begin{figure*}[ht]
    %trim=left bottom right top
     \centering
     \includegraphics[trim=2.2cm 2.5cm 2.2cm 2.5cm,clip,width=0.8\textwidth]{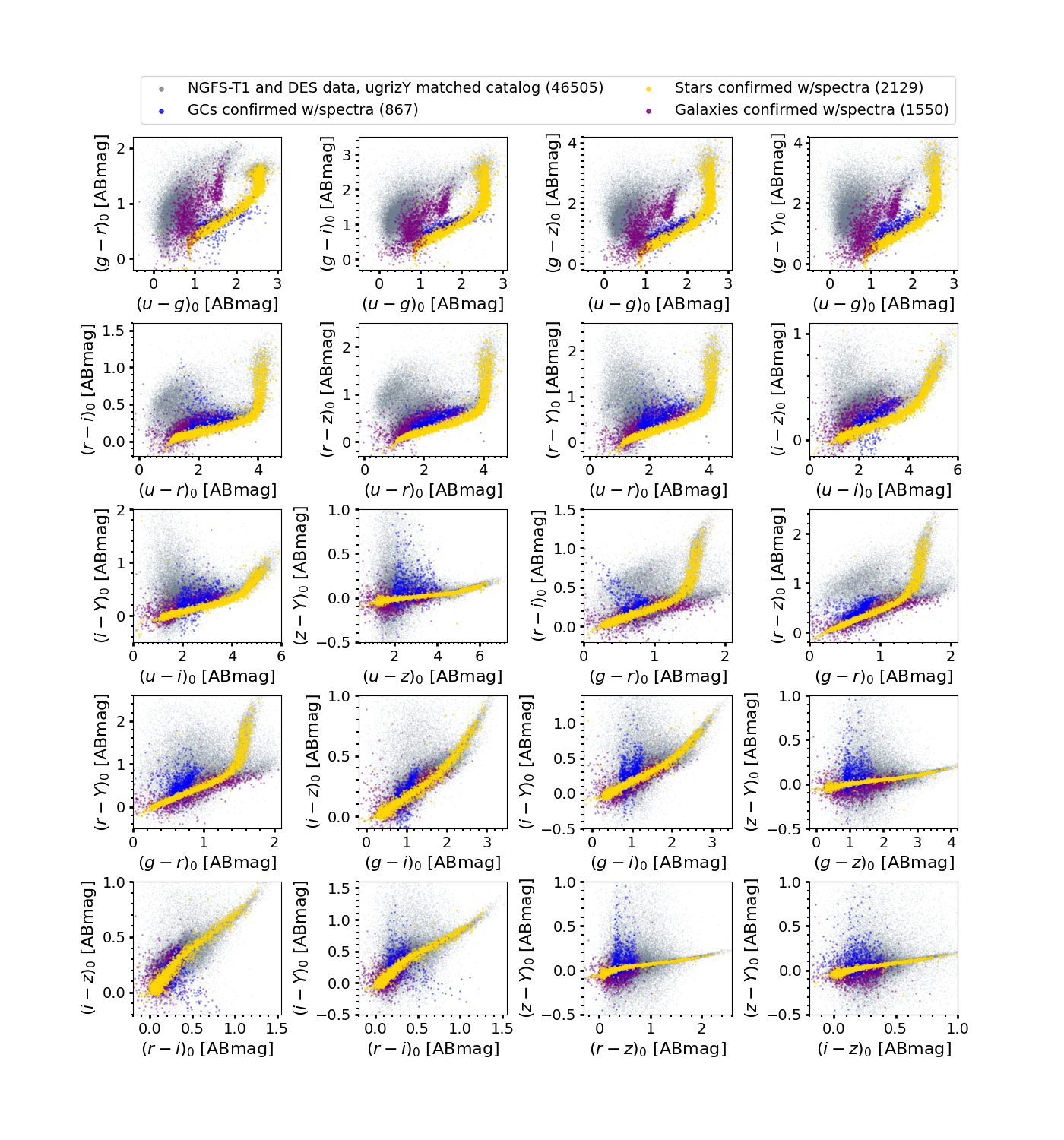}
     \caption{Color–color diagrams for all sources with multi-wavelength photometry in the core region of the Fornax galaxy cluster. Gray points represent the full photometric sample, while colored points highlight spectroscopically confirmed GCs (blue), stars (gold) and galaxies (purple) (see Sect. \ref{sect:traindataset}). All diagrams display the same set of sources, from the NGFS + DES DR2 matched catalog with photometric coverage in the $u'g'r'i'z'Y$ bands (see Sect. \ref{sect:lsst}).}
     \label{fig:f_cc_ugrizY}
\end{figure*}

The cross-matched catalog between NGFS-T1 and DES-DR2 contains a total of 46,505 objects. The cc-diagram combinations using the $u'g'r'i'z'Y$ bands are shown in Figure \ref{fig:f_cc_ugrizY}, which demonstrates the diagnostic capability of a broad SED coverage from $u'$ to $Y$ for distinguishing between different object types present in the future LSST deep imaging data.

\begin{figure*}[ht]
%trim=left bottom right top
 \centering
 \includegraphics[trim=1cm 0.8cm 0.8cm 0.8cm,clip,width=0.7\textwidth]{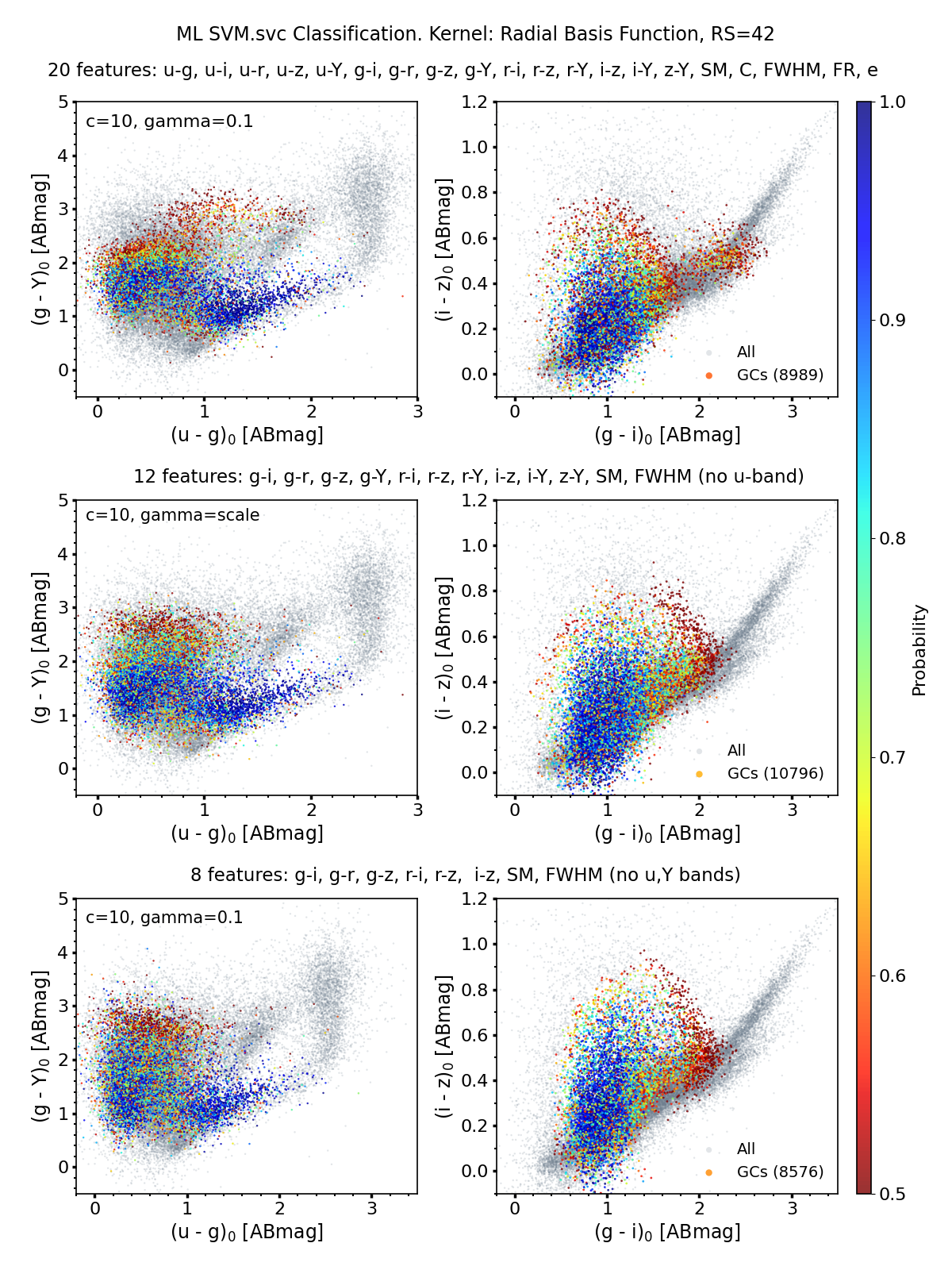} 
 \caption{Results of the \texttt{svm.SVC} model applied to the LSST filter system test. The top-row panels display results from the 20 feature model (full $u'g'r'i'z'Y$ coverage). The middle-row panels show the 12 feature model, excluding the u'-band. The bottom-row panels show the 8 feature model, excluding both $u'$- and $Y$- bands. Each row represents in the left: $u'g'Y$ cc-diagram and in the right: $g'i'z'$ cc-diagram, representative of configuration without $u'$- and $Y$- bands. The color scale indicates the classification probability assigned by the model for each source, with higher values reflecting greater confidence in the predicted object class.}
 \label{fig:result_cc_lsst}
\end{figure*}

Among the combinations, the $(u'-g')$ vs. $(g'-Y)$ diagram (top-right panel in Fig.\ref{fig:f_cc_ugrizY}) provides the clearest separation between the three primary sequences (GCs, stars, and galaxies). However, GCs still remain closely associated with the stellar locus. In the absence of $u'$-band (near-UV) data, this separation deteriorates significantly, highlighting the critical role of the $u'$-band in effective photometric classification.
Using the same \texttt{svm.SVC} configuration applied to the $u'g'i'JK_s$ photometry, we cross-matched the labeled (training/test) sample with the $u'g'r'i'z'Y$ bands to obtain the dataset in the LSST filter system. We present the classification results for three model configurations:

\begin{enumerate}[label=\roman*.]
\item 20 feature model (20F): full filter coverage, including all $u'g'r'i'z'Y$ and morpho-parameters (Top panel of Figure \ref{fig:result_cc_lsst} and its classification report in Table \ref{tab:clreport_20F_lsst}).
\item 12 feature model (12F), same as above but without the $u'$-band information (Middle-panel of Figure \ref{fig:result_cc_lsst} and its classification report in Table \ref{tab:clreport_12F_lsst}).
\item 8 feature model (8F): Lacking both $u'$ and $Y$ bands (bottom-panel of Figure \ref{fig:result_cc_lsst} and its classification report in Table \ref{tab:cl_report_8F_lsst}).
\end{enumerate}

In Figure \ref{fig:result_cc_lsst}, the first column shows the $u'g'Y$ cc- diagram, which provides the best discrimination between sources for the LSST-like filters. The second column presents the $g'i'z'$ diagram, representing the case where both the $u'$ and $Y$ bands are unavailable. The 20F model yields better classification performance than the 12F and 8F models. The precision (or purity) parameter for GCs - the objects most affected by selection bias - is systematically low in the LSST test, indicating a higher number of false positives compared to cases that include NIR filters. For the 20F model, the false positive rate is 7.8\%, increasing to 8.3\% and 10\% for the 12F and 8F models, respectively. The 12F model without the $u'$ band (middle panels) classified more sources than necessary, increasing the number of objects with high probability in the galaxy and star regions. When the classification relies solely on $g'r'i'z'$ data, the color-color diagrams in the bottom-right panel reveal significant source mixing within the galaxy region. It is worth noting that when $u'$ and $Y$ band data are unavailable, only the right-column panels would be used for interpretation, where no clear separation among the three object classes is possible. This illustrates the limitations imposed by the reduced filter coverage.

These results underscore that robust photometric classification using the LSST filter system alone critically depends on the availability and depth of the $u'$- and $Y$-bands. The full six-filter set significantly enhances class separation, particularly when complemented by NIR data. For example, from space-based missions, Euclid, which is already operational, and the upcoming Nancy Grace Roman Space Telescope, will provide essential NIR coverage. Euclid is expected to overlap with LSST across approximately $7000 deg^2$ of the Southern sky \citep{euclid0}, offering an excellent opportunity to combine optical and NIR data.
 
Such joint datasets will enable classification methodologies like the one presented here to perform substantially better, particularly in distinguishing between compact stellar systems and background galaxies. It is also worth noting that this overlapping survey area will play a key role in enhancing the scientific return of photometric classification efforts in large-scale extragalactic surveys.

%%%%%%%%%%%%%%%%%%%%%%%%%%%%%%%%%%%%%%%%%
\section{Summary \& future work} 
\label{sect:con}

We have developed a supervised machine learning classifier based on Support vector machine (SVM; implemented in using \texttt{scikit-learn}, \texttt{svm.SVC}) to distinguish between GCs, stars, and galaxies using deep photometric data from the central tile of the Next Generation Fornax Survey \citep[NGFS;][]{Munoz2015, eigethaler2018}. The model leverages both optical ($u'g'i'$) and near-infrared ($JK_s$) filters to construct color indices. In terms of color-color diagrams, the most discriminating combinations being $(u'-g')$ vs $(g'-K_s)$, referred to as $u'g'K_s$, as well as $u'i'K_s$, $u'JK_s$, and $g'JK_s$, to identify the three objects type. We use individual color indices, in addition to morphological parameters such as $FWHM$, spread model ($SM$), concentration index, ellipticity and $FLUX\_RADIUS$ to enhance class separation (see Sect. \ref{sect:f_selection}).

The model was trained and tested using a set of spectroscopically confirmed sources within the same field of view: 1,209 confirmed GCs, 2,151 foreground stars and 1,587 galaxies (see Sect. \ref{sect:traindataset}). Our results demonstrate that broad spectral energy distribution coverage, particularly spanning the near-UV to NIR, is essential for achieving high classification accuracy and minimizing confusion among compact stellar systems and background galaxies.

The optimized 7F model, which incorporates key color indices and structural parameters, demonstrates the most reliable performance, achieving 96.6\% accuracy and a misclassifications rate of 10.4\%. Even if the full feature 15F obtained higher scores in the classification report, we prove that some of the features are irrelevant for the models and few more are redundant, among them $(u'-i')$ color, which is correlated with $(u'-g')$ and $(g'-i')$ that have a higher importance for the \texttt{svm.SVC} model. This difference in scores is likely coming from overfitting, the 7F model offers a computational efficiency, and a more practical choice for large-scale applications.

In contrast, models trained on reduced filter sets, particularly those lacking near-UV ($u'$) and NIR ($JK_s$) data, exhibit substantially degraded performance. Simulations using LSST-like filters ($u'g'r'i'z'Y$), indicate that the $u'$ and $Y$ bands are essential for acceptable classification accuracy, although even with their inclusion, performance remains inferior to models that incorporate NIR coverage.

These results highlight the diagnostic power of broad spectral energy distribution coverage, spanning the near-UV to the near-IR, especially when combined with basic morphological parameters. The final classification catalog produced by our model will serve as the foundation for a detailed statistical analysis of the globular cluster population in the core of the Fornax cluster (Ordenes-Brice\~no et al. (\textit{in prep.})).

Looking ahead, the integration of data from upcoming space-based missions, such as Euclid and the Nancy Grace Roman Space Telescope, will further enhance photometric classification capabilities across large sky areas, especially in synergy with ground-based surveys like LSST.

%%%%%%%%%%%%%%%%%%%%%%%%%%%%%%%%%%%%%%%%%%%%%%%%%%%%%%%%%%%%%%%%%%%%%%%%%%%%%%%%%%%%
\begin{acknowledgements}
Y.~Ordenes-Briceño acknowledges support from the FONDECYT Postdoctorado 2021 No.~3210442 and ESO comit\'e mixto 2024. T.H.~Puzia gratefully acknowledges support through FONDECYT Regular No. 1201016. T.H.~Puzia, E.J.~Johnston and P.K.~Nayak acknowledge the support from the ANID CATA-BASAL project FB210003. J.P.~Carvajal and R.~Rahatgaonkar gratefully acknowledge support from ANID Beca Doctorado Nacional. This research has made use of the NASA/IPAC Extragalactic Database, which is funded by the National Aeronautics and Space Administration and operated by the California Institute of Technology. The authors deeply thank the citizens of Chile for their tax contributions to the national development of science and this project in these difficult post-pandemic times. The authors extend their gratitude to the researchers whose studies have been instrumental for this work.\\

{\it Facilities:} CTIO (4m Blanco/DECam), ESO:VISTA.\\
{\it Software:}\\
\texttt{NumPY/Python3} v2.1.0; \texttt{Pandas/Python3} v2.2.2; \texttt{Scipy/Python3} v1.14.1; \texttt{Sklearn/Python3} \citep[v.1.5.1][]{pedregosa2011}
\texttt{Astropy/Python3} v6.1.2 \cite[v6.1.2][]{astropy1,astropy2,astropy3};
\texttt{Matplotlib/Python3} \citep[v3.9.2][]{Hunter2007};
\texttt{Seaborn/Python3}\citep[v0.13.2][]{Waskom2021};
\texttt{TopCat} \citep[][]{taylor2005}

\end{acknowledgements}

\bibliographystyle{aa} % style aa.bst
\bibliography{references} % your references Yourfile.bib

\begin{thebibliography}{108}
\expandafter\ifx\csname natexlab\endcsname\relax\def\natexlab#1{#1}\fi

\bibitem[{{Abbott} {et~al.}(2021){Abbott}, {Adam{\'o}w}, {Aguena}, {Allam}, {Amon}, {Annis}, {Avila}, {Bacon}, {Banerji}, {Bechtol}, {Becker}, {Bernstein}, {Bertin}, {Bhargava}, {Bridle}, {Brooks}, {Burke}, {Carnero Rosell}, {Carrasco Kind}, {Carretero}, {Castander}, {Cawthon}, {Chang}, {Choi}, {Conselice}, {Costanzi}, {Crocce}, {da Costa}, {Davis}, {De Vicente}, {DeRose}, {Desai}, {Diehl}, {Dietrich}, {Drlica-Wagner}, {Eckert}, {Elvin-Poole}, {Everett}, {Evrard}, {Ferrero}, {Fert{\'e}}, {Flaugher}, {Fosalba}, {Friedel}, {Frieman}, {Garc{\'\i}a-Bellido}, {Gaztanaga}, {Gelman}, {Gerdes}, {Giannantonio}, {Gill}, {Gruen}, {Gruendl}, {Gschwend}, {Gutierrez}, {Hartley}, {Hinton}, {Hollowood}, {Honscheid}, {Huterer}, {James}, {Jeltema}, {Johnson}, {Kent}, {Kron}, {Kuehn}, {Kuropatkin}, {Lahav}, {Li}, {Lidman}, {Lin}, {MacCrann}, {Maia}, {Manning}, {Maloney}, {March}, {Marshall}, {Martini}, {Melchior}, {Menanteau}, {Miquel}, {Morgan}, {Myles}, {Neilsen}, {Ogando}, {Palmese}, {Paz-Chinch{\'o}n}, {Petravick},
  {Pieres}, {Plazas}, {Pond}, {Rodriguez-Monroy}, {Romer}, {Roodman}, {Rykoff}, {Sako}, {Sanchez}, {Santiago}, {Scarpine}, {Serrano}, {Sevilla-Noarbe}, {Smith}, {Smith}, {Soares-Santos}, {Suchyta}, {Swanson}, {Tarle}, {Thomas}, {To}, {Tremblay}, {Troxel}, {Tucker}, {Turner}, {Varga}, {Walker}, {Wechsler}, {Weller}, {Wester}, {Wilkinson}, {Yanny}, {Zhang}, {Nikutta}, {Fitzpatrick}, {Jacques}, {Scott}, {Olsen}, {Huang}, {Herrera}, {Juneau}, {Nidever}, {Weaver}, {Adean}, {Correia}, {de Freitas}, {Freitas}, {Singulani}, {Vila-Verde}, \& {Linea Science Server}}]{DESdr2}
{Abbott}, T.~M.~C., {Adam{\'o}w}, M., {Aguena}, M., {et~al.} 2021, \apjs, 255, 20

\bibitem[{{Alamo-Mart{\'\i}nez} {et~al.}(2013){Alamo-Mart{\'\i}nez}, {Blakeslee}, {Jee}, {C{\^o}t{\'e}}, {Ferrarese}, {Gonz{\'a}lez-L{\'o}pezlira}, {Jord{\'a}n}, {Meurer}, {Peng}, \& {West}}]{alamo2013}
{Alamo-Mart{\'\i}nez}, K.~A., {Blakeslee}, J.~P., {Jee}, M.~J., {et~al.} 2013, \apj, 775, 20

\bibitem[{{Anand} {et~al.}(2024){Anand}, {Tully}, {Cohen}, {Makarov}, {Makarova}, {Jensen}, {Blakeslee}, {Cantiello}, {Kourkchi}, \& {Raimondo}}]{Anand2024}
{Anand}, G.~S., {Tully}, R.~B., {Cohen}, Y., {et~al.} 2024, \apj, 973, 83

\bibitem[{{Angora} {et~al.}(2019){Angora}, {Brescia}, {Cavuoti}, {Paolillo}, {Longo}, {Cantiello}, {Capaccioli}, {D'Abrusco}, {D'Ago}, {Hilker}, {Iodice}, {Mieske}, {Napolitano}, {Peletier}, {Pota}, {Puzia}, {Riccio}, \& {Spavone}}]{Angora2019}
{Angora}, G., {Brescia}, M., {Cavuoti}, S., {et~al.} 2019, \mnras, 490, 4080

\bibitem[{{Ashman} \& {Zepf}(1992)}]{ashman92}
{Ashman}, K.~M. \& {Zepf}, S.~E. 1992, \apj, 384, 50

\bibitem[{{Astropy Collaboration} {et~al.}(2022){Astropy Collaboration}, {Price-Whelan}, {Lim}, {Earl}, {Starkman}, {Bradley}, {Shupe}, {Patil}, {Corrales}, {Brasseur}, {N{\"o}the}, {Donath}, {Tollerud}, {Morris}, {Ginsburg}, {Vaher}, {Weaver}, {Tocknell}, {Jamieson}, {van Kerkwijk}, {Robitaille}, {Merry}, {Bachetti}, {G{\"u}nther}, {Aldcroft}, {Alvarado-Montes}, {Archibald}, {B{\'o}di}, {Bapat}, {Barentsen}, {Baz{\'a}n}, {Biswas}, {Boquien}, {Burke}, {Cara}, {Cara}, {Conroy}, {Conseil}, {Craig}, {Cross}, {Cruz}, {D'Eugenio}, {Dencheva}, {Devillepoix}, {Dietrich}, {Eigenbrot}, {Erben}, {Ferreira}, {Foreman-Mackey}, {Fox}, {Freij}, {Garg}, {Geda}, {Glattly}, {Gondhalekar}, {Gordon}, {Grant}, {Greenfield}, {Groener}, {Guest}, {Gurovich}, {Handberg}, {Hart}, {Hatfield-Dodds}, {Homeier}, {Hosseinzadeh}, {Jenness}, {Jones}, {Joseph}, {Kalmbach}, {Karamehmetoglu}, {Ka{\l}uszy{\'n}ski}, {Kelley}, {Kern}, {Kerzendorf}, {Koch}, {Kulumani}, {Lee}, {Ly}, {Ma}, {MacBride}, {Maljaars}, {Muna}, {Murphy}, {Norman},
  {O'Steen}, {Oman}, {Pacifici}, {Pascual}, {Pascual-Granado}, {Patil}, {Perren}, {Pickering}, {Rastogi}, {Roulston}, {Ryan}, {Rykoff}, {Sabater}, {Sakurikar}, {Salgado}, {Sanghi}, {Saunders}, {Savchenko}, {Schwardt}, {Seifert-Eckert}, {Shih}, {Jain}, {Shukla}, {Sick}, {Simpson}, {Singanamalla}, {Singer}, {Singhal}, {Sinha}, {Sip{\H{o}}cz}, {Spitler}, {Stansby}, {Streicher}, {{\v{S}}umak}, {Swinbank}, {Taranu}, {Tewary}, {Tremblay}, {de Val-Borro}, {Van Kooten}, {Vasovi{\'c}}, {Verma}, {de Miranda Cardoso}, {Williams}, {Wilson}, {Winkel}, {Wood-Vasey}, {Xue}, {Yoachim}, {Zhang}, {Zonca}, \& {Astropy Project Contributors}}]{astropy3}
{Astropy Collaboration}, {Price-Whelan}, A.~M., {Lim}, P.~L., {et~al.} 2022, \apj, 935, 167

\bibitem[{{Astropy Collaboration} {et~al.}(2018){Astropy Collaboration}, {Price-Whelan}, {Sip{\H{o}}cz}, {G{\"u}nther}, {Lim}, {Crawford}, {Conseil}, {Shupe}, {Craig}, {Dencheva}, {Ginsburg}, {VanderPlas}, {Bradley}, {P{\'e}rez-Su{\'a}rez}, {de Val-Borro}, {Aldcroft}, {Cruz}, {Robitaille}, {Tollerud}, {Ardelean}, {Babej}, {Bach}, {Bachetti}, {Bakanov}, {Bamford}, {Barentsen}, {Barmby}, {Baumbach}, {Berry}, {Biscani}, {Boquien}, {Bostroem}, {Bouma}, {Brammer}, {Bray}, {Breytenbach}, {Buddelmeijer}, {Burke}, {Calderone}, {Cano Rodr{\'\i}guez}, {Cara}, {Cardoso}, {Cheedella}, {Copin}, {Corrales}, {Crichton}, {D'Avella}, {Deil}, {Depagne}, {Dietrich}, {Donath}, {Droettboom}, {Earl}, {Erben}, {Fabbro}, {Ferreira}, {Finethy}, {Fox}, {Garrison}, {Gibbons}, {Goldstein}, {Gommers}, {Greco}, {Greenfield}, {Groener}, {Grollier}, {Hagen}, {Hirst}, {Homeier}, {Horton}, {Hosseinzadeh}, {Hu}, {Hunkeler}, {Ivezi{\'c}}, {Jain}, {Jenness}, {Kanarek}, {Kendrew}, {Kern}, {Kerzendorf}, {Khvalko}, {King}, {Kirkby}, {Kulkarni},
  {Kumar}, {Lee}, {Lenz}, {Littlefair}, {Ma}, {Macleod}, {Mastropietro}, {McCully}, {Montagnac}, {Morris}, {Mueller}, {Mumford}, {Muna}, {Murphy}, {Nelson}, {Nguyen}, {Ninan}, {N{\"o}the}, {Ogaz}, {Oh}, {Parejko}, {Parley}, {Pascual}, {Patil}, {Patil}, {Plunkett}, {Prochaska}, {Rastogi}, {Reddy Janga}, {Sabater}, {Sakurikar}, {Seifert}, {Sherbert}, {Sherwood-Taylor}, {Shih}, {Sick}, {Silbiger}, {Singanamalla}, {Singer}, {Sladen}, {Sooley}, {Sornarajah}, {Streicher}, {Teuben}, {Thomas}, {Tremblay}, {Turner}, {Terr{\'o}n}, {van Kerkwijk}, {de la Vega}, {Watkins}, {Weaver}, {Whitmore}, {Woillez}, {Zabalza}, \& {Astropy Contributors}}]{astropy2}
{Astropy Collaboration}, {Price-Whelan}, A.~M., {Sip{\H{o}}cz}, B.~M., {et~al.} 2018, \aj, 156, 123

\bibitem[{{Astropy Collaboration} {et~al.}(2013){Astropy Collaboration}, {Robitaille}, {Tollerud}, {Greenfield}, {Droettboom}, {Bray}, {Aldcroft}, {Davis}, {Ginsburg}, {Price-Whelan}, {Kerzendorf}, {Conley}, {Crighton}, {Barbary}, {Muna}, {Ferguson}, {Grollier}, {Parikh}, {Nair}, {Unther}, {Deil}, {Woillez}, {Conseil}, {Kramer}, {Turner}, {Singer}, {Fox}, {Weaver}, {Zabalza}, {Edwards}, {Azalee Bostroem}, {Burke}, {Casey}, {Crawford}, {Dencheva}, {Ely}, {Jenness}, {Labrie}, {Lim}, {Pierfederici}, {Pontzen}, {Ptak}, {Refsdal}, {Servillat}, \& {Streicher}}]{astropy1}
{Astropy Collaboration}, {Robitaille}, T.~P., {Tollerud}, E.~J., {et~al.} 2013, \aap, 558, A33

\bibitem[{{Barbisan} {et~al.}(2022){Barbisan}, {Huang}, {Dage}, {Haggard}, {Arnason}, {Bahramian}, {Clarkson}, {Kundu}, \& {Zepf}}]{Barbisan2022}
{Barbisan}, E., {Huang}, J., {Dage}, K.~C., {et~al.} 2022, \mnras, 514, 943

\bibitem[{{Bergond} {et~al.}(2007){Bergond}, {Athanassoula}, {Leon}, {Balkowski}, {Cayatte}, {Chemin}, {Guzm{\'a}n}, {Meylan}, \& {Prugniel}}]{bergond2007}
{Bergond}, G., {Athanassoula}, E., {Leon}, S., {et~al.} 2007, \aap, 464, L21

\bibitem[{{Bertin}(2006)}]{Bertin06}
{Bertin}, E. 2006, in Astronomical Society of the Pacific Conference Series, Vol. 351, Astronomical Data Analysis Software and Systems XV, ed. C.~{Gabriel}, C.~{Arviset}, D.~{Ponz}, \& S.~{Enrique}, 112

\bibitem[{{Bertin}(2011)}]{Bertin2011}
{Bertin}, E. 2011, in Astronomical Society of the Pacific Conference Series, Vol. 442, Astronomical Data Analysis Software and Systems XX, ed. I.~N. {Evans}, A.~{Accomazzi}, D.~J. {Mink}, \& A.~H. {Rots}, 435

\bibitem[{{Bertin} \& {Arnouts}(1996)}]{Bertin96}
{Bertin}, E. \& {Arnouts}, S. 1996, \aaps, 117, 393

\bibitem[{{Bertin} {et~al.}(2002){Bertin}, {Mellier}, {Radovich}, {Missonnier}, {Didelon}, \& {Morin}}]{Bertin02}
{Bertin}, E., {Mellier}, Y., {Radovich}, M., {et~al.} 2002, in Astronomical Society of the Pacific Conference Series, Vol. 281, Astronomical Data Analysis Software and Systems XI, ed. D.~A. {Bohlender}, D.~{Durand}, \& T.~H. {Handley}, 228

\bibitem[{Blanton \& Roweis(2007)}]{Blan07}
Blanton, M.~R. \& Roweis, S. 2007, The Astronomical Journal, 133, 734

\bibitem[{{Breiman}(2001)}]{Breiman2001}
{Breiman}, L. 2001, Machine Learning, 45, 5

\bibitem[{{Brodie} \& {Strader}(2006)}]{brodie2006}
{Brodie}, J.~P. \& {Strader}, J. 2006, \araa, 44, 193

\bibitem[{{Cantiello} {et~al.}(2020){Cantiello}, {Venhola}, {Grado}, {Paolillo}, {D'Abrusco}, {Raimondo}, {Quintini}, {Hilker}, {Mieske}, {Tortora}, {Spavone}, {Capaccioli}, {Iodice}, {Peletier}, {Barroso}, {Limatola}, {Napolitano}, {Schipani}, {van de Ven}, {Gentile}, \& {Covone}}]{cantiello2020}
{Cantiello}, M., {Venhola}, A., {Grado}, A., {et~al.} 2020, \aap, 639, A136

\bibitem[{{Cenarro} {et~al.}(2019){Cenarro}, {Moles}, {Crist{\'o}bal-Hornillos}, {Mar{\'\i}n-Franch}, {Ederoclite}, {Varela}, {L{\'o}pez-Sanjuan}, {Hern{\'a}ndez-Monteagudo}, {Angulo}, {V{\'a}zquez Rami{\'o}}, {Viironen}, {Bonoli}, {Orsi}, {Hurier}, {San Roman}, {Greisel}, {Vilella-Rojo}, {D{\'\i}az-Garc{\'\i}a}, {Logro{\~n}o-Garc{\'\i}a}, {Gurung-L{\'o}pez}, {Spinoso}, {Izquierdo-Villalba}, {Aguerri}, {Allende Prieto}, {Bonatto}, {Carvano}, {Chies-Santos}, {Daflon}, {Dupke}, {Falc{\'o}n-Barroso}, {Gon{\c{c}}alves}, {Jim{\'e}nez-Teja}, {Molino}, {Placco}, {Solano}, {Whitten}, {Abril}, {Ant{\'o}n}, {Bello}, {Bielsa de Toledo}, {Castillo-Ram{\'\i}rez}, {Chueca}, {Civera}, {D{\'\i}az-Mart{\'\i}n}, {Dom{\'\i}nguez-Mart{\'\i}nez}, {Garzar{\'a}n-Calderaro}, {Hern{\'a}ndez-Fuertes}, {Iglesias-Marzoa}, {I{\~n}iguez}, {Jim{\'e}nez Ruiz}, {Kruuse}, {Lamadrid}, {Lasso-Cabrera}, {L{\'o}pez-Alegre}, {L{\'o}pez-Sainz}, {Ma{\'\i}cas}, {Moreno-Signes}, {Muniesa}, {Rodr{\'\i}guez-Llano}, {Rueda-Teruel}, {Rueda-Teruel},
  {Soriano-Lagu{\'\i}a}, {Tilve}, {Valdivielso}, {Yanes-D{\'\i}az}, {Alcaniz}, {Mendes de Oliveira}, {Sodr{\'e}}, {Coelho}, {Lopes de Oliveira}, {Tamm}, {Xavier}, {Abramo}, {Akras}, {Alfaro}, {Alvarez-Candal}, {Ascaso}, {Beasley}, {Beers}, {Borges Fernandes}, {Bruzual}, {Buzzo}, {Carrasco}, {Cepa}, {Cortesi}, {Costa-Duarte}, {De Pr{\'a}}, {Favole}, {Galarza}, {Galbany}, {Garcia}, {Gonz{\'a}lez Delgado}, {Gonz{\'a}lez-Serrano}, {Guti{\'e}rrez-Soto}, {Hernandez-Jimenez}, {Kanaan}, {Kuncarayakti}, {Landim}, {Laur}, {Licandro}, {Lima Neto}, {Lyman}, {Ma{\'\i}z Apell{\'a}niz}, {Miralda-Escud{\'e}}, {Morate}, {Nogueira-Cavalcante}, {Novais}, {Oncins}, {Oteo}, {Overzier}, {Pereira}, {Rebassa-Mansergas}, {Reis}, {Roig}, {Sako}, {Salvador-Rusi{\~n}ol}, {Sampedro}, {S{\'a}nchez-Bl{\'a}zquez}, {Santos}, {Schmidtobreick}, {Siffert}, {Telles}, \& {Vilchez}}]{Cenarro2019}
{Cenarro}, A.~J., {Moles}, M., {Crist{\'o}bal-Hornillos}, D., {et~al.} 2019, \aap, 622, A176

\bibitem[{Chapelle {et~al.}(2002)Chapelle, Vapnik, Bousquet, \& Mukherjee}]{Chapelle2002}
Chapelle, O., Vapnik, V., Bousquet, O., \& Mukherjee, S. 2002, Machine Learning, 46, 131 – 159, cited by: 2114; All Open Access, Bronze Open Access

\bibitem[{{Chaturvedi} {et~al.}(2022){Chaturvedi}, {Hilker}, {Cantiello}, {Napolitano}, {van de Ven}, {Spiniello}, {Fahrion}, {Paolillo}, {Gatto}, \& {Puzia}}]{chaturvedi22}
{Chaturvedi}, A., {Hilker}, M., {Cantiello}, M., {et~al.} 2022, \aap, 657, A93

\bibitem[{{Chen} {et~al.}(2025){Chen}, {Mo}, \& {Wang}}]{chen2025}
{Chen}, Y., {Mo}, H., \& {Wang}, H. 2025, \mnras, 540, 1235

\bibitem[{{Chies-Santos} {et~al.}(2022){Chies-Santos}, {de Souza}, {Caso}, {Ennis}, {de Souza}, {Barbosa}, {Chen}, {Cenarro}, {Ederoclite}, {Crist{\'o}bal-Hornillos}, {Hern{\'a}ndez-Monteagudo}, {L{\'o}pez-Sanjuan}, {Mar{\'\i}n-Franch}, {Moles}, {Varela}, {V{\'a}zquez Rami{\'o}}, {Dupke}, {Sodr{\'e}}, \& {Angulo}}]{chiessantos2022}
{Chies-Santos}, A.~L., {de Souza}, R.~S., {Caso}, J.~P., {et~al.} 2022, \mnras, 516, 1320

\bibitem[{{Chilingarian} {et~al.}(2011){Chilingarian}, {Mieske}, {Hilker}, \& {Infante}}]{chilingarian2011}
{Chilingarian}, I.~V., {Mieske}, S., {Hilker}, M., \& {Infante}, L. 2011, \mnras, 412, 1627

\bibitem[{{Cooper} {et~al.}(2025){Cooper}, {Frenk}, {Hellwing}, \& {Bose}}]{Cooper2025}
{Cooper}, A.~P., {Frenk}, C.~S., {Hellwing}, W.~A., \& {Bose}, S. 2025, \mnras, 540, 2049

\bibitem[{Cortes \& Vapnik(1995)}]{Cortes1995}
Cortes, C. \& Vapnik, V. 1995, Machine learning, 20, 273

\bibitem[{Crammer \& Singer(2002)}]{crammer2001}
Crammer, K. \& Singer, Y. 2002, J. Mach. Learn. Res., 2, 265–292

\bibitem[{{Cristiani} {et~al.}(2001){Cristiani}, {Grazian}, {Omizzolo}, \& {Corbally}}]{cristiani2001}
{Cristiani}, S., {Grazian}, A., {Omizzolo}, A., \& {Corbally}, C. 2001, in Mining the Sky, ed. A.~J. {Banday}, S.~{Zaroubi}, \& M.~{Bartelmann}, 154

\bibitem[{{Drinkwater} {et~al.}(2001){Drinkwater}, {Gregg}, \& {Colless}}]{drinkwater2001}
{Drinkwater}, M.~J., {Gregg}, M.~D., \& {Colless}, M. 2001, \apjl, 548, L139

\bibitem[{{Drinkwater} {et~al.}(2000){Drinkwater}, {Phillipps}, {Jones}, {Gregg}, {Deady}, {Davies}, {Parker}, {Sadler}, \& {Smith}}]{drinkwater2000}
{Drinkwater}, M.~J., {Phillipps}, S., {Jones}, J.~B., {et~al.} 2000, \aap, 355, 900

\bibitem[{{Durrell} {et~al.}(2014){Durrell}, {C{\^o}t{\'e}}, {Peng}, {Blakeslee}, {Ferrarese}, {Mihos}, {Puzia}, {Lan{\c{c}}on}, {Liu}, {Zhang}, {Cuillandre}, {McConnachie}, {Jord{\'a}n}, {Accetta}, {Boissier}, {Boselli}, {Courteau}, {Duc}, {Emsellem}, {Gwyn}, {Mei}, \& {Taylor}}]{durrell2014}
{Durrell}, P.~R., {C{\^o}t{\'e}}, P., {Peng}, E.~W., {et~al.} 2014, \apj, 794, 103

\bibitem[{{Eigenthaler} {et~al.}(2018){Eigenthaler}, {Puzia}, {Taylor}, {Ordenes-Brice{\~n}o}, {Mu{\~n}oz}, {Ribbeck}, {Alamo-Mart{\'\i}nez}, {Zhang}, {{\'A}ngel}, {Capaccioli}, {C{\^o}t{\'e}}, {Ferrarese}, {Galaz}, {Grebel}, {Hempel}, {Hilker}, {Lan{\c{c}}on}, {Mieske}, {Miller}, {Paolillo}, {Powalka}, {Richtler}, {Roediger}, {Rong}, {S{\'a}nchez-Janssen}, \& {Spengler}}]{eigethaler2018}
{Eigenthaler}, P., {Puzia}, T.~H., {Taylor}, M.~A., {et~al.} 2018, \apj, 855, 142

\bibitem[{{Euclid Collaboration} {et~al.}(2025{\natexlab{a}}){Euclid Collaboration}, {Mellier}, {Abdurro'uf}, {Acevedo Barroso}, {Ach{\'u}carro}, {Adamek}, {Adam}, {Addison}, {Aghanim}, {Aguena}, {Ajani}, {Akrami}, {Al-Bahlawan}, {Alavi}, {Albuquerque}, {Alestas}, {Alguero}, {Allaoui}, {Allen}, {Allevato}, {Alonso-Tetilla}, {Altieri}, {Alvarez-Candal}, {Alvi}, {Amara}, {Amendola}, {Amiaux}, {Andika}, {Andreon}, {Andrews}, {Angora}, {Angulo}, {Annibali}, {Anselmi}, {Anselmi}, {Arcari}, {Archidiacono}, {Aric{\`o}}, {Arnaud}, {Arnouts}, {Asgari}, {Asorey}, {Atayde}, {Atek}, {Atrio-Barandela}, {Aubert}, {Aubourg}, {Auphan}, {Auricchio}, {Aussel}, {Aussel}, {Avelino}, {Avgoustidis}, {Avila}, {Awan}, {Azzollini}, {Baccigalupi}, {Bachelet}, {Bacon}, {Baes}, {Bagley}, {Bahr-Kalus}, {Balaguera-Antolinez}, {Balbinot}, {Balcells}, {Baldi}, {Baldry}, {Balestra}, {Ballardini}, {Ballester}, {Balogh}, {Ba{\~n}ados}, {Barbier}, {Bardelli}, {Baron}, {Barreiro}, {Barrena}, {Barriere}, {Barros}, {Barthelemy}, {Bartolo},
  {Basset}, {Battaglia}, {Battisti}, {Baugh}, {Baumont}, {Bazzanini}, {Beaulieu}, {Beckmann}, {Belikov}, {Bel}, {Bellagamba}, {Bella}, {Bellini}, {Benabed}, {Bender}, {Benevento}, {Bennett}, {Benson}, {Bergamini}, {Bermejo-Climent}, {Bernardeau}, {Bertacca}, {Berthe}, {Berthier}, {Bethermin}, {Beutler}, {Bevillon}, {Bhargava}, {Bhatawdekar}, {Bianchi}, {Bisigello}, {Biviano}, {Blake}, {Blanchard}, {Blazek}, {Blot}, {Bosco}, {Bodendorf}, {Boenke}, {B{\"o}hringer}, {Boldrini}, {Bolzonella}, {Bonchi}, {Bonici}, {Bonino}, {Bonino}, {Bonvin}, {Bon}, {Booth}, {Borgani}, {Borlaff}, {Borsato}, {Bose}, {Botticella}, {Boucaud}, {Bouche}, {Boucher}, {Boutigny}, {Bouvard}, {Bouwens}, {Bouy}, {Bowler}, {Bozza}, {Bozzo}, {Branchini}, {Brando}, {Brau-Nogue}, {Brekke}, {Bremer}, {Brescia}, {Breton}, {Brinchmann}, {Brinckmann}, {Brockley-Blatt}, {Brodwin}, {Brouard}, {Brown}, {Bruton}, {Bucko}, {Buddelmeijer}, {Buenadicha}, {Buitrago}, {Burger}, {Burigana}, {Busillo}, {Busonero}, {Cabanac}, {Cabayol-Garcia}, {Cagliari},
  {Caillat}, {Caillat}, {Calabrese}, {Calabro}, {Calderone}, {Calura}, {Camacho Quevedo}, {Camera}, {Campos}, {Ca{\~n}as-Herrera}, {Candini}, {Cantiello}, {Capobianco}, {Cappellaro}, {Cappelluti}, {Cappi}, {Caputi}, {Cara}, {Carbone}, {Cardone}, {Carella}, {Carlberg}, {Carle}, {Carminati}, {Caro}, {Carrasco}, {Carretero}, {Carrilho}, {Carron Duque}, \& {Carry}}]{euclid0}
{Euclid Collaboration}, {Mellier}, Y., {Abdurro'uf}, {et~al.} 2025{\natexlab{a}}, \aap, 697, A1

\bibitem[{{Euclid Collaboration} {et~al.}(2025{\natexlab{b}}){Euclid Collaboration}, {Voggel}, {Lan{\c{c}}on}, {Saifollahi}, {Larsen}, {Cantiello}, {Rejkuba}, {Cuillandre}, {Hudelot}, {Nucita}, {Urbano}, {Romelli}, {Raj}, {Schirmer}, {Tortora}, {Abdurro'uf}, {Annibali}, {Baes}, {Boldrini}, {Cabanac}, {Carollo}, {Conselice}, {Duc}, {Ferguson}, {Hunt}, {Knapen}, {Lonare}, {Marleau}, {Paolillo}, {Poulain}, {S{\'a}nchez-Janssen}, {Sola}, {Andreon}, {Auricchio}, {Baccigalupi}, {Baldi}, {Bardelli}, {Bodendorf}, {Bonino}, {Branchini}, {Brescia}, {Brinchmann}, {Camera}, {Capobianco}, {Carbone}, {Carlberg}, {Carretero}, {Casas}, {Castellano}, {Castignani}, {Cavuoti}, {Cimatti}, {Colodro-Conde}, {Congedo}, {Conversi}, {Copin}, {Courbin}, {Courtois}, {Cropper}, {Da Silva}, {Degaudenzi}, {De Lucia}, {Di Giorgio}, {Dinis}, {Dubath}, {Dupac}, {Dusini}, {Farina}, {Farrens}, {Ferriol}, {Fotopoulou}, {Frailis}, {Franceschi}, {Fumana}, {Galeotta}, {George}, {Gillard}, {Gillis}, {Giocoli}, {G{\'o}mez-Alvarez}, {Grazian},
  {Grupp}, {Haugan}, {Hoekstra}, {Holmes}, {Hook}, {Hormuth}, {Hornstrup}, {Jahnke}, {Keih{\"a}nen}, {Kermiche}, {Kiessling}, {Kilbinger}, {Kohley}, {Kubik}, {K{\"u}mmel}, {Kunz}, {Kurki-Suonio}, {Laureijs}, {Liebing}, {Ligori}, {Lilje}, {Lindholm}, {Lloro}, {Maino}, {Maiorano}, {Mansutti}, {Marggraf}, {Markovic}, {Martinelli}, {Martinet}, {Marulli}, {Massey}, {Maurogordato}, {Medinaceli}, {Mei}, {Mellier}, {Meneghetti}, {Merlin}, {Meylan}, {Moresco}, {Moscardini}, {Munari}, {Nakajima}, {Neissner}, {Nichol}, {Niemi}, {Nightingale}, {Padilla}, {Paltani}, {Pasian}, {Pedersen}, {Pettorino}, {Pires}, {Polenta}, {Poncet}, {Popa}, {Pozzetti}, {Raison}, {Rebolo}, {Renzi}, {Rhodes}, {Riccio}, {Roncarelli}, {Rossetti}, {Saglia}, {Sakr}, {Sapone}, {Sartoris}, {Scaramella}, {Schneider}, {Schrabback}, {Secroun}, {Sefusatti}, {Seidel}, {Serrano}, {Sirignano}, {Sirri}, {Stanco}, {Steinwagner}, {Surace}, {Tallada-Cresp{\'\i}}, {Teplitz}, {Tereno}, {Toledo-Moreo}, {Torradeflot}, {Tutusaus}, {Valentijn}, {Valenziano},
  {Vassallo}, {Veropalumbo}, {Wang}, {Weller}, {Zamorani}, {Zucca}, {Biviano}, {Bolzonella}, {Bozzo}, {Burigana}, {Calabrese}, {Di Ferdinando}, {Escartin Vigo}, {Farinelli}, {Gracia-Carpio}, {Mauri}, {Scottez}, {Tenti}, {Viel}, {Wiesmann}, {Akrami}, {Allevato}, {Anselmi}, {Ballardini}, {Bethermin}, {Blanchard}, {Blot}, {Borgani}, {Borlaff}, {Bruton}, \& {Calabro}}]{Euclid2025}
{Euclid Collaboration}, {Voggel}, K., {Lan{\c{c}}on}, A., {et~al.} 2025{\natexlab{b}}, \aap, 693, A251

\bibitem[{{Fahrion} {et~al.}(2020){Fahrion}, {Lyubenova}, {Hilker}, {van de Ven}, {Falc{\'o}n-Barroso}, {Leaman}, {Mart{\'\i}n-Navarro}, {Bittner}, {Coccato}, {Corsini}, {Gadotti}, {Iodice}, {McDermid}, {Pinna}, {Sarzi}, {Viaene}, {de Zeeuw}, \& {Zhu}}]{fahrion2020}
{Fahrion}, K., {Lyubenova}, M., {Hilker}, M., {et~al.} 2020, \aap, 637, A26

\bibitem[{Fawcett(2006)}]{faw2006}
Fawcett, T. 2006, Pattern Recognition Letters, 27, 861, rOC Analysis in Pattern Recognition

\bibitem[{{Ferguson}(1989)}]{ferguson1989}
{Ferguson}, H.~C. 1989, \aj, 98, 367

\bibitem[{{Ferrarese} {et~al.}(2012){Ferrarese}, {C{\^o}t{\'e}}, {Cuillandre}, {Gwyn}, {Peng}, {MacArthur}, {Duc}, {Boselli}, {Mei}, {Erben}, {McConnachie}, {Durrell}, {Mihos}, {Jord{\'a}n}, {Lan{\c{c}}on}, {Puzia}, {Emsellem}, {Balogh}, {Blakeslee}, {van Waerbeke}, {Gavazzi}, {Vollmer}, {Kavelaars}, {Woods}, {Ball}, {Boissier}, {Courteau}, {Ferriere}, {Gavazzi}, {Hildebrandt}, {Hudelot}, {Huertas-Company}, {Liu}, {McLaughlin}, {Mellier}, {Milkeraitis}, {Schade}, {Balkowski}, {Bournaud}, {Carlberg}, {Chapman}, {Hoekstra}, {Peng}, {Sawicki}, {Simard}, {Taylor}, {Tully}, {van Driel}, {Wilson}, {Burdullis}, {Mahoney}, \& {Manset}}]{Ferrarese2012}
{Ferrarese}, L., {C{\^o}t{\'e}}, P., {Cuillandre}, J.-C., {et~al.} 2012, \apjs, 200, 4

\bibitem[{{Fioc} \& {Rocca-Volmerange}(1997)}]{fioc97}
{Fioc}, M. \& {Rocca-Volmerange}, B. 1997, \aap, 326, 950

\bibitem[{{Fitzpatrick}(1999)}]{Fitz99}
{Fitzpatrick}, E.~L. 1999, \pasp, 111, 63

\bibitem[{{Flaugher} {et~al.}(2015){Flaugher}, {Diehl}, {Honscheid}, {Abbott}, {Alvarez}, {Angstadt}, {Annis}, {Antonik}, {Ballester}, {Beaufore}, {Bernstein}, {Bernstein}, {Bigelow}, {Bonati}, {Boprie}, {Brooks}, {Buckley-Geer}, {Campa}, {Cardiel-Sas}, {Castander}, {Castilla}, {Cease}, {Cela-Ruiz}, {Chappa}, {Chi}, {Cooper}, {da Costa}, {Dede}, {Derylo}, {DePoy}, {de Vicente}, {Doel}, {Drlica-Wagner}, {Eiting}, {Elliott}, {Emes}, {Estrada}, {Fausti Neto}, {Finley}, {Flores}, {Frieman}, {Gerdes}, {Gladders}, {Gregory}, {Gutierrez}, {Hao}, {Holland}, {Holm}, {Huffman}, {Jackson}, {James}, {Jonas}, {Karcher}, {Karliner}, {Kent}, {Kessler}, {Kozlovsky}, {Kron}, {Kubik}, {Kuehn}, {Kuhlmann}, {Kuk}, {Lahav}, {Lathrop}, {Lee}, {Levi}, {Lewis}, {Li}, {Mandrichenko}, {Marshall}, {Martinez}, {Merritt}, {Miquel}, {Mu{\~n}oz}, {Neilsen}, {Nichol}, {Nord}, {Ogando}, {Olsen}, {Palaio}, {Patton}, {Peoples}, {Plazas}, {Rauch}, {Reil}, {Rheault}, {Roe}, {Rogers}, {Roodman}, {Sanchez}, {Scarpine}, {Schindler}, {Schmidt},
  {Schmitt}, {Schubnell}, {Schultz}, {Schurter}, {Scott}, {Serrano}, {Shaw}, {Smith}, {Soares-Santos}, {Stefanik}, {Stuermer}, {Suchyta}, {Sypniewski}, {Tarle}, {Thaler}, {Tighe}, {Tran}, {Tucker}, {Walker}, {Wang}, {Watson}, {Weaverdyck}, {Wester}, {Woods}, {Yanny}, \& {DES Collaboration}}]{Flaugher15}
{Flaugher}, B., {Diehl}, H.~T., {Honscheid}, K., {et~al.} 2015, \aj, 150, 150

\bibitem[{{Forbes} {et~al.}(1997){Forbes}, {Brodie}, \& {Grillmair}}]{Forbes97}
{Forbes}, D.~A., {Brodie}, J.~P., \& {Grillmair}, C.~J. 1997, \aj, 113, 1652

\bibitem[{{Forbes} {et~al.}(2018){Forbes}, {Read}, {Gieles}, \& {Collins}}]{forbes2018}
{Forbes}, D.~A., {Read}, J.~I., {Gieles}, M., \& {Collins}, M. L.~M. 2018, \mnras, 481, 5592

\bibitem[{{Gaia Collaboration} {et~al.}(2021){Gaia Collaboration}, {Brown}, {Vallenari}, {Prusti}, {de Bruijne}, {Babusiaux}, {Biermann}, {Creevey}, {Evans}, {Eyer}, {Hutton}, {Jansen}, {Jordi}, {Klioner}, {Lammers}, {Lindegren}, {Luri}, {Mignard}, {Panem}, {Pourbaix}, {Randich}, {Sartoretti}, {Soubiran}, {Walton}, {Arenou}, {Bailer-Jones}, {Bastian}, {Cropper}, {Drimmel}, {Katz}, {Lattanzi}, {van Leeuwen}, {Bakker}, {Cacciari}, {Casta{\~n}eda}, {De Angeli}, {Ducourant}, {Fabricius}, {Fouesneau}, {Fr{\'e}mat}, {Guerra}, {Guerrier}, {Guiraud}, {Jean-Antoine Piccolo}, {Masana}, {Messineo}, {Mowlavi}, {Nicolas}, {Nienartowicz}, {Pailler}, {Panuzzo}, {Riclet}, {Roux}, {Seabroke}, {Sordo}, {Tanga}, {Th{\'e}venin}, {Gracia-Abril}, {Portell}, {Teyssier}, {Altmann}, {Andrae}, {Bellas-Velidis}, {Benson}, {Berthier}, {Blomme}, {Brugaletta}, {Burgess}, {Busso}, {Carry}, {Cellino}, {Cheek}, {Clementini}, {Damerdji}, {Davidson}, {Delchambre}, {Dell'Oro}, {Fern{\'a}ndez-Hern{\'a}ndez}, {Galluccio}, {Garc{\'\i}a-Lario},
  {Garcia-Reinaldos}, {Gonz{\'a}lez-N{\'u}{\~n}ez}, {Gosset}, {Haigron}, {Halbwachs}, {Hambly}, {Harrison}, {Hatzidimitriou}, {Heiter}, {Hern{\'a}ndez}, {Hestroffer}, {Hodgkin}, {Holl}, {Jan{\ss}en}, {Jevardat de Fombelle}, {Jordan}, {Krone-Martins}, {Lanzafame}, {L{\"o}ffler}, {Lorca}, {Manteiga}, {Marchal}, {Marrese}, {Moitinho}, {Mora}, {Muinonen}, {Osborne}, {Pancino}, {Pauwels}, {Petit}, {Recio-Blanco}, {Richards}, {Riello}, {Rimoldini}, {Robin}, {Roegiers}, {Rybizki}, {Sarro}, {Siopis}, {Smith}, {Sozzetti}, {Ulla}, {Utrilla}, {van Leeuwen}, {van Reeven}, {Abbas}, {Abreu Aramburu}, {Accart}, {Aerts}, {Aguado}, {Ajaj}, {Altavilla}, {{\'A}lvarez}, {{\'A}lvarez Cid-Fuentes}, {Alves}, {Anderson}, {Anglada Varela}, {Antoja}, {Audard}, {Baines}, {Baker}, {Balaguer-N{\'u}{\~n}ez}, {Balbinot}, {Balog}, {Barache}, {Barbato}, {Barros}, {Barstow}, {Bartolom{\'e}}, {Bassilana}, {Bauchet}, {Baudesson-Stella}, {Becciani}, {Bellazzini}, {Bernet}, {Bertone}, {Bianchi}, {Blanco-Cuaresma}, {Boch}, {Bombrun}, {Bossini},
  {Bouquillon}, {Bragaglia}, {Bramante}, {Breedt}, {Bressan}, {Brouillet}, {Bucciarelli}, {Burlacu}, {Busonero}, {Butkevich}, {Buzzi}, {Caffau}, {Cancelliere}, {C{\'a}novas}, {Cantat-Gaudin}, {Carballo}, {Carlucci}, {Carnerero}, {Carrasco}, {Casamiquela}, {Castellani}, {Castro-Ginard}, {Castro Sampol}, {Chaoul}, {Charlot}, {Chemin}, {Chiavassa}, {Cioni}, {Comoretto}, {Cooper}, {Cornez}, {Cowell}, {Crifo}, {Crosta}, {Crowley}, {Dafonte}, {Dapergolas}, {David}, \& {David}}]{gaia2021}
{Gaia Collaboration}, {Brown}, A.~G.~A., {Vallenari}, A., {et~al.} 2021, \aap, 649, A1

\bibitem[{{Georgiev} {et~al.}(2010){Georgiev}, {Puzia}, {Goudfrooij}, \& {Hilker}}]{Georgiev2010}
{Georgiev}, I.~Y., {Puzia}, T.~H., {Goudfrooij}, P., \& {Hilker}, M. 2010, \mnras, 406, 1967

\bibitem[{{Gonz{\'a}lez-L{\'o}pezlira} {et~al.}(2017){Gonz{\'a}lez-L{\'o}pezlira}, {Lomel{\'\i}-N{\'u}{\~n}ez}, {{\'A}lamo-Mart{\'\i}nez}, {{\'O}rdenes-Brice{\~n}o}, {Loinard}, {Georgiev}, {Mu{\~n}oz}, {Puzia}, {Bruzual A.}, \& {Gwyn}}]{Gonzalez2017}
{Gonz{\'a}lez-L{\'o}pezlira}, R.~A., {Lomel{\'\i}-N{\'u}{\~n}ez}, L., {{\'A}lamo-Mart{\'\i}nez}, K., {et~al.} 2017, \apj, 835, 184

\bibitem[{{Gonz{\'a}lez-L{\'o}pezlira} {et~al.}(2019){Gonz{\'a}lez-L{\'o}pezlira}, {Mayya}, {Loinard}, {{\'A}lamo-Mart{\'\i}nez}, {Heald}, {Georgiev}, {{\'O}rdenes-Brice{\~n}o}, {Lan{\c{c}}on}, {Lara-L{\'o}pez}, {Lomel{\'\i}-N{\'u}{\~n}ez}, {Bruzual}, \& {Puzia}}]{Gonzalez2019}
{Gonz{\'a}lez-L{\'o}pezlira}, R.~A., {Mayya}, Y.~D., {Loinard}, L., {et~al.} 2019, \apj, 876, 39

\bibitem[{{Gregg} {et~al.}(2009){Gregg}, {Drinkwater}, {Evstigneeva}, {Jurek}, {Karick}, {Phillipps}, {Bridges}, {Jones}, {Bekki}, \& {Couch}}]{gregg2009}
{Gregg}, M.~D., {Drinkwater}, M.~J., {Evstigneeva}, E., {et~al.} 2009, \aj, 137, 498

\bibitem[{Guyon {et~al.}(2002)Guyon, Weston, Barnhill, \& Vapnik}]{Guyon2002}
Guyon, I., Weston, J., Barnhill, S., \& Vapnik, V. 2002, Machine Learning, 46, 389 – 422, cited by: 8253; All Open Access, Bronze Open Access

\bibitem[{{Harris} {et~al.}(2013){Harris}, {Harris}, \& {Alessi}}]{harris2013}
{Harris}, W.~E., {Harris}, G. L.~H., \& {Alessi}, M. 2013, \apj, 772, 82

\bibitem[{{Harris} \& {Reina-Campos}(2024)}]{Harris2024}
{Harris}, W.~E. \& {Reina-Campos}, M. 2024, \apj, 971, 155

\bibitem[{{Hilker} {et~al.}(2007){Hilker}, {Baumgardt}, {Infante}, {Drinkwater}, {Evstigneeva}, \& {Gregg}}]{hilker2007}
{Hilker}, M., {Baumgardt}, H., {Infante}, L., {et~al.} 2007, \aap, 463, 119

\bibitem[{{Hilker} {et~al.}(1999){Hilker}, {Infante}, {Vieira}, {Kissler-Patig}, \& {Richtler}}]{hilker1999}
{Hilker}, M., {Infante}, L., {Vieira}, G., {Kissler-Patig}, M., \& {Richtler}, T. 1999, \aaps, 134, 75

\bibitem[{{Huertas-Company} {et~al.}(2008){Huertas-Company}, {Rouan}, {Tasca}, {Soucail}, \& {Le F{\`e}vre}}]{huertascompany2008}
{Huertas-Company}, M., {Rouan}, D., {Tasca}, L., {Soucail}, G., \& {Le F{\`e}vre}, O. 2008, \aap, 478, 971

\bibitem[{Hunter(2007)}]{Hunter2007}
Hunter, J.~D. 2007, Computing in Science \& Engineering, 9, 90

\bibitem[{{Ivezi{\'c}} {et~al.}(2019){Ivezi{\'c}}, {Kahn}, {Tyson}, {Abel}, {Acosta}, {Allsman}, {Alonso}, {AlSayyad}, {Anderson}, {Andrew}, {Angel}, {Angeli}, {Ansari}, {Antilogus}, {Araujo}, {Armstrong}, {Arndt}, {Astier}, {Aubourg}, {Auza}, {Axelrod}, {Bard}, {Barr}, {Barrau}, {Bartlett}, {Bauer}, {Bauman}, {Baumont}, {Bechtol}, {Bechtol}, {Becker}, {Becla}, {Beldica}, {Bellavia}, {Bianco}, {Biswas}, {Blanc}, {Blazek}, {Blandford}, {Bloom}, {Bogart}, {Bond}, {Booth}, {Borgland}, {Borne}, {Bosch}, {Boutigny}, {Brackett}, {Bradshaw}, {Brandt}, {Brown}, {Bullock}, {Burchat}, {Burke}, {Cagnoli}, {Calabrese}, {Callahan}, {Callen}, {Carlin}, {Carlson}, {Chandrasekharan}, {Charles-Emerson}, {Chesley}, {Cheu}, {Chiang}, {Chiang}, {Chirino}, {Chow}, {Ciardi}, {Claver}, {Cohen-Tanugi}, {Cockrum}, {Coles}, {Connolly}, {Cook}, {Cooray}, {Covey}, {Cribbs}, {Cui}, {Cutri}, {Daly}, {Daniel}, {Daruich}, {Daubard}, {Daues}, {Dawson}, {Delgado}, {Dellapenna}, {de Peyster}, {de Val-Borro}, {Digel}, {Doherty}, {Dubois},
  {Dubois-Felsmann}, {Durech}, {Economou}, {Eifler}, {Eracleous}, {Emmons}, {Fausti Neto}, {Ferguson}, {Figueroa}, {Fisher-Levine}, {Focke}, {Foss}, {Frank}, {Freemon}, {Gangler}, {Gawiser}, {Geary}, {Gee}, {Geha}, {Gessner}, {Gibson}, {Gilmore}, {Glanzman}, {Glick}, {Goldina}, {Goldstein}, {Goodenow}, {Graham}, {Gressler}, {Gris}, {Guy}, {Guyonnet}, {Haller}, {Harris}, {Hascall}, {Haupt}, {Hernandez}, {Herrmann}, {Hileman}, {Hoblitt}, {Hodgson}, {Hogan}, {Howard}, {Huang}, {Huffer}, {Ingraham}, {Innes}, {Jacoby}, {Jain}, {Jammes}, {Jee}, {Jenness}, {Jernigan}, {Jevremovi{\'c}}, {Johns}, {Johnson}, {Johnson}, {Jones}, {Juramy-Gilles}, {Juri{\'c}}, {Kalirai}, {Kallivayalil}, {Kalmbach}, {Kantor}, {Karst}, {Kasliwal}, {Kelly}, {Kessler}, {Kinnison}, {Kirkby}, {Knox}, {Kotov}, {Krabbendam}, {Krughoff}, {Kub{\'a}nek}, {Kuczewski}, {Kulkarni}, {Ku}, {Kurita}, {Lage}, {Lambert}, {Lange}, {Langton}, {Le Guillou}, {Levine}, {Liang}, {Lim}, {Lintott}, {Long}, {Lopez}, {Lotz}, {Lupton}, {Lust}, {MacArthur}, {Mahabal},
  {Mandelbaum}, {Markiewicz}, {Marsh}, {Marshall}, {Marshall}, {May}, {McKercher}, {McQueen}, {Meyers}, {Migliore}, {Miller}, \& {Mills}}]{Ivezic2019}
{Ivezi{\'c}}, {\v{Z}}., {Kahn}, S.~M., {Tyson}, J.~A., {et~al.} 2019, \apj, 873, 111

\bibitem[{Joachims(1998)}]{Joachims1998}
Joachims, T. 1998, Lecture Notes in Computer Science (including subseries Lecture Notes in Artificial Intelligence and Lecture Notes in Bioinformatics), 1398, 137 – 142, cited by: 4970

\bibitem[{{Jord{\'a}n} {et~al.}(2015){Jord{\'a}n}, {Peng}, {Blakeslee}, {C{\^o}t{\'e}}, {Eyheramendy}, \& {Ferrarese}}]{jordan2015}
{Jord{\'a}n}, A., {Peng}, E.~W., {Blakeslee}, J.~P., {et~al.} 2015, \apjs, 221, 13

\bibitem[{{Kissler-Patig} {et~al.}(1999){Kissler-Patig}, {Grillmair}, {Meylan}, {Brodie}, {Minniti}, \& {Goudfrooij}}]{Kissler-Patig1999}
{Kissler-Patig}, M., {Grillmair}, C.~J., {Meylan}, G., {et~al.} 1999, \aj, 117, 1206

\bibitem[{{Kluge} {et~al.}(2025){Kluge}, {Hatch}, {Montes}, {Golden-Marx}, {Gonzalez}, {Cuillandre}, {Bolzonella}, {Lan{\c{c}}on}, {Laureijs}, {Saifollahi}, {Schirmer}, {Stone}, {Boselli}, {Cantiello}, {Sorce}, {Marleau}, {Duc}, {Sola}, {Urbano}, {Ahad}, {Bah{\'e}}, {Bamford}, {Bellhouse}, {Buitrago}, {Dimauro}, {Durret}, {Ellien}, {Jimenez-Teja}, {Slezak}, {Aghanim}, {Altieri}, {Andreon}, {Auricchio}, {Baldi}, {Balestra}, {Bardelli}, {Bender}, {Bonino}, {Branchini}, {Brescia}, {Brinchmann}, {Camera}, {Candini}, {Capobianco}, {Carbone}, {Carretero}, {Casas}, {Castellano}, {Cavuoti}, {Cimatti}, {Congedo}, {Conselice}, {Conversi}, {Copin}, {Courbin}, {Courtois}, {Cropper}, {Da Silva}, {Degaudenzi}, {Dinis}, {Duncan}, {Dupac}, {Dusini}, {Farina}, {Farrens}, {Ferriol}, {Fosalba}, {Frailis}, {Franceschi}, {Fumana}, {Galeotta}, {Garilli}, {Gillard}, {Gillis}, {Giocoli}, {G{\'o}mez-Alvarez}, {Granett}, {Grazian}, {Grupp}, {Guzzo}, {Haugan}, {Hoar}, {Hoekstra}, {Holmes}, {Hook}, {Hormuth}, {Hornstrup}, {Hudelot},
  {Jahnke}, {Keih{\"a}nen}, {Kermiche}, {Kiessling}, {Kitching}, {Kohley}, {Kubik}, {K{\"u}mmel}, {Kunz}, {Kurki-Suonio}, {Lahav}, {Ligori}, {Lilje}, {Lindholm}, {Lloro}, {Maiorano}, {Mansutti}, {Marggraf}, {Markovic}, {Martinet}, {Marulli}, {Massey}, {Maurogordato}, {McCracken}, {Medinaceli}, {Mei}, {Melchior}, {Mellier}, {Meneghetti}, {Merlin}, {Meylan}, {Moresco}, {Moscardini}, {Munari}, {Nichol}, {Niemi}, {Nightingale}, {Padilla}, {Paltani}, {Pasian}, {Pedersen}, {Percival}, {Pettorino}, {Pires}, {Polenta}, {Poncet}, {Popa}, {Pozzetti}, {Racca}, {Raison}, {Rebolo}, {Renzi}, {Rhodes}, {Riccio}, {Rix}, {Romelli}, {Roncarelli}, {Rossetti}, {Saglia}, {Sapone}, {Sartoris}, {Sauvage}, {Scaramella}, {Schneider}, {Schrabback}, {Secroun}, {Seidel}, {Seiffert}, {Serrano}, {Sirignano}, {Sirri}, {Skottfelt}, {Stanco}, {Tallada-Cresp{\'\i}}, {Taylor}, {Teplitz}, {Tereno}, {Toledo-Moreo}, {Torradeflot}, {Tutusaus}, {Valentijn}, {Valenziano}, {Vassallo}, {Verdoes Kleijn}, {Veropalumbo}, {Wang}, {Weller}, {Williams},
  {Zamorani}, {Zucca}, {Biviano}, {Burigana}, {De Lucia}, {George}, {Scottez}, {Simon}, {Mora}, {Mart{\'\i}n-Fleitas}, {Ruppin}, \& {Scott}}]{kluge2025}
{Kluge}, M., {Hatch}, N.~A., {Montes}, M., {et~al.} 2025, \aap, 697, A13

\bibitem[{{Li} {et~al.}(2025){Li}, {Lu}, {Wang}, \& {Wang}}]{Li2025}
{Li}, G., {Lu}, Z., {Wang}, J., \& {Wang}, Z. 2025, arXiv e-prints, arXiv:2502.15300

\bibitem[{{Lim} {et~al.}(2024){Lim}, {Peng}, {C{\^o}t{\'e}}, {Ferrarese}, {Roediger}, {Liu}, {Spengler}, {Sola}, {Duc}, {Sales}, {Blakeslee}, {Cuillandre}, {Durrell}, {Emsellem}, {Gwyn}, {Lan{\c{c}}on}, {Marleau}, {Mihos}, {M{\"u}ller}, {Puzia}, \& {S{\'a}nchez-Janssen}}]{Lim2024}
{Lim}, S., {Peng}, E.~W., {C{\^o}t{\'e}}, P., {et~al.} 2024, \apj, 966, 168

\bibitem[{{Lim} {et~al.}(2025){Lim}, {Peng}, {C{\^o}t{\'e}}, {Ferrarese}, {Roediger}, {Liu}, {Spengler}, {Sola}, {Duc}, {Sales}, {Blakeslee}, {Cuillandre}, {Durrell}, {Emsellem}, {Gwyn}, {Lan{\c{c}}on}, {Marleau}, {Mihos}, {M{\"u}ller}, {Puzia}, \& {S{\'a}nchez-Janssen}}]{Lim2025}
{Lim}, S., {Peng}, E.~W., {C{\^o}t{\'e}}, P., {et~al.} 2025, \apjs, 276, 34

\bibitem[{Liu {et~al.}(2011)Liu, Chen, Zhang, \& Hu}]{liu2011}
Liu, Q., Chen, C., Zhang, Y., \& Hu, Z. 2011, Artif. Intell. Rev., 36, 99–115

\bibitem[{{Maddox} {et~al.}(2019){Maddox}, {Serra}, {Venhola}, {Peletier}, {Loubser}, \& {Iodice}}]{maddox2019}
{Maddox}, N., {Serra}, P., {Venhola}, A., {et~al.} 2019, \mnras, 490, 1666

\bibitem[{{Madrid} {et~al.}(2018){Madrid}, {O'Neill}, {Gagliano}, \& {Marvil}}]{madrid2018}
{Madrid}, J.~P., {O'Neill}, C.~R., {Gagliano}, A.~T., \& {Marvil}, J.~R. 2018, \apj, 867, 144

\bibitem[{{Ma{\l}ek} {et~al.}(2013){Ma{\l}ek}, {Solarz}, {Pollo}, {Fritz}, {Garilli}, {Scodeggio}, {Iovino}, {Granett}, {Abbas}, {Adami}, {Arnouts}, {Bel}, {Bolzonella}, {Bottini}, {Branchini}, {Cappi}, {Coupon}, {Cucciati}, {Davidzon}, {De Lucia}, {de la Torre}, {Franzetti}, {Fumana}, {Guzzo}, {Ilbert}, {Krywult}, {Le Brun}, {Le Fevre}, {Maccagni}, {Marulli}, {McCracken}, {Paioro}, {Polletta}, {Schlagenhaufer}, {Tasca}, {Tojeiro}, {Vergani}, {Zanichelli}, {Burden}, {Di Porto}, {Marchetti}, {Marinoni}, {Mellier}, {Moscardini}, {Nichol}, {Peacock}, {Percival}, {Phleps}, {Wolk}, \& {Zamorani}}]{Malek2013}
{Ma{\l}ek}, K., {Solarz}, A., {Pollo}, A., {et~al.} 2013, \aap, 557, A16

\bibitem[{{Maschmann} {et~al.}(2024){Maschmann}, {Lee}, {Thilker}, {Whitmore}, {Deger}, {Boquien}, {Chandar}, {Dale}, {Wofford}, {Hannon}, {Larson}, {Leroy}, {Schinnerer}, {Rosolowsky}, {{\'U}beda}, {Barnes}, {Emsellem}, {Grasha}, {Groves}, {Indebetouw}, {Kim}, {Klessen}, {Kreckel}, {Levy}, {Pinna}, {Rodr{\'\i}guez}, {Tian}, \& {Williams}}]{Maschmann2024}
{Maschmann}, D., {Lee}, J.~C., {Thilker}, D.~A., {et~al.} 2024, \apjs, 273, 14

\bibitem[{{Mieske} {et~al.}(2002){Mieske}, {Hilker}, \& {Infante}}]{mieske2002}
{Mieske}, S., {Hilker}, M., \& {Infante}, L. 2002, \aap, 383, 823

\bibitem[{{Mieske} {et~al.}(2004){Mieske}, {Hilker}, \& {Infante}}]{mieske2004}
{Mieske}, S., {Hilker}, M., \& {Infante}, L. 2004, \aap, 418, 445

\bibitem[{{Mieske} {et~al.}(2008){Mieske}, {Hilker}, {Jord{\'a}n}, {Infante}, {Kissler-Patig}, {Rejkuba}, {Richtler}, {C{\^o}t{\'e}}, {Baumgardt}, {West}, {Ferrarese}, \& {Peng}}]{mieske2008}
{Mieske}, S., {Hilker}, M., {Jord{\'a}n}, A., {et~al.} 2008, \aap, 487, 921

\bibitem[{{Mohammadi} {et~al.}(2022){Mohammadi}, {Mutatiina}, {Saifollahi}, \& {Bunte}}]{Mohammadi2022}
{Mohammadi}, M., {Mutatiina}, J., {Saifollahi}, T., \& {Bunte}, K. 2022, Astronomy and Computing, 39, 100555

\bibitem[{{Mu{\~n}oz} {et~al.}(2015){Mu{\~n}oz}, {Eigenthaler}, {Puzia}, {Taylor}, {Ordenes-Brice{\~n}o}, {Alamo-Mart{\'\i}nez}, {Ribbeck}, {{\'A}ngel}, {Capaccioli}, {C{\^o}t{\'e}}, {Ferrarese}, {Galaz}, {Hempel}, {Hilker}, {Jord{\'a}n}, {Lan{\c{c}}on}, {Mieske}, {Paolillo}, {Richtler}, {S{\'a}nchez-Janssen}, \& {Zhang}}]{Munoz2015}
{Mu{\~n}oz}, R.~P., {Eigenthaler}, P., {Puzia}, T.~H., {et~al.} 2015, \apjl, 813, L15

\bibitem[{{Mu{\~n}oz} {et~al.}(2014){Mu{\~n}oz}, {Puzia}, {Lan{\c{c}}on}, {Peng}, {C{\^o}t{\'e}}, {Ferrarese}, {Blakeslee}, {Mei}, {Cuillandre}, {Hudelot}, {Courteau}, {Duc}, {Balogh}, {Boselli}, {Bournaud}, {Carlberg}, {Chapman}, {Durrell}, {Eigenthaler}, {Emsellem}, {Gavazzi}, {Gwyn}, {Huertas-Company}, {Ilbert}, {Jord{\'a}n}, {L{\"a}sker}, {Licitra}, {Liu}, {MacArthur}, {McConnachie}, {McCracken}, {Mellier}, {Peng}, {Raichoor}, {Taylor}, {Tonry}, {Tully}, \& {Zhang}}]{Munoz2014}
{Mu{\~n}oz}, R.~P., {Puzia}, T.~H., {Lan{\c{c}}on}, A., {et~al.} 2014, \apjs, 210, 4

\bibitem[{{Ordenes Brice{\~n}o}(2018)}]{ordenes2018_thesis}
{Ordenes Brice{\~n}o}, Y. 2018, PhD thesis, Ruprecht-Karls University of Heidelberg, Germany

\bibitem[{{Ordenes-Brice{\~n}o} {et~al.}(2018){Ordenes-Brice{\~n}o}, {Eigenthaler}, {Taylor}, {Puzia}, {Alamo-Mart{\'\i}nez}, {Ribbeck}, {Mu{\~n}oz}, {Zhang}, {Grebel}, {{\'A}ngel}, {C{\^o}t{\'e}}, {Ferrarese}, {Hilker}, {Lan{\c{c}}on}, {Mieske}, {Miller}, {Rong}, \& {S{\'a}nchez-Janssen}}]{Ordenes2018}
{Ordenes-Brice{\~n}o}, Y., {Eigenthaler}, P., {Taylor}, M.~A., {et~al.} 2018, \apj, 859, 52

\bibitem[{{Pedregosa} {et~al.}(2011){Pedregosa}, {Varoquaux}, {Gramfort}, {Michel}, {Thirion}, {Grisel}, {Blondel}, {M{\"u}ller}, {Nothman}, {Louppe}, {Prettenhofer}, {Weiss}, {Dubourg}, {Vanderplas}, {Passos}, {Cournapeau}, {Brucher}, {Perrot}, \& {Duchesnay}}]{pedregosa2011}
{Pedregosa}, F., {Varoquaux}, G., {Gramfort}, A., {et~al.} 2011, Journal of Machine Learning Research, 12, 2825

\bibitem[{{Peng} {et~al.}(2011){Peng}, {Ferguson}, {Goudfrooij}, {Hammer}, {Lucey}, {Marzke}, {Puzia}, {Carter}, {Balcells}, {Bridges}, {Chiboucas}, {del Burgo}, {Graham}, {Guzm{\'a}n}, {Hudson}, {Matkovi{\'c}}, {Merritt}, {Miller}, {Mouhcine}, {Phillipps}, {Sharples}, {Smith}, {Tully}, \& {Verdoes Kleijn}}]{peng2011}
{Peng}, E.~W., {Ferguson}, H.~C., {Goudfrooij}, P., {et~al.} 2011, \apj, 730, 23

\bibitem[{{Peng} {et~al.}(2006){Peng}, {Jord{\'a}n}, {C{\^o}t{\'e}}, {Blakeslee}, {Ferrarese}, {Mei}, {West}, {Merritt}, {Milosavljevi{\'c}}, \& {Tonry}}]{Peng2006}
{Peng}, E.~W., {Jord{\'a}n}, A., {C{\^o}t{\'e}}, P., {et~al.} 2006, \apj, 639, 95

\bibitem[{{Pfeffer} {et~al.}(2018){Pfeffer}, {Kruijssen}, {Crain}, \& {Bastian}}]{Pfeffer2018}
{Pfeffer}, J., {Kruijssen}, J.~M.~D., {Crain}, R.~A., \& {Bastian}, N. 2018, \mnras, 475, 4309

\bibitem[{Platt(1999{\natexlab{a}})}]{platt99a}
Platt, J.~C. 1999{\natexlab{a}}, Fast training of support vector machines using sequential minimal optimization (Cambridge, MA, USA: MIT Press), 185–208

\bibitem[{Platt(1999{\natexlab{b}})}]{Platt1999b}
Platt, J.~C. 1999{\natexlab{b}}, in Advances in Large Margin Classifiers (MIT Press), 61--74

\bibitem[{{Pota} {et~al.}(2018){Pota}, {Napolitano}, {Hilker}, {Spavone}, {Schulz}, {Cantiello}, {Tortora}, {Iodice}, {Paolillo}, {D'Abrusco}, {Capaccioli}, {Puzia}, {Peletier}, {Romanowsky}, {van de Ven}, {Spiniello}, {Norris}, {Lisker}, {Munoz}, {Schipani}, {Eigenthaler}, {Taylor}, {S{\'a}nchez-Janssen}, \& {Ordenes-Brice{\~n}o}}]{pota2018}
{Pota}, V., {Napolitano}, N.~R., {Hilker}, M., {et~al.} 2018, \mnras, 481, 1744

\bibitem[{{Powalka} {et~al.}(2016){Powalka}, {Lan{\c{c}}on}, {Puzia}, {Peng}, {Liu}, {Mu{\~n}oz}, {Blakeslee}, {C{\^o}t{\'e}}, {Ferrarese}, {Roediger}, {S{\'a}nchez-Janssen}, {Zhang}, {Durrell}, {Cuillandre}, {Duc}, {Guhathakurta}, {Gwyn}, {Hudelot}, {Mei}, \& {Toloba}}]{powalka2016}
{Powalka}, M., {Lan{\c{c}}on}, A., {Puzia}, T.~H., {et~al.} 2016, \apjs, 227, 12

\bibitem[{{Puzia} {et~al.}(2006){Puzia}, {Kissler-Patig}, \& {Goudfrooij}}]{Puzia2006}
{Puzia}, T.~H., {Kissler-Patig}, M., \& {Goudfrooij}, P. 2006, \apj, 648, 383

\bibitem[{{Puzia} {et~al.}(2005){Puzia}, {Kissler-Patig}, {Thomas}, {Maraston}, {Saglia}, {Bender}, {Goudfrooij}, \& {Hempel}}]{Puzia2005}
{Puzia}, T.~H., {Kissler-Patig}, M., {Thomas}, D., {et~al.} 2005, \aap, 439, 997

\bibitem[{{Puzia} {et~al.}(2014){Puzia}, {Paolillo}, {Goudfrooij}, {Maccarone}, {Fabbiano}, \& {Angelini}}]{puzia2014}
{Puzia}, T.~H., {Paolillo}, M., {Goudfrooij}, P., {et~al.} 2014, \apj, 786, 78

\bibitem[{{Saifollahi} {et~al.}(2021){Saifollahi}, {Janz}, {Peletier}, {Cantiello}, {Hilker}, {Mieske}, {Valentijn}, {Venhola}, \& {Verdoes Kleijn}}]{Saifollahi2021}
{Saifollahi}, T., {Janz}, J., {Peletier}, R.~F., {et~al.} 2021, \mnras, 504, 3580

\bibitem[{{Saifollahi} {et~al.}(2025){Saifollahi}, {Voggel}, {Lan{\c{c}}on}, {Cantiello}, {Raj}, {Cuillandre}, {Larsen}, {Marleau}, {Venhola}, {Schirmer}, {Carollo}, {Duc}, {Ferguson}, {Hunt}, {K{\"u}mmel}, {Laureijs}, {Marchal}, {Nucita}, {Peletier}, {Poulain}, {Rejkuba}, {S{\'a}nchez-Janssen}, {Urbano}, {Abdurro'uf}, {Altieri}, {Baes}, {Bolzonella}, {Conselice}, {Cote}, {Dimauro}, {Gonzalez}, {Habas}, {Hudelot}, {Kluge}, {Lonare}, {Massari}, {Romelli}, {Scaramella}, {Sola}, {Stone}, {Tortora}, {van Mierlo}, {Knapen}, {Mart{\'\i}n-Fleitas}, {Mora}, {Rom{\'a}n}, {Aghanim}, {Amara}, {Andreon}, {Auricchio}, {Baldi}, {Balestra}, {Bardelli}, {Basset}, {Bender}, {Bonino}, {Branchini}, {Brescia}, {Brinchmann}, {Camera}, {Capobianco}, {Carbone}, {Carretero}, {Casas}, {Castellano}, {Cavuoti}, {Cimatti}, {Congedo}, {Conversi}, {Copin}, {Courbin}, {Courtois}, {Cropper}, {Da Silva}, {Degaudenzi}, {Di Giorgio}, {Dinis}, {Dubath}, {Dupac}, {Dusini}, {Fabricius}, {Farina}, {Farrens}, {Ferriol}, {Fosalba}, {Frailis},
  {Franceschi}, {Fumana}, {Galeotta}, {Garilli}, {Gillard}, {Gillis}, {Giocoli}, {G{\'o}mez-Alvarez}, {Granett}, {Grazian}, {Grupp}, {Guzzo}, {Haugan}, {Hoar}, {Hoekstra}, {Holmes}, {Hook}, {Hormuth}, {Hornstrup}, {Jahnke}, {Jhabvala}, {Keih{\"a}nen}, {Kermiche}, {Kiessling}, {Kitching}, {Kohley}, {Kubik}, {Kuijken}, {Kunz}, {Kurki-Suonio}, {Lahav}, {Le Mignant}, {Ligori}, {Lilje}, {Lindholm}, {Lloro}, {Maino}, {Maiorano}, {Mansutti}, {Marggraf}, {Markovic}, {Martinet}, {Marulli}, {Massey}, {Maurogordato}, {McCracken}, {Medinaceli}, {Mei}, {Melchior}, {Mellier}, {Meneghetti}, {Meylan}, {Moresco}, {Moscardini}, {Munari}, {Nakajima}, {Nichol}, {Niemi}, {Padilla}, {Paltani}, {Pasian}, {Pedersen}, {Percival}, {Pettorino}, {Pires}, {Polenta}, {Poncet}, {Popa}, {Pozzetti}, {Racca}, {Raison}, {Rebolo}, {Refregier}, {Renzi}, {Rhodes}, {Riccio}, {Roncarelli}, {Rossetti}, {Saglia}, {Sapone}, {Sartoris}, {Schneider}, {Schrabback}, {Secroun}, {Seidel}, {Serrano}, {Sirignano}, {Sirri}, {Stanco}, {Tallada-Cresp{\'\i}},
  {Taylor}, {Teplitz}, {Tereno}, {Toledo-Moreo}, {Torradeflot}, {Tsyganov}, {Tutusaus}, {Valentijn}, {Valenziano}, {Vassallo}, {Verdoes Kleijn}, {Veropalumbo}, {Wang}, {Weller}, {Williams}, {Zamorani}, {Zucca}, {Biviano}, {Burigana}, {Scottez}, {Simon}, {Balogh}, \& {Scott}}]{Saifollahi2025}
{Saifollahi}, T., {Voggel}, K., {Lan{\c{c}}on}, A., {et~al.} 2025, \aap, 697, A10

\bibitem[{{Schlafly} \& {Finkbeiner}(2011)}]{Schf11}
{Schlafly}, E.~F. \& {Finkbeiner}, D.~P. 2011, \apj, 737, 103

\bibitem[{{Schlegel} {et~al.}(1998){Schlegel}, {Finkbeiner}, \& {Davis}}]{Schl98}
{Schlegel}, D.~J., {Finkbeiner}, D.~P., \& {Davis}, M. 1998, \apj, 500, 525

\bibitem[{{Schuberth} {et~al.}(2010){Schuberth}, {Richtler}, {Hilker}, {Dirsch}, {Bassino}, {Romanowsky}, \& {Infante}}]{schuberth2010}
{Schuberth}, Y., {Richtler}, T., {Hilker}, M., {et~al.} 2010, \aap, 513, A52

\bibitem[{{Shi} {et~al.}(2015){Shi}, {Liu}, {Sun}, {Li}, {Lei}, \& {Wang}}]{shi2015}
{Shi}, F., {Liu}, Y.-Y., {Sun}, G.-L., {et~al.} 2015, \mnras, 453, 122

\bibitem[{{Skrutskie} {et~al.}(2006){Skrutskie}, {Cutri}, {Stiening}, {Weinberg}, {Schneider}, {Carpenter}, {Beichman}, {Capps}, {Chester}, {Elias}, {Huchra}, {Liebert}, {Lonsdale}, {Monet}, {Price}, {Seitzer}, {Jarrett}, {Kirkpatrick}, {Gizis}, {Howard}, {Evans}, {Fowler}, {Fullmer}, {Hurt}, {Light}, {Kopan}, {Marsh}, {McCallon}, {Tam}, {Van Dyk}, \& {Wheelock}}]{Skru06}
{Skrutskie}, M.~F., {Cutri}, R.~M., {Stiening}, R., {et~al.} 2006, \aj, 131, 1163

\bibitem[{{Smith} {et~al.}(2015){Smith}, {S{\'a}nchez-Janssen}, {Beasley}, {Candlish}, {Gibson}, {Puzia}, {Janz}, {Knebe}, {Aguerri}, {Lisker}, {Hensler}, {Fellhauer}, {Ferrarese}, \& {Yi}}]{Smith2015}
{Smith}, R., {S{\'a}nchez-Janssen}, R., {Beasley}, M.~A., {et~al.} 2015, \mnras, 454, 2502

\bibitem[{Sokolova \& Lapalme(2009)}]{sokolova2009}
Sokolova, M. \& Lapalme, G. 2009, Information Processing \& Management, 45, 427

\bibitem[{{Sutherland} {et~al.}(2015){Sutherland}, {Emerson}, {Dalton}, {Atad-Ettedgui}, {Beard}, {Bennett}, {Bezawada}, {Born}, {Caldwell}, {Clark}, {Craig}, {Henry}, {Jeffers}, {Little}, {McPherson}, {Murray}, {Stewart}, {Stobie}, {Terrett}, {Ward}, {Whalley}, \& {Woodhouse}}]{Sutherland15}
{Sutherland}, W., {Emerson}, J., {Dalton}, G., {et~al.} 2015, \aap, 575, A25

\bibitem[{{Taylor} {et~al.}(2017){Taylor}, {Puzia}, {Mu{\~n}oz}, {Mieske}, {Lan{\c{c}}on}, {Zhang}, {Eigenthaler}, \& {Bovill}}]{Taylor2017}
{Taylor}, M.~A., {Puzia}, T.~H., {Mu{\~n}oz}, R.~P., {et~al.} 2017, \mnras, 469, 3444

\bibitem[{{Taylor}(2005)}]{taylor2005}
{Taylor}, M.~B. 2005, in Astronomical Society of the Pacific Conference Series, Vol. 347, Astronomical Data Analysis Software and Systems XIV, ed. P.~{Shopbell}, M.~{Britton}, \& R.~{Ebert}, 29

\bibitem[{{Ting} {et~al.}(2025){Ting}, {Nguyen}, {Ghosal}, {Pan}, {Arora}, {Sun}, {de Haan}, {Ramachandra}, {Wells}, {Madireddy}, \& {Accomazzi}}]{Ting2025}
{Ting}, Y.~S., {Nguyen}, T.~D., {Ghosal}, T., {et~al.} 2025, Astronomy and Computing, 51, 100893

\bibitem[{{Usher} {et~al.}(2023){Usher}, {Dage}, {Girardi}, {Barmby}, {Bonatto}, {Chies-Santos}, {Clarkson}, {G{\'o}mez Camus}, {Hartmann}, {Ferguson}, {Pieres}, {Prisinzano}, {Rhode}, {Rich}, {Ripepi}, {Santiago}, {Stassun}, {Street}, {Szab{\'o}}, {Venuti}, {Zaggia}, {Canossa}, {Floriano}, {Lopes}, {Miranda}, {Oliveira}, {Reina-Campos}, {Roman-Lopes}, \& {Sobeck}}]{usher2023}
{Usher}, C., {Dage}, K.~C., {Girardi}, L., {et~al.} 2023, \pasp, 135, 074201

\bibitem[{{Valdes} {et~al.}(2014){Valdes}, {Gruendl}, \& {DES Project}}]{Valdes14}
{Valdes}, F., {Gruendl}, R., \& {DES Project}. 2014, in Astronomical Society of the Pacific Conference Series, Vol. 485, Astronomical Data Analysis Software and Systems XXIII, ed. N.~{Manset} \& P.~{Forshay}, 379

\bibitem[{{Vandenberg} {et~al.}(1996){Vandenberg}, {Bolte}, \& {Stetson}}]{Vandenberg1996}
{Vandenberg}, D.~A., {Bolte}, M., \& {Stetson}, P.~B. 1996, \araa, 34, 461

\bibitem[{{Vavilova} {et~al.}(2021){Vavilova}, {Dobrycheva}, {Vasylenko}, {Elyiv}, {Melnyk}, \& {Khramtsov}}]{vavilova2021}
{Vavilova}, I.~B., {Dobrycheva}, D.~V., {Vasylenko}, M.~Y., {et~al.} 2021, \aap, 648, A122

\bibitem[{{Verro} {et~al.}(2022){Verro}, {Trager}, {Peletier}, {Lan{\c{c}}on}, {Arentsen}, {Chen}, {Coelho}, {Dries}, {Falc{\'o}n-Barroso}, {Gonneau}, {Lyubenova}, {Martins}, {Prugniel}, {S{\'a}nchez-Bl{\'a}zquez}, \& {Vazdekis}}]{verro2022}
{Verro}, K., {Trager}, S.~C., {Peletier}, R.~F., {et~al.} 2022, \aap, 661, A50

\bibitem[{Wang {et~al.}(2022)Wang, Bai, L{\'{o}}pez-Sanjuan, Yuan, Wang, Liu, Sobral, Baqui, Mart{\'{i}}n, {Andres Galarza}, Alcaniz, Angulo, Cenarro, Crist{\'{o}}bal-Hornillos, Dupke, Ederoclite, Hern{\'{a}}ndez-Monteagudo, Mar{\'{i}}n-Franch, Moles, Sodr{\'{e}}, {V{\'{a}}zquez Rami{\'{o}}}, \& Varela}]{Wang2022a}
Wang, C., Bai, Y., L{\'{o}}pez-Sanjuan, C., {et~al.} 2022, A{\&}A, 659, A144

\bibitem[{Waskom(2021)}]{Waskom2021}
Waskom, M.~L. 2021, Journal of Open Source Software, 6, 3021

\bibitem[{{Willman} \& {Strader}(2012)}]{Willman2012}
{Willman}, B. \& {Strader}, J. 2012, \aj, 144, 76

\end{thebibliography}

%%%%%%%%%%%%%%%%%%%%%%%%%%%%%%%%%%%%%%%%%
\begin{appendix}

\section{Photometric Calibration for NGFS-T1}
\label{sect:A1_phot_cal}

In this section we present a brief explanation about the data reduction process for the central Fornax region ($3deg^2$ in the DECam FoV), for the complete NGFS survey a dedicated paper will be published.

\subsection{Data Process}
For the optical data, the raw DECam images are processed by the DECam Community pipeline \citep[v3.4.0][]{Valdes14}, taking care of bias, flat fielding and image crosstalk correction, thus only for removal of the instrumental signal (InstCal fits files).~We use our custom background subtraction strategy, followed by the astrometry, photometric calibration and stacking processes performed with SCAMP \citep[v2.2.6,][]{Bertin06}, Source Extractor \citep[SE, v2.19.5,][]{Bertin96}, and SWARP \citep[v2.19.5][]{Bertin02}, respectively.~For the astrometric calibration, we use the SDSS stripe 82 standard star frames, and color term correction were applied for each source after photometry. We cross-checked our optical-band photometry in $g'i'$ bands with the Dark Energy Survey DR2 \citep{DESdr2}. In addition we use the photometry from the Fornax Deep Survey \citep{cantiello2020}.

The central region (VIRCam T1 and T2 -- overlapped) was processed in \cite{ordenes2018_thesis}, PhD Thesis. The VISTA data in $JK_s$ filters was processed from scratch, using a similar wrapped-up pipeline as for the optical data, but modeling the sky with different function in order to find the best compromise for a highly variable sky in the NIR, as it varies rapidly due to atmospheric emission, mainly from water vapor and OH molecules. These fluctuations occur on short timescales and make sky subtraction a critical step in NIR image processing. For the astrometric and photometric calibration, we use the reference stars from the 2MASS Point Source Catalog \citep{Skru06}.

\begin{figure}[ht]
%trim=left bottom right top
 \centering
 \includegraphics[trim=0.2cm 1cm 0cm 1cm,clip,width=1.02\linewidth]{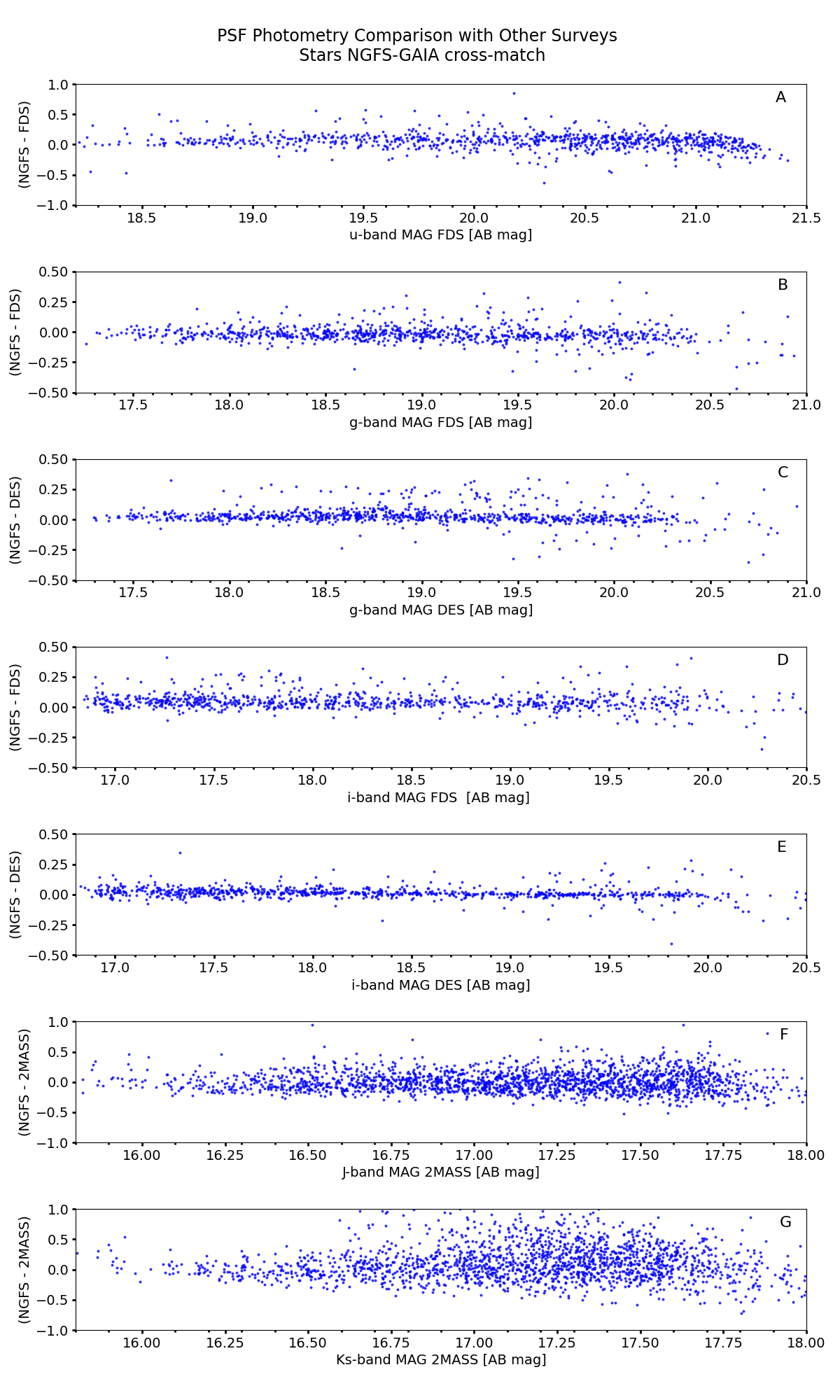} 
 \caption{NGFS-GAIA DR3 cross-match. PSF photometry comparison with other surveys in the literature: Fornax Deep Survey \citep{cantiello2020} with $u'-$band (panel A), $g'-$band (panel B) and $i'-$ band(panel D), Dark Energy Survey \citep{DESdr2} with $g'-$band (panel C) and $i'-$ band (panel E) and 2MASS \citep{Skru06} with $J-$band (panel F) and $K_s-$ band (panel G).}
 \label{fig:stars_calib}
\end{figure}

Figure \ref{fig:stars_calib} shows the results of our photometric calibration in comparison with the optical and NIR surveys mentioned above. We cross-matched the NGFS catalog ($u'g'i'JK_s$) with Gaia DR3 \citep{gaia2021}. For the Gaia catalog, we imposed a declination error of $\le 0.5$ miliarcsec. For our optical catalog, we applied the following criteria: SPREAD\_MODEL error (in the $u'$ band) $<$ 0.00015 and $FLAGS = 0$. The FWHM in the $u'$ band shows the lowest image quality, mainly due to the filter throughput characteristics.

\subsection{Point-like Source Detection Image}
\label{sec:point_source_detection}

The central region of the Fornax Cluster is very rich in galaxies (see Fig. \ref{fig:t1_rgb}). Their extended surface brightness profiles hamper source detection, as many faint sources are hidden behind the diffuse galaxy light. To detect point sources in the science images, we developed an iterative procedure using \texttt{SExtractor} and its check images to construct a detection image that is as free as possible from the main galaxy light components.

\begin{figure*}[htb!]
%trim=left bottom right top
 \centering
 \includegraphics[trim=0cm 0cm 0cm 0cm,clip,width=0.8\textwidth]{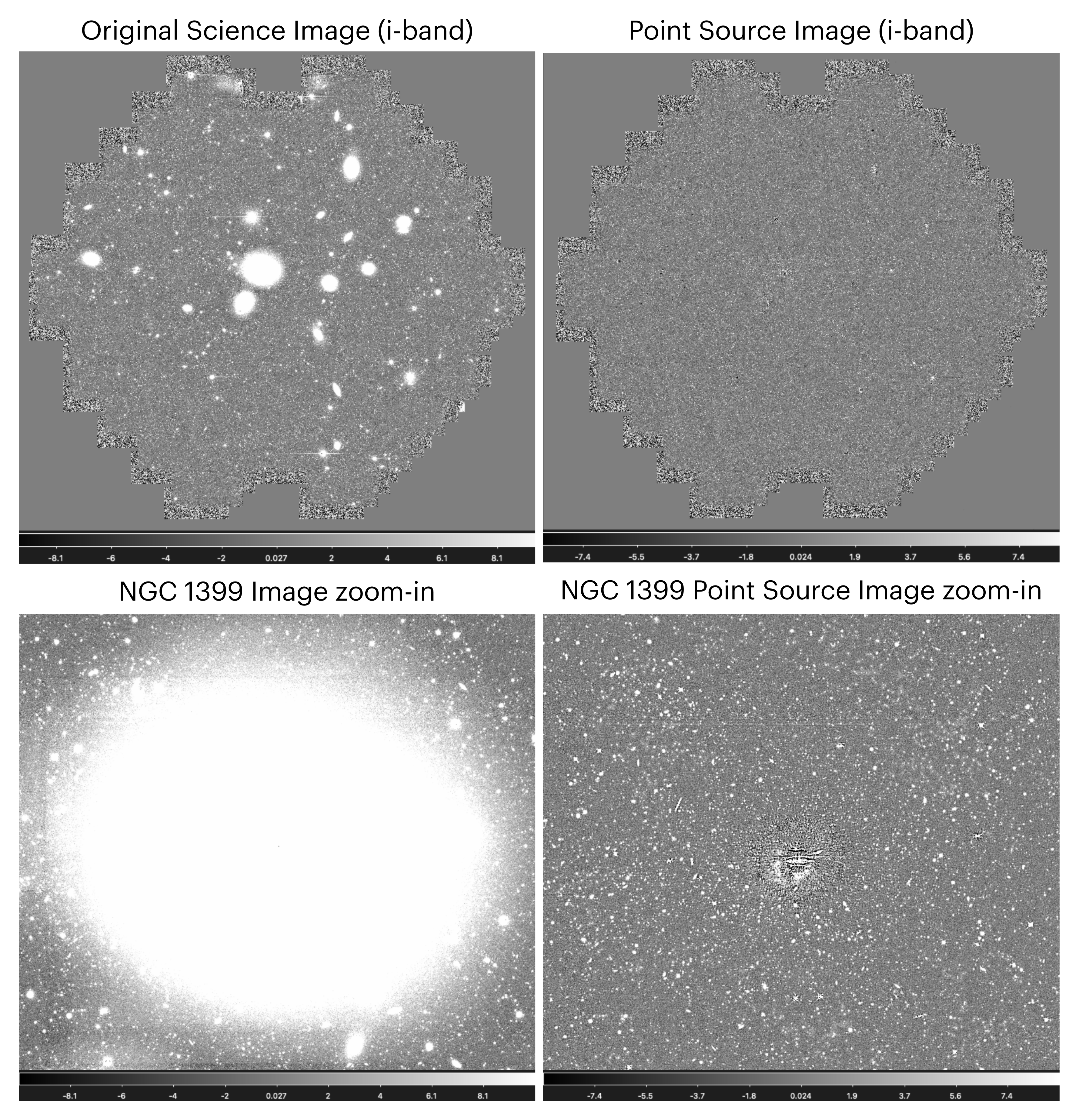} 
 \caption{NGFS-T1 i-band image. Left panels show the science image and right panels the result of the "point source" detection image.}
 \label{fig:GC_detection}
\end{figure*}

The procedure consists of two phases. In the first phase, we run \texttt{SExtractor} on the original image and use the \texttt{BACKGROUND} check image to identify the diffuse light from the brightest galaxies and other extended components. This allows us to build a diffuse light map, which is then subtracted from the original image. Several iterations are required; in our case, five iterations were sufficient to obtain homogeneous sky models around sources and significantly reduced galaxy halos.

The second phase involves a median-filtering iteration. Using the output image from the first phase and the \texttt{OBJECT} check image, masking everything except detected point sources. We subtract these two images, leaving only the extended structures unmasked and thus ready for median filtering. We apply a median filter of $21$ pixels, which provides a good balance between filtering performance and computational time, given the large size of the DECam images ($\sim 3\ \mathrm{deg}^2$). The process is repeated until only point sources remain (i.e., with no residual extended halos) in each filter. We then run \texttt{SExtractor} on these processed images and use the resulting catalogs as input for the science image analysis (see Section \ref{sect:data}).

%%%%%%%%%%%%%%%%%%%%%%%%%%%%%%%%%%%%%%%%%
\section{Literature Comparison with the \texttt{svm.SVC} Methodology and GC Sample}
\label{sect:literature_comparison}

\subsection{Machine Learning Methodologies in the Literature}
\label{sect:ML_literature}
We have shown that Machine Learning methodologies are essential for identifying extragalactic globular clusters (GCs) in wide-field surveys, overcoming the inherent difficulty of distinguishing them from contaminants, which can result in 30–70\% contamination in purely optical samples. Studies emphasize the necessity of broad spectral coverage, integrating optical and near-infrared (NIR) filters, such as $u'$ and $K_s$-bands, to maximize separation power.

Our own work, utilizing the SVM method on NGFS-T1 data, was trained on 1,209 confirmed GCs, 2,151 foreground stars and 1,587 galaxies. The resulting optimal 7F SVM model achieved outstanding performance metrics on the test set, including a GC completeness (recall) of 96.1\%, a GC purity (precision) of 93.3\%, and an overall Accuracy of 96.6\%. With a \texttt{svm.SVC} selection (3,960 total) at 80\% of 1,717 GC candidates and together with RV confirmed objects, a final catalog of 2,916 objects.

Focusing on the Fornax cluster environment, previous ML efforts have included unsupervised clustering via Growing Neural Gas (GNG) \citep{Angora2019}, which utilized a training sample of 357 GCs to achieve a Completeness (recall) of 90.8\% and purity (precision) of 80.0\%, with an overall Accuracy of 86.5\% (Average Efficiency or AE) in a 3-class problem. This effort classified approximately 522 common GC candidates. Subsequent supervised techniques like K-Nearest Neighbors ($\text{KNN}+100$) \citep{Saifollahi2021} used 137 pre-selected GCs for training, achieving 81\% Recall and 77\% Precision for the GC class during cross-validation, and subsequently identified 1,155 UCD/GC candidates in total. Another works utilizing supervised methods, such as Localized Generalized Matrix Learning Vector Quantization (LGMLVQ) and Random Forest (RF) \citep{Mohammadi2022}, trained on 512 GCs and achieved strong performance exceeding a recall of 96.3\% and a precision of 93.5\%, with an overall Accuracy of 98.2\%.

Other classification methods include Random Forest (RF) and Neural Networks (NN) used in the Virgo cluster \citep{Barbisan2022}, which trained on 1,243 confirmed UCDs and GCs and achieved up to 99.4\% overall accuracy with a GCs+UCDs precision of 98.9\% and recall of 99.2\%. Separately, a specialized statistical pipeline involving Principal Component Analysis (PCA) applied to the M81/M82 group \citep{chiessantos2022} used only 73 known GCs for training and identified 642 new GC candidates.

These studies universally address the challenge of highly imbalanced datasets through strategies like oversampling, using two-step filtering/pre-selection to balance the training sample, or trimming majority classes, while here we used the undersampling of the majority classes approach. Moreover, unlike some preceding studies that leverage distance-dependent absolute magnitudes for applications across galaxies of similar distances or utilize apparent magnitude variants to distinguish some contaminants, our methodology relies exclusively on distance-independent colors and morphological parameters. This advantageous approach ensures robust generality for surveys like LSST, as colors robustly trace intrinsic stellar population properties such as age and metallicity.

\subsection{\texttt{svm.SVC} GC Sample and Comparison with Existing Photometric Catalogs}
\label{sect:cat_comparison}

In the Fornax Cluster region, several photometric catalogs have been produced by different surveys, each employing distinct methodologies and filter sets for selecting GC candidates. For reference, our \texttt{svm.SVC} GC catalog contains 3,960 objects, of which 1,717 have a probability $\ge 80\%$. Our full catalog (combining \texttt{svm.SVC} predictions and RV-confirmed GCs) comprises 5,169 objects, with 2,926 sources having $prob \ge 80\%$. Below, we present the comparison:

\begin{enumerate}[label=\roman*.]
\item \cite{cantiello2020}: The Fornax Deep Survey, which selected GC candidates using $u'g'r'i'$ photometry, provides the catalog FDS\_master\_gc-ucd.dat containing 3,331 sources. Within the same central region covered in this work, 2,411 of these are GC candidates (before cross-matching). The cross-match yields 375 sources in common with our \texttt{svm.SVC} catalog and 1,369 sources in common with our full catalog with $prob \ge 80\%$.

\item \cite{Saifollahi2021}: The Fornax Deep Survey combined with NIR imaging ($u'g'r'i'JK_s$), and using an ML method (KNN+100; see Sect. \ref{sect:ML_literature}), provides the catalog ucd\_gc\_candidates.fits containing 1,155 sources, 453 of which lie within the NGFS-T1 FoV. The comparison with our \texttt{svm.SVC} catalog and with the full catalog yields 186 and 251 sources in common, respectively.

\item \cite{jordan2015}: The HST/ACS Fornax Cluster Survey observed 43 galaxies and measured magnitudes in the F475W (Sloan g) and F850LP (Sloan z) filters. In addition, the superb spatial resolution of HST/ACS allows reliable measurements of half-light radii, a key discriminator for separating GCs from contaminants when only a few filters are available. For each source, they estimated a GC probability using a model-based mixture approach. The dataset file ACSFCS\_gz.dat contains a total of 9,136 sources, which represents a highly trustworthy photometric catalog. Within the field of view of this work, only 14 of the ACSFCS galaxies are covered. Applying the same probability threshold of 80\% used in our analysis, 3,052 sources satisfy this requirement. The cross-match between the ACSFCS catalog and our catalogs depends on the astrometric offset, as we detected a systematic coordinate shift between ACSFCS and NGFS-T1/FDS. This was verified by overlaying our catalog with the FDS and ACSFCS sources on the NGFS-T1 $i'-$band image. To account for this offset, we adopted a $2''$ separation tolerance instead of the standard $1''$.
The resulting cross-match includes 346 sources in common with our \texttt{svm.SVC} catalog and 564 with the full catalog. The relatively small overlap (3,052 reduced to fewer than 600) is likely driven by their selection being based solely on g and z photometry, whereas our selection uses $u'g'i'JK_s$ bands. This limits our ability to recover ACSFCS sources in the shallowest bands, particularly $u'$ and $K_s$. Many objects well detected in HST/ACS appear faint in our data, leading to low-quality photometry and their exclusion from the \texttt{svm.SVC} model. Another contributing factor is that ACSFCS detects sources very close to galaxy centers, while in our ground-based data the innermost regions suffer from saturation and cannot be reliably used.
\end{enumerate}

\section{Additional Tables and Plots for Testing \texttt{svm.SVC} Model}

\subsection{Testing \texttt{svm.SVC} Model with different Split Train/Test Samples - continued}
\label{sect:A2_testsplit}

\begin{figure*}[ht]
%trim=left bottom right top
\centering
     \includegraphics[trim=1.3cm 0.6cm 1.3cm 0.4cm,clip,width=0.7\textwidth]{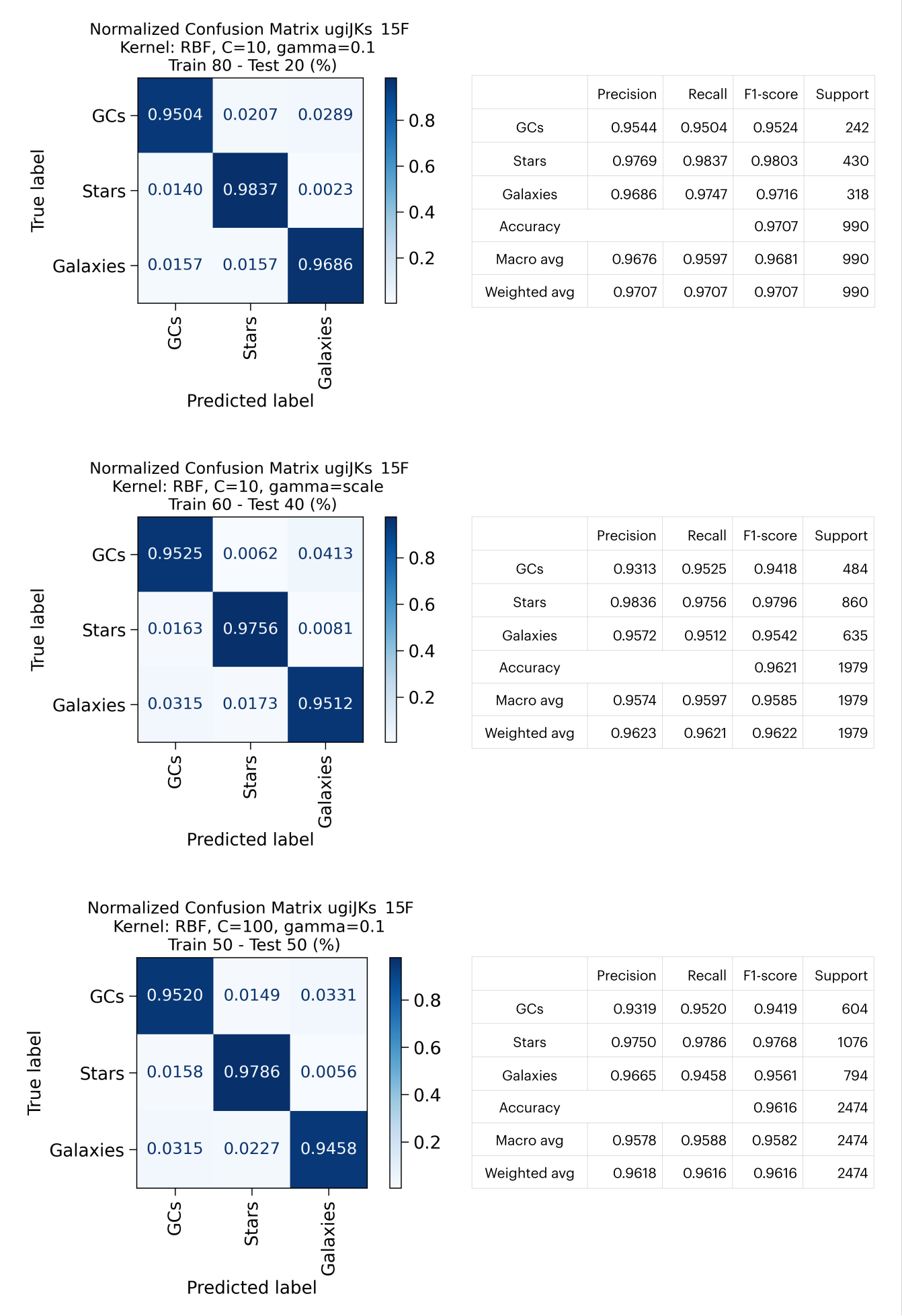}   
    \caption{Different split of the train-test labeled sample. Left Panels show the normalized confusion matrix and right panels the classification report. The splits are 80-20 \% (top row), 60-40 \% (middle row) and 50-50 \% (bottom row). Subtitles of left panel contain the model information of the kernel and parameters.}
    \label{fig:CM_poi_other_split}
\end{figure*}

In Sect. \ref{sect:impl_svm_model}, we discussed the importance of achieving optimal validation performance and highlighted that one key aspect is the random division of the labeled sample into training and testing subsets (using \texttt{train\_test\_split}). We explored several split configurations and found that the best performance is obtained with a 70\% training and 30\% testing split. In this Section we present the other combinations, such as 80/20, 60/40, and 50/50. They are shown in Fig. \ref{fig:CM_poi_other_split}.

\subsection{Testing the \texttt{svm.SVC} Classifier Using a Magnitude Constrained Train/Test Sample - continued}
\label{sect:A2_test_magcut}
\begin{figure*}[ht]
%trim=left bottom right top
\centering
     \includegraphics[trim=0.1cm 0.1cm 0.1cm 0.1cm,clip,width=0.8\textwidth]{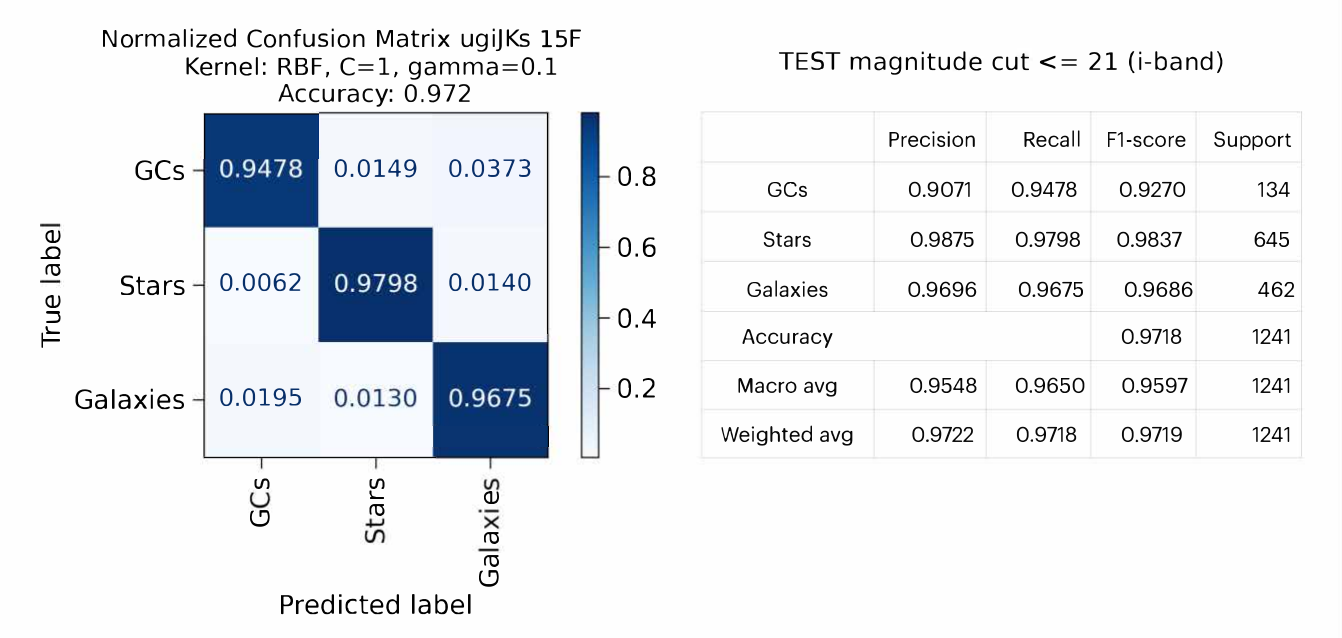}   
    \caption{Performance of the 15F model evaluated on the train-test labeled sample restricted to $mag_i \le 21$ mag. The left panels display the normalized confusion matrix, while the right panels present the classification report. The subtitles in the left panels specify the kernel and parameter configuration of the model.}
    \label{fig:CM_magcut}
\end{figure*}

This section presents the results of testing the hypothetical scenario in which the labeled sample reaches only $mag_i = 21$ mag. The performance of the 15F model under this magnitude constraint is shown in Figure \ref{fig:CM_magcut}, see Sect. \ref{sect:test_magcut} for more details.

\subsection{Testing \texttt{svm.SVC} model with fewer filter information - continued}

\begin{figure*}[ht!]
%trim=left bottom right top
     \centering
     \includegraphics[trim=3cm 0.1cm 0.8cm 0.2cm,clip,width=0.45\linewidth]{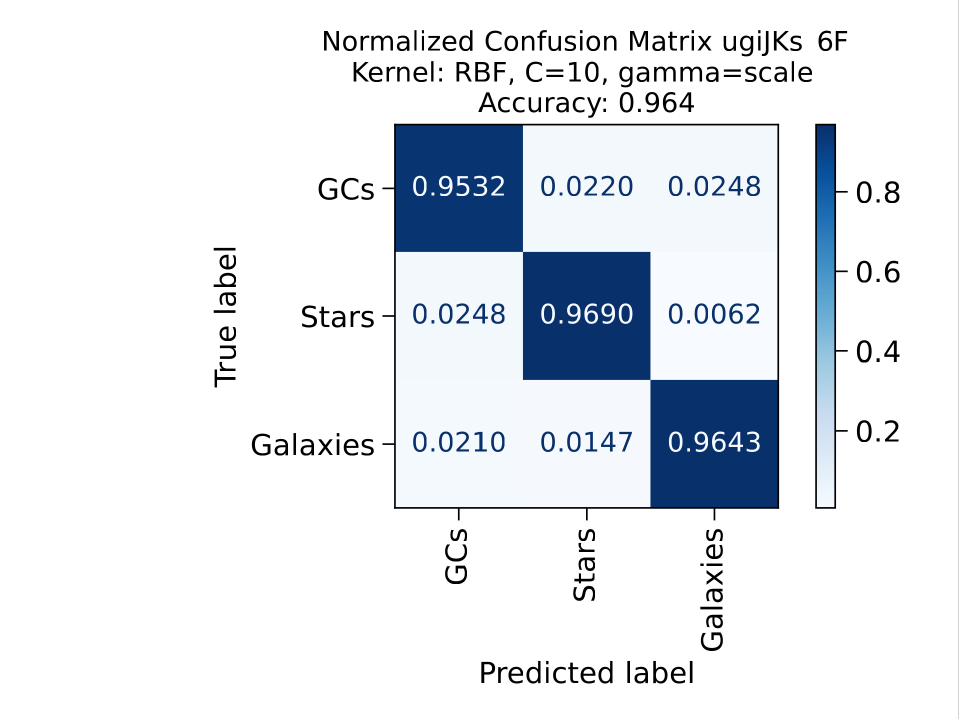} 
     \includegraphics[trim=3cm 0.1cm 0.8cm 0.2cm,clip,width=0.45\linewidth]{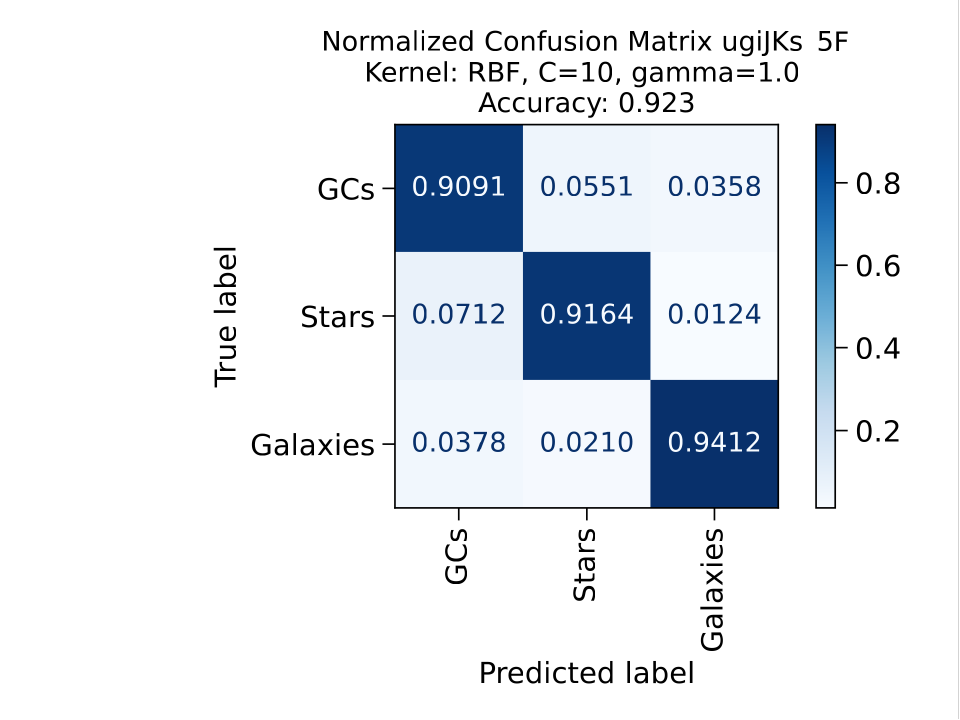} 
    \caption{Normalized confusion matrix for the test model 6F (no $u'$-band) and 5F (no NIR) shown in the top and bottom panel, respectively. See Section \ref{sect:svm_fewerfilters}}
     \label{fig:CM_6F_5F}
\end{figure*}

This section shows the complementary information for the cases where fewer filters are used in the \texttt{svm.SVC} model, see Section \ref{sect:svm_fewerfilters} for details. The confusion matrix (Fig. \ref{fig:CM_6F_5F}) and the classification report tables \ref{tab:A.1.1_cl_report_6f}, \ref{tab:A.1.2_cl_report_5f}.

\begin{table}[ht!]
\centering
\caption{Classification Report for the 6 features (6F) model, no $u'$-band: $(g'-i')$, $(i'-J)$, $(i'-Ks)$, $(J-K_s)$, $SM$, $FWHM$. See Sect. \ref{sect:svm_fewerfilters} and Fig. \ref{fig:svm_fewerfilter}, top-row.}
\label{tab:A.1.1_cl_report_6f}
\begin{tabular}{|l|c|c|c|c|}
\hline
\textbf{Class} & \textbf{Precision} & \textbf{Recall} & \textbf{F1-score} & \textbf{Support} \\
\hline
GCs       & 0.9301 & 0.9532 & 0.9415 & 363  \\
Stars     & 0.9766 & 0.9690 & 0.9728 & 646 \\
Galaxies  & 0.9725 & 0.9643 & 0.9684 & 476 \\
\hline
\textbf{Accuracy}     & \multicolumn{4}{c|}{0.9636} \\
\textbf{Macro Avg}    & 0.9597 & 0.9622 & 0.9609 & 1485 \\
\textbf{Weighted Avg} & 0.9639 & 0.9636 & 0.9637 & 1485 \\
\hline
\end{tabular}
\vspace{0.05\linewidth}
\caption{Classification Report for the 5 features (5F) model, no NIR: $(u'-g')$, $(u'-i')$, $(g'-i')$, $SM$, $FWHM$. See Sect. \ref{sect:svm_fewerfilters} and Fig. \ref{fig:svm_fewerfilter}, bottom-row.}
\label{tab:A.1.2_cl_report_5f}
\begin{tabular}{|l|c|c|c|c|}
\hline
\textbf{Class} & \textbf{Precision} & \textbf{Recall} & \textbf{F1-score} & \textbf{Support} \\
\hline
GCs       & 0.8376 & 0.9091 & 0.8719 & 363  \\
Stars     & 0.9518 & 0.9164 & 0.9338 & 646  \\
Galaxies  & 0.9552 & 0.9412  & 0.9481 & 476 \\
\hline
\textbf{Accuracy}     & \multicolumn{4}{c|}{0.9226} \\
\textbf{Macro Avg}    & 0.9149 & 0.9222 & 0.9179 & 1485 \\
\textbf{Weighted Avg} & 0.9250 & 0.9226 & 0.9232 & 1485 \\
\hline
\end{tabular}
\end{table}

\subsection{Application: testing the method for LSST filter system - continued}

The following tables \ref{tab:clreport_20F_lsst}, \ref{tab:clreport_12F_lsst} and \ref{tab:cl_report_8F_lsst} show the classification report for the three models with different features, see Section \ref{sect:lsst}.

\begin{table}[ht!]
\centering
\caption{Classification Report for the LSST filter system - 20F model (sect.\ref{sect:lsst}). See Fig. \ref{fig:result_cc_lsst}, top-row panels}.
\vspace{-0.06\linewidth}
\label{tab:clreport_20F_lsst}
\begin{tabular}{|l|c|c|c|c|}
\hline
\textbf{Class} & \textbf{Precision} & \textbf{Recall} & \textbf{F1-score} & \textbf{Support} \\
\hline
GCs       & 0.9209 & 0.9846 & 0.9517 & 260  \\
Stars     & 0.9825 & 0.9640 & 0.9731 & 639  \\
Galaxies  & 0.9717 & 0.9591 & 0.9654 & 465 \\
\hline
\textbf{Accuracy}     & \multicolumn{4}{c|}{0.9663} \\
\textbf{Macro Avg}    & 0.9583 & 0.9693 & 0.9634 & 1364 \\
\textbf{Weighted Avg} & 0.9670 & 0.9663 & 0.9664 & 1364 \\
\hline
\end{tabular}
\vspace{0.05\linewidth}
\caption{Classification Report for the LSST filter system - 12F model, no $u'$-band (sect.\ref{sect:lsst}). See Fig. \ref{fig:result_cc_lsst}, middle-row panels.}
\vspace{-0.03\linewidth}
\label{tab:clreport_12F_lsst}
\begin{tabular}{|l|c|c|c|c|}
\hline
\textbf{Class} & \textbf{Precision} & \textbf{Recall} & \textbf{F1-score} & \textbf{Support} \\
\hline
GCs       & 0.9170 & 0.9769 & 0.9460 & 260  \\
Stars     & 0.9690 & 0.9781 & 0.9735 & 639  \\
Galaxies  & 0.9774 & 0.9290 & 0.9526 & 465 \\
\hline
\textbf{Accuracy}     & \multicolumn{4}{c|}{0.9611} \\
\textbf{Macro Avg}    & 0.9544 & 0.9613 & 0.9574 & 1364 \\
\textbf{Weighted Avg} & 0.9619 & 0.9611 & 0.9611 & 1364 \\
\hline
\end{tabular}
\vspace{0.05\linewidth}
\caption{Classification Report for the LSST filter system - 8F model, no $u'$-band and $Y$-band (sect.\ref{sect:lsst}). See Fig. \ref{fig:result_cc_lsst} bottom-row panels.}
\label{tab:cl_report_8F_lsst}
\begin{tabular}{|l|c|c|c|c|}
\hline
\textbf{Class} & \textbf{Precision} & \textbf{Recall} & \textbf{F1-score} & \textbf{Support} \\
\hline
GCs       & 0.8996 & 0.9654 & 0.9314 & 260  \\
Stars     & 0.9596 & 0.9671 & 0.9634 & 639  \\
Galaxies  & 0.9751 & 0.9247 & 0.9492 & 465 \\
\hline
\textbf{Accuracy}     & \multicolumn{4}{c|}{0.9523} \\
\textbf{Macro Avg}    & 0.9448 & 0.9524 & 0.9480 & 1364 \\
\textbf{Weighted Avg} & 0.9535 & 0.9523 & 0.9524 & 1364 \\
\hline
\end{tabular}
\vspace{0.05\linewidth}
\end{table}

\end{appendix}

\end{document}